\RequirePackage{fix-cm}
\documentclass[smallextended]{svjour3}       
\smartqed  
\usepackage{graphicx}

\usepackage{amsmath}
\usepackage{amsthm}
\usepackage{algpseudocode}
\usepackage[ruled, linesnumbered]{algorithm2e} 

\SetCommentSty{mycommfont}
\usepackage{amssymb}
\usepackage{hyperref}
\usepackage{hypcap}
\usepackage{multirow}
\usepackage{subcaption}
\usepackage{color}
\usepackage{booktabs} 
\usepackage{rotating}
\usepackage{pdflscape}
\begin{document}
	
	\title{Mining user interaction patterns in the darkweb to predict enterprise cyber incidents}

	
	\author{Soumajyoti Sarkar         \and Mohammad Almukaynizi \and Jana Shakarian \and Paulo Shakarian}
	
	
	\institute{S. Sarkar, M. Almukaynizi, P. Shakarian \at
		Arizona State University \\
		\email{ssarka18@asu.edu}           \\
		\email{malmukay@asu.edu}            \\
		\email{shak@asu.edu}            \\
		\and
		J. Shakarian \at
		Cyber Reconnaissance Inc. \\
		\email{jana@cyr3con.ai} 
	}
	
	\date{Received: date / Accepted: date}

	\maketitle
	
	\begin{abstract}
		With rise in security breaches over the past few years, there has been an increasing need to mine insights from social media platforms to raise alerts of possible attacks in an attempt to defend conflict during competition. In this study, we attempt to build a framework that utilizes unconventional signals from the darkweb forums by leveraging the reply network structure of user interactions with the goal of predicting enterprise related external cyber attacks. We use both unsupervised and supervised learning models that address the challenges that come with the lack of enterprise attack metadata for ground truth validation as well as insufficient data for training the models. We validate our models on a binary classification problem that attempts to predict cyber attacks on a daily basis for an organization. Using several controlled studies on features leveraging the network structure, we measure the extent to which the indicators from the darkweb forums can be successfully used to predict attacks.  We use information from 53 forums in the darkweb over a span of 17 months for the task. Our framework to predict real world organization cyber attacks of 3 different security events, suggest that focusing on the reply path structure between groups of users based on random walk transitions and community structures has an advantage in terms of better performance solely relying on forum or user posting statistics prior to attacks.  
	\end{abstract}
	
	\section{Introduction}
	
	\label{sec:intro}
	With recent data breaches such as those of Yahoo, Uber, Equifax \footnote{https://www.consumer.ftc.gov/blog/2017/09/equifax-data-breach-what-do, https://www.consumer.ftc.gov/blog/2016/09/yahoo-breach-watch} among several others that emphasize the increasing financial and social impact of cyber attacks, there has been an enormous requirement for technologies that could provide such organizations with prior alerts on such data breach possibilities. Such security threat intelligence information would help address the following: (1) while organizations spend a lot of money to secure network systems that could avoid such data breaches \cite{liu_breach}, it is not devoid of exposures to vulnerabilities specially as such platforms depend on a large number of third party software systems. (2) an alert for a possible intrusion into technology platforms like email servers or malware injection into softwares could actually help organizations focus on a specific set of components in a short time, thereby allowing faster security tightening to avoid being exploited on a regular basis \cite{risk_analysis}. 
	
	The total number of data breaches in 2017 crossed 1000\footnote{https://www.wombatsecurity.com/blog/scary-data-breach-statistics-of-2017} across all sectors which is a record high, considering previous years and exposing over a billion records containing sensitive data. On the vulnerability front, the Risk Based Security's VulnDB database \footnote{https://www.riskbasedsecurity.com/2017/05/29-increase-in-vulnerabilities-already-disclosed-in-2017/} published a total of 4837 vulnerabilities in a quarter of 2017 which was around 30\% higher than previous year. This motivates the need for an extensive application that can track vulnerability based information from external sources to raise alerts on such data breaches. While the darkweb is one such place on the internet where users can share information on software vulnerabilities and ways to exploit them \cite{chen_top,sagar} and where it might be difficult to track the actual identity of those users, what they leave behind are the footprints of their posting and interaction patterns in forums. In this paper, as one of contributions in the field of cyber attack prediction, we leverage the information obtained from evolving reply networks of discussions in the darkweb forums while also capturing the user and thread posting statistics in these forums to understand the extent to which the darkweb information can be useful as signals for predicting real world target specific enterprise cyber attacks.
	
	In the vulnerability lifecycle, a vulnerability goes through multiple stages. It starts with being undisclosed when the general public does not know about it and attackers can identify them, develop exploits and use them for ``zero-day" attacks. However, once a vulnerability is identified, an indexing is done with an ID assigned to it - that is where the vendor starts working on a patch. Once a patch is released, is when hackers would try to reverse engineer the patch and develop exploits. The last stage would generally entail using metasploit modules to launch these attacks via these exploits \cite{almuk1}. So in this vulnerability lifecycle, the significance of discussions in darkweb forums or most social media platforms can appear in two phases: once when the vulnerability is undisclosed and there are discussions revolving around related vulnerabilities or exploits and second, after the patch is released but before an exploit is materialized for an attack. So our goal is to leverage the discussions in these two phases to be able to predict cyber attacks before the exploit is weaponized.
	
	We attempt to build an integrated approach utilizing unconventional signals from the darkweb discussions for predicting attacks on a target organization  - here ``unconventional" means that the information from the darkweb might not necessarily be observables of the actual attacks on the target organization. This is in contrary to traditional studies where authors use system level features within the target organization to predict attacks in future for the same or related organization \cite{liu_cyber,2017riskteller}.   With this in mind, we hypothesize that the interaction dynamics focused on a set of specialized users and the attention broadcast by them to other posts in these underground platforms can be one of ways to generate warnings for future attacks.  We mine patterns of anomalous behavior  from these forums and use them directly for cyber attack prediction on external enterprise host systems. 
	We note that we \textit{do not} consider whether vulnerabilities mentioned in forum discussions, have been exploited or not as the basis for attacks since a lot of zero day attacks \cite{Bilge:2012:BWK:2382196.2382284} might occur before such vulnerabilities are even indexed and their gravity might lie hidden in discussions related to other associated vulnerabilities or some discussion on exploits. The premise on which this research is setup is based on evaluating the dynamics of all kinds of discussions in the darkweb forums but we attempt to filter out the noise to mine important patterns by studying whether a piece of information gains traction within important communities. So in this sense we do not explicitly focus on discussions relating vulnerabilities exclusively nor their exploits to predict real world cyber attacks.
	
	We try to quantify the correlation between the pattern of replies by a specific group of users we term $experts$ who engage more frequently with popular vulnerability mentions in their posts over time and which gain attention from other users, and a real world cyber attack in the near future as a first challenge in this study. A second research opportunity in this direction is to see whether we can use company agnostic unsupervised models that overcome the lack of company specific metadata from the attack ground truth. We investigate the extent to which we can correlate anomalies from these darkweb network interactions to near term cyber attacks and how well they materialize for different companies. To this end, the major contributions of this research investigation are as follows:
	
	\begin{itemize}
		\item We create a novel network mining technique using the directed reply network of users to extract a set of specialized users we term $experts$ whose posts with popular vulnerability mentions gain attention from other users in a specific time frame. Following this, we  generate several time series of features that capture the dynamics of interactions centered around these $experts$ across individual forums of the darkweb.
		\item We apply a widely used unsupervised anomaly detection technique that uses residual analysis to detect anomalies and propose an anomaly based attack prediction technique on a daily basis. Additionally, we also train a supervised learning model based on logistic regression with attack labels from an organization to predict daily attacks.
		\item Empirical evidence from our unsupervised anomaly detector suggests that a feature based on graph conductance that measures the random walk transition probability between groups of users is a useful indicator for attack occurrences  given that it achieved the best AUC score of 0.69 for one type of attacks. We obtain similar best results for the supervised model having the best F1 score of 0.53 for the same feature and attack type compared to the random (without prior probabilities) F1 score of 0.37. We additionally investigate the performance of the models in weeks where frequency of attacks are higher and find the superior performance of community structures in networks in predicting these attacks.
	\end{itemize}
	
	To the best of our knowledge, this is a first attempt in creating a framework that investigates the network structure of the darkweb forums data as an external source of information to generate alerts and predict real world cyber attacks without having the need to monitor vulnerability prioritization or exploitation. 
	
	\begin{table}[!t]
		\centering
		\begin{tabular}{|p{1.1cm}|p{10cm}|}
			\hline
			\textbf{Symbols}           & \textbf{Definition}                                                                                                        \\ \hline \hline
			f, F                       & a particular (single) forum, set of forums considered in study                                                             \\ \hline
			t                          & discrete time point instance  \\ \hline
			A                          & set of attack types: malicious-email, endpoint malware and malware destination  \\ \hline                                                                                                     
			$\Gamma$                   & global time range of  points in our study $\{t_1, t_2 \ldots t_k\}$                                                        \\ \hline
			$\tau$                     & an ordered subsequence of time points $\in \Gamma$                                                                                  \\ \hline
			x                          & feature in our study of machine learning based prediction models                                                           \\ \hline
			h                          & a thread in a forum ( a thread is a series of posts on a particular topic initiated by a user)                             \\ \hline
			$p_{h, i}$                         & in a chronologically ordered set of posts  in thread $h$, it denotes the $i^{th}$ post (from beginning) in $h$                     \\ \hline
			$\mathcal{T}_{x, f}$       & a time series data for feature $x$ from discussion posts in forum $f$                                                      \\ \hline
			$\mathcal{R}_{x}$       & residual vector time series data for feature $x$ from discussion posts aggregated over all forums                                                    \\ \hline
			$H_{\tau}$                 & historical time period (prior to $\tau$) w.r.t. $\tau$, $\forall t' \in H_\tau$ and for any $ t \in \tau$, $t'< t$ and there is a time gap between the start of $\tau$ and end of $H_\tau$                                                                      \\ \hline
			$V^f_{\tau}$, $E^f_{\tau}$ & set of nodes in reply network from forum $f$ discussions and in time period $\tau$, set of edges with the same constraints (we drop $f$ when we generalize for all forums) \\ \hline
			$G_{H_{\tau}}$             & Reply network induced by discussion in $H_{\tau}$                                                                          \\ \hline
			$exp_{\tau}$               & Experts in the time subsequence $\tau$                                                                                     \\ \hline
			\textbf{X}                 & \#features $\times$ T matrix (T denotes the time dimesnion)                                                                \\ \hline
			\textbf{Y}                 & T $\times$ F  matrix                                                                                                       \\ \hline
			\textbf{y}                 & 1 $\times$ F vector                                                                                                        \\ \hline
			$\beta$                    & weight for a feature in the logistic regression model                                                                      \\ \hline
			$\eta$                     & time window for feature selection in $\mathcal{T}_{x, f}$                                                                  \\ \hline
			$\delta$                   & time gap between attack prediction at a time point and the feature window                                                  \\ \hline
			$\zeta$                   & anomaly to attack prediction (anomaly) count threshold parameter                                                  \\ \hline
		\end{tabular}
		\caption{Table of notations}
		\label{tab:tab_symbols}
	\end{table}
	
	\section{Related Work and Motivation}
	In this work, we discuss some  of the past and ongoing research in the domain of cyber security analytics that also caters to the general area of predicting future cyber breach incidents in real world systems. Most of the work on vulnerability discussions on trading, exploitation in the underground forums \cite{allodi2017economic,edkrantz2015predicting,miller2007legitimate}  and related social media platforms like Twitter\cite{sapienza2018discover,Khandpur,sabottke2015vulnerability} have focused on two aspects: (1) analyzing the dynamics of the underground forums and the markets that drive it, thereby focusing on mechanisms that enable the market activity, and giving rise to the belief that the ``lifecycle of vulnerabilities" in these forums and marketplaces have significant impact on real world cyber attacks \cite{kotenko2005analyzing,Bilge:2012:BWK:2382196.2382284} (2)  prioritization of vulnerabilities using these social media platforms or binary file appearance logs of machines and using them to predict the risk state of machines or systems through exploitation of these vulnerabilities \cite{2017riskteller}. So, the two components in majority of these studies that have been repeatedly worked upon in silos are analysis of vulnerabilities and their likelihood of exploitation in these forums or platforms and, then vulnerability exploitation severity based prediction to associate them to real world cyber breach incidents \cite{almuk1,sabottke2015vulnerability}. In this paper, we ignore the gap between vulnerability exploit analysis and the final task of real world cyber attack prediction by removing the preconceived notions used in earlier studies where vulnerability exploitation is considered a precursor towards attack prediction.  \\
	
	The rapid expansion of the cyber-threat landscape is augmented by the presence of underground platforms that support the discussion, proliferation of exploit awareness, deployment and monetization of such exploits leading to cyber-attacks \cite{grier2012manufacturing,herley2010nobody,allodi2016then,allodi2017economic}. However, despite the existing literature that studies the economies of these underground forums and markets present in the darkweb, there has been very few studies that focus on filtering the markets and forums that actually contribute to the threat scenario \cite{yip2013forums,sood2013crime,shakarian2016exploring}. One of the ways to understand the indicators surrounding these underground platforms, that could lead to potentially malicious attempts to breach systems at scale is to monitor the interactions that receive attention in these platforms. \\
	
	We discuss 3 areas within which our work falls when we discuss the landscape of cyber attack prediction based on signals from social media and attacks on an organization. However, we point out the main differences that bring out the significance and novelty of our approach and the problem we attempt to solve in the following:
	\begin{enumerate}
		\item \textit{Cyber attack with within-organization system signals}: Cyber attack prediction on external organizations have recently been studied in the context of feature engineering for gathering predictive signals. Some of the most related works in this area include a study \cite{liu_breach}, where features are gathered from the network systems and the log files of a target organization. These features are then used for training a classifier to predict future attacks for the same organization, and where the ground truth for the attacks are collected from reported cyber incidents from Web Hacking Database, Hackmageddon. Contrary to this, we use unconventional signals from the darkweb that are not necessarily observables of the attacks for the organization but we try to measure the extent to which they can perform well over other measures. Similar to this study, there have already been attempts to develop systems at scale that could predict the risk of systems by analyzing various sensors such as binary appearance of log files \cite{2017riskteller}. \\
		
		\item \textit{Cyber attack prediction using social media data}: There have been several attempts to use external social media data sources to predict real world cyber attacks \cite{liu_breach,liu_cyber,Khandpur,colbaugh2011proactive}. However, the problem these studies focus on is to build predictive models to correlate the social media signals to attacks in the real world that are not observed for a specific organization. Our attack prediction problem specifically proposes to build models specific to an external organization using external sensors not obtained from the internal system data for the same organization. One of the closest works in this area is done bu authors in \cite{okutan_2018}, where the authors use signals using GDELT, Twitter and OTX based on keywords related to the organization. One of the challenges related to our dataset is that we did not find any keywords directly related to the name of our target organization in the darkweb - similar issues are reported in \cite{palash} where the authors relied on some curated keyword search from Twitter and blogs and the darkweb for attack prediction Our work has a slight advantage in that our selection of forums and the features including vulnerability information does not depend on human engineered knowledge, rather it focuses on the trends in time - so in a sense our streaming nature of prediction is scalable.\\

		\item \textit{Social Network analysis for cyber security}:  Using network analysis to understand the topology of darkweb forums has been studied at breadth in \cite{philips_dark} where the authors use social network analysis techniques on the reply networks of forums in order to identify members of Islamic community within the darkweb. Similarly in \cite{topics_sna}, the authors use topic modeling and the network structure of the darkweb forums in order to understand the interactions between extremist groups. However, such analysis of reply networks have been conducted on static networks \cite{almuk2} where authors devised network features of users for predictive modeling. A recent study done in \cite{sarkarcyber} show how to leverage the network structure of these reply networks for cyber attack prediction. These studies suggest that the nature of interactions can unveil important actors in darkweb forums and their activity regarding discussions can provide us with signals for cyber attacks.  One of our contributions in this paper is that we use evolving networks of the users with certain constraints that can now be leveraged for streaming prediction on  a daily basis in an automated manner. Our hypothesis lies on the premise that the attention broadcast by these users towards other posts are in fact sensors for impending cyber attacks. Such studies of separating specialized users have been studied before in the context of trading financial information in carding forums \cite{haslebacher2017all}. \\
	\end{enumerate}
	The rest of the paper is organized as follows: we first introduce a few security terminologies relevant to our work and the dataset sources and attributes in Section~\ref{sec:prelim}, following which we formally define the prediction problem attempted in this paper in Section~\ref{sec:pred_problem}. We then discuss the technical details of our attack prediction framework including the feature engineering and the model learning components in Section~\ref{sec:framework}. We discuss the experimental settings and the results in Section~\ref{sec:exp} and finally we end this work with some discussion and case studies in Section~\ref{sec:discuss}.
	
	\section{Cyber Security terms and Dataset}
	\label{sec:prelim}
	We first introduce a few terms commonly used in the cyber security domain and that we would use in this paper frequently. Vulnerability is a weakness in a software system that can be exploited by an attacker to compromise the confidentiality, integrity or availability of the system to cause harm \cite{Pfleeger}.\\
	\noindent \textit{Common Vulnerabilities and Exposures (CVE) :} The database of Common Vulnerabilities and Exposures maintained on a platform operated by the MITRE corporation\footnote{https://www.mitre.org/} provides an identity mapping for publicly known information-security vulnerabilities and exposures. \\
	\noindent \textit{Common Platform Enumeration (CPE): } A CPE is a structured naming scheme for identifying and grouping clusters of information technology systems, software and packages maintained in a platform NVD (National Vulnerability Database) operated by NIST\footnote{https://www.nist.gov/}. \\
	\noindent \textit{CVE - CPE mapping: } Each CVE can be assigned to different CPE groups based on the naming system of CPE families as described in \cite{almuk2}. Similarly, each CPE family can have several CVEs that conform to its vendors and products that the specific CPE caters to. For the purpose of this paper, we form a simplified grouping hierarchy to cluster the CVEs by their CPE levels which we describe in Section~\ref{sec:CPE groups}. \\
	\noindent \textit{Forum topic: }  Each darkweb forum or site $f$ consists of several threads $h$ initiated by a specific user and over time, several users post and reply in these threads. We note that one user can appear multiple times in the sequence of posts depending on when and how many times the user posted in that thread. Since each thread is associated with a topic (or a title), we would often use the terms topic to refer to a particular thread $h$ comprising all posts in the relevant forum. We denote the set of these 53 forums used in this dataset using the symbol $F$.
	
	The ground truth and the darkweb data have been collected from two different sources as will be described in the following sections and although we validate our prediction models based on the available ground truth, we perform extensive case studies to show the significance of our prediction models in the real world.
	
	\begin{figure*}[!t]
		\centering
		\minipage{0.5\textwidth}
		\includegraphics[width=6cm, height=4cm]{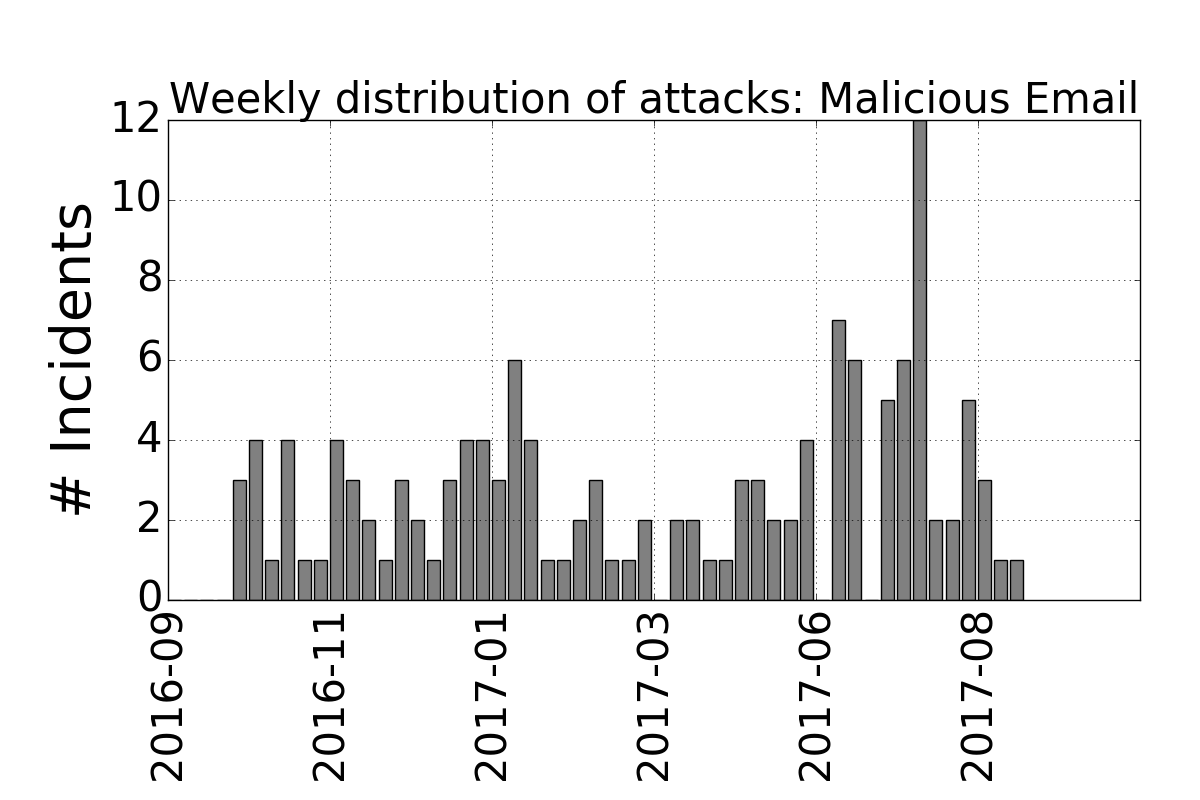}
		\subcaption{}
		\endminipage
		\hfill
		\minipage{0.5\textwidth}
		\includegraphics[width=6cm, height=4cm]{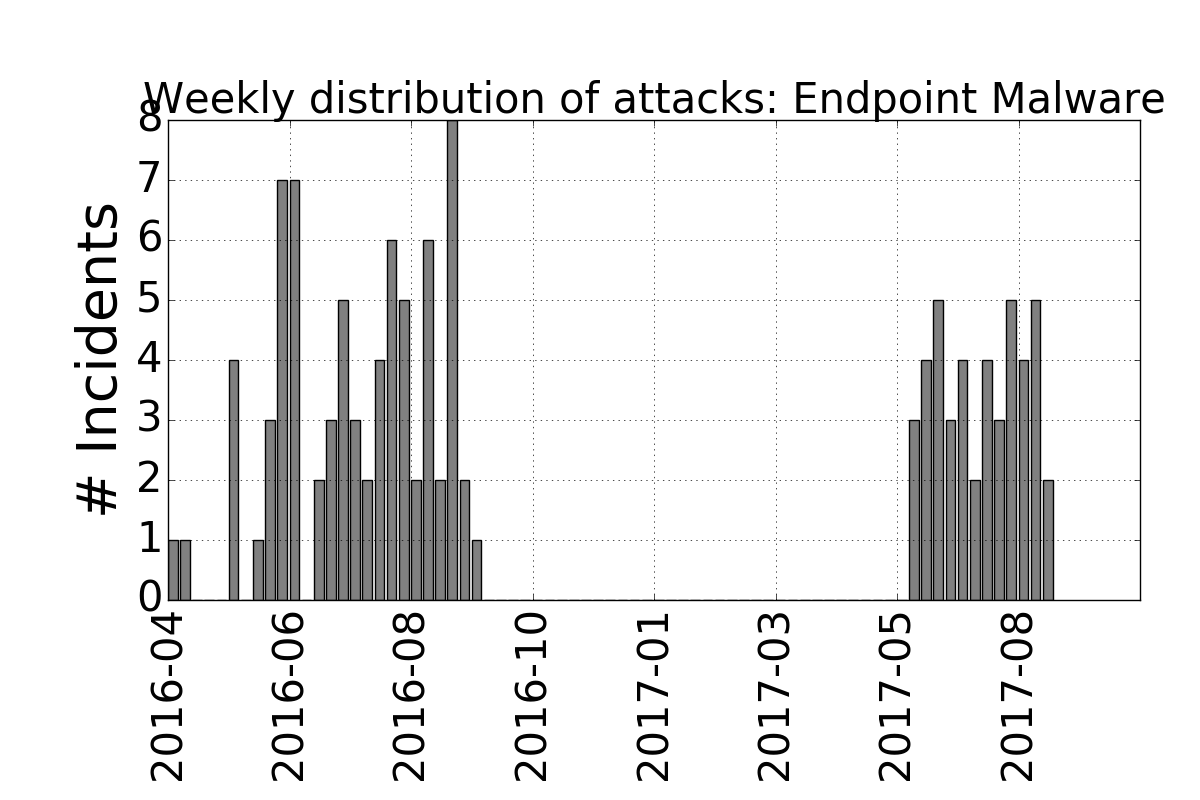}
		\subcaption{}
		\endminipage
		\hfill
		\\
		\minipage{0.5\textwidth}
		\includegraphics[width=6cm, height=4cm]{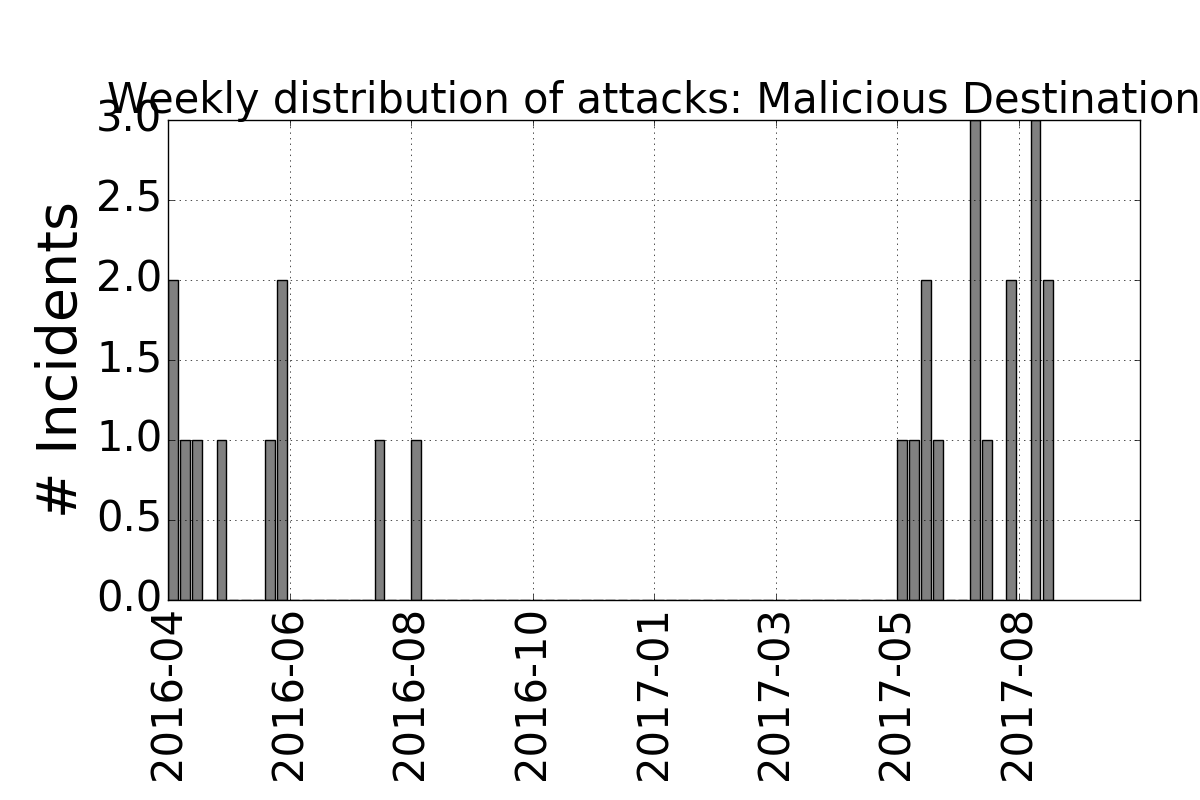}
		\subcaption{}
		\endminipage
		\hfill
		\caption{Weekly occurrence of security breach incidents of different types (a) Malicious email (b) Endpoint Malware (c) Malicious destination}
		\label{fig:types_events}
	\end{figure*}

	\subsection{Enterprise-Relevant External Threats (GT)}
	We use the cyber attacks Ground Truth (GT) from the data provided from a corporate entity to funders of this work\footnote{https://www.iarpa.gov/index.php/research-programs/cause}. The corporate entity is \textit{Armstrong  Corporation}  to conceal the actual identity.  The data contains information on cyber attacks on their systems in the period of April 2016 to September 2017. Each data point is a record of a detected deliberate malicious attempt to gain unauthorized access, alter or destroy data, or interrupt services or resources in the environment of the participating organization. Those malicious attempts were real-world events detected in the wild, in uncontrolled environment, and by different attack detectors such as anti-virus and IDS software and hardware products. The data contains the following relevant attributes: \textbf{$\{$} \textit{event-type}: The type of attack which are categorized as malicious-email, endpoint-malware and malicious-destination, \textit{event occurred date}: Date on which there was an attack of particular event-type, \textit{event reported date}: Date on which the attack was reported, \textit{detector}: the software service that detected the system intrusion attempting to break into their systems, \textit{threat\_designation\_family}: the categories of threats from among a Threat Family Dictionary. $\mathbf{\}}$. The \textit{event-types} that are used in this study are:
	
	\begin{itemize}
		\item \textit{Malicious Email}: A malicious attempt is identified as a Malicious Email event if an email is received by the organization, and it either contains a malicious email attachment, or a link (embedded URL or IP address) to a known malicious destination.
		\item \textit{Malicious Destination}: A malicious attempt is identified	as visit to a Malicious Destination if the visited URL or IP address hosts malicious content.
		\item \textit{Endpoint Malware}: A Malware on Endpoint event is identified if malware is discovered on an endpoint device. This includes, but not limited to, ransomware, spyware, and adware.
	\end{itemize}
	
	We denote these set of attack types as $A$. Here the term ``malicious" means that the end goal of all these 3 attempts were to intrude the systems of the host enterprise and exploit them, however to what extent are they successful is not known and is not a matter of concern for an incident to be qualified as a cyber-attack. In our research, we use the categories: \textit{event-type} and \textit{attack occurred date} as our ground truth (GT) for validation and avoid the use of other attributes present in the dataset as they are metadata provided by third party software services which are not available for all security incident reports. Additionally, since our research is focused on using the darkweb as an external source of data to capture the behavioral patterns of user interactions, we only use the \textit{event-type} and \textit{event-occurred-date} as our ground truth. We note that the absence of information that can accurately provide us with information regarding vulnerabilities and exploits that caused the attacks, for our model validation makes the problem more challenging.
	As shown in Figure~\ref{fig:types_events}, the distribution of attacks over time is different for the 3 events. Additionally, we also observe that for the events \textit{endpoint-malware} shown in Figure~\ref{fig:types_events}(b) and \textit{malicious-destination} shown in Figure~\ref{fig:types_events}(c), the weekly occurrence has not been captured consistently and there is missing information for these events in few time intervals. We take note of this while building our learning models to predict the occurrence of an attack. The total number of incidents reported for the events are as follows: 26 incidents tagged as \textit{malicious-destination}, 119 tagged as \textit{endpoint-malware} and 135 for \textit{malicious-email} events resulting in a total of 280 incidents over a span of 17 months that were considered in our study.

	\subsection{Darkweb data}

	The entire focus of this research has been to disentangle the interactions centered around a few users over time and the noise that is present in the form of random discussions in different forums. It helps us at assessing whether they could be used as indicators for external cyber attacks in future \textit{without} having any knowledge of which CVEs or a group of CVEs might cause them. Of course, in retrospective causal analysis, one can analyze the features that led to the predictions of an attack or an alarm for an attack in the future.
	
	However, since the attack data from Armstrong had no references to any CVEs nor was it possible to trace any CVEs given the metadata information, we resort to using the time frame of the GT attacks for gathering the darkweb forum information and computing features based on this time frame so as to train the models using data having close temporal associations. However, we later provide a comprehensive discussion related to how our predictions correlate with important time events of attacks in the real world and that also correlate the attack ground truth data provided. We obtain darkweb and deepweb data through a commercial platform\footnote{Data is provided by Cyber Reconaissance, Inc., www.cyr3con.ai}. \\
	
	\begin{figure}[!t]
		\centering
		\includegraphics[width=6cm, height=4cm]{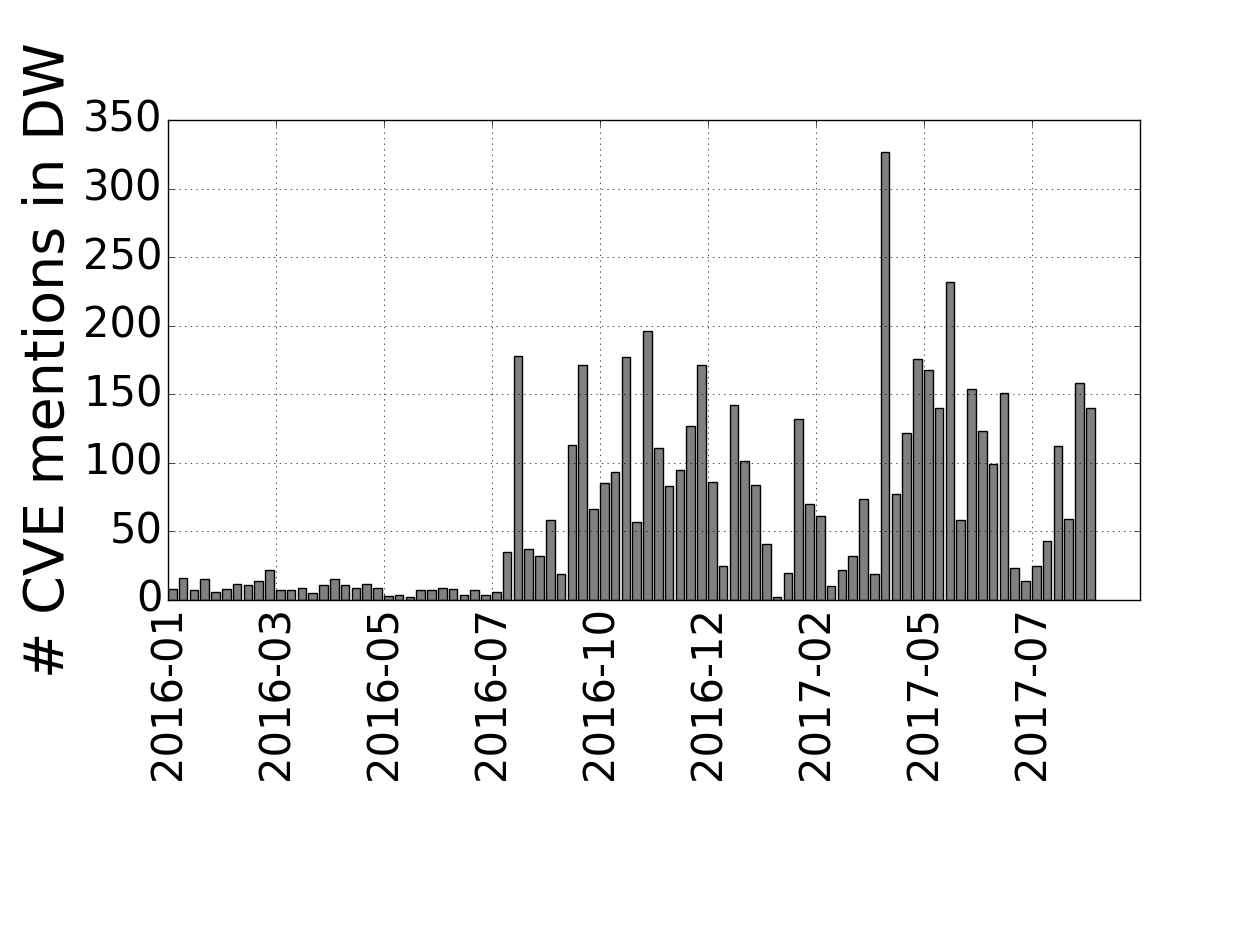}
		\caption{Weekly distribution of unique vulnerability mentions in darkweb forums posts across all forums.}
		\label{fig:vulnMention}
	\end{figure}
	
	\noindent \textit{CVE data:} Using the API, we collect all the information regarding the vulnerability mentions in the darkweb forums in the period from January 2016 to October 2017. The total number of unique CVE mentions in this period are 3553 across all forums that are scraped by the SDK and the weekly distribution of the number of vulnerability mentions in the forums is shown in Figure~\ref{fig:vulnMention}. We realize that for the months from January 2016 to May 2016, there may be a collection bias in the vulnerability mentions in forum posts but since we train our models over multiple months using these mentions, we hope to overcome this collection bias error over time. \\
	
	\begin{figure}[!h]
		\centering
		\includegraphics[width=6cm, height=4cm]{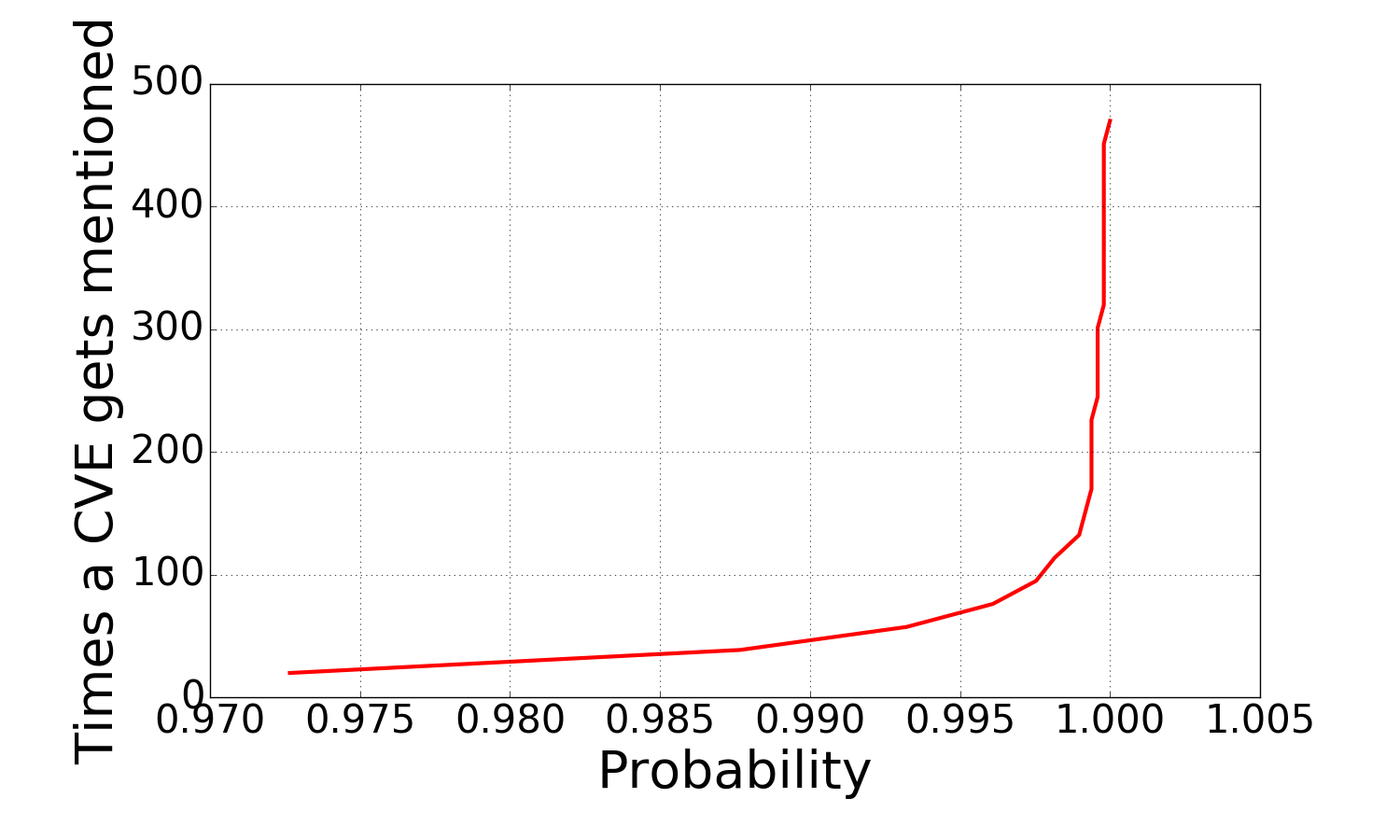}
		\caption{Cumulative Distribution  Function (cdf) showing the number of times each CVE is mentioned in posts in the darkweb.}
		\label{fig:vulnMention_cdf}
	\end{figure}
	
	In fact, when we look at the distribution of the number of times one CVE is mentioned in the darkweb (over the span of the time period of our study that we considered), we found that on average a CVE would be mentioned 3.5 times in the darkweb forums, with a median value of 1. Figure~\ref{fig:vulnMention_cdf} shows that the probability of less mentions is very huge thus making the problem of selecting discussions surrounding vulnerabilities even more difficult since prioritizing vulnerabilities without looking at the content or the user-user network is very difficult.

	\begin{figure}[!h]
		\centering
		\includegraphics[width=6cm, height=4cm]{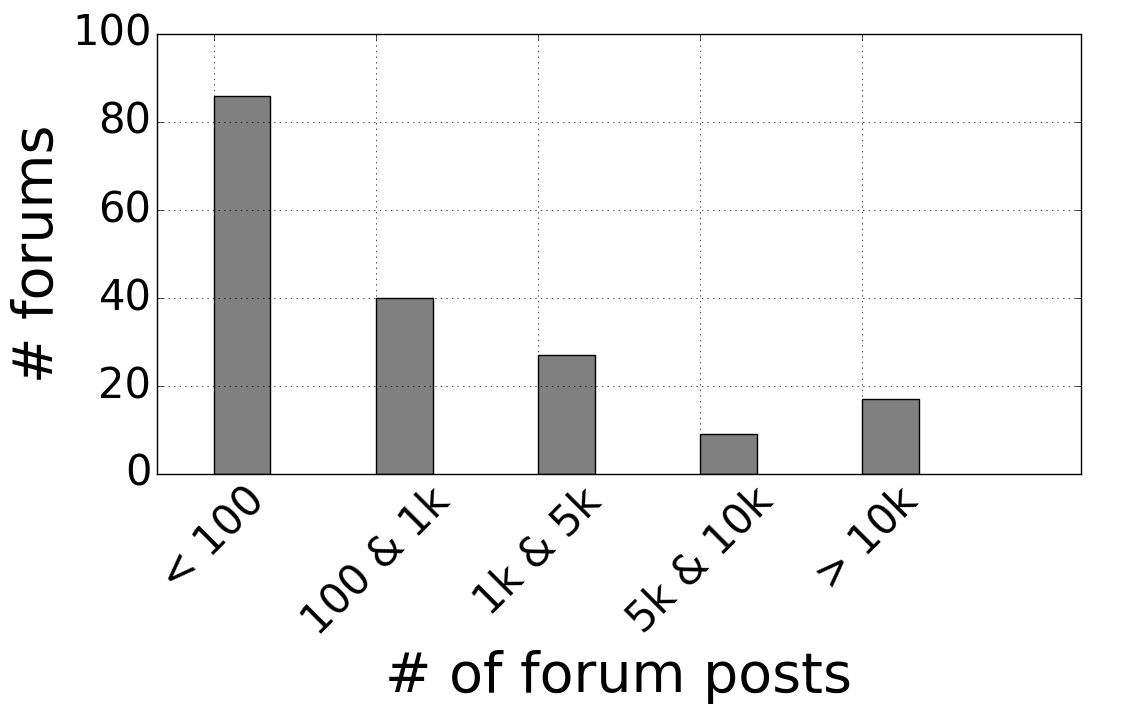}
		\caption{Distribution of the number of forums posts across all forums}
		\label{fig:postsDist}
	\end{figure}
	
	\noindent \textit{Forums data:} In this paper, we consider the dynamics of interactions in darkweb forums and for that we filter out forums based on a threshold number of posts that were created in the timeframe of January 2016 to September 2017. We gathered data from 179 forums in that time period where the total number of unique posts irrespective of the thread that they belonged to, were 557689. As shown in Figure~\ref{fig:postsDist}, the number of forums with less than 100 posts is large and therefore we only consider forums which have greater than 5000 posts in that time period which gave us a total of 53 forums. As will be described later, we rely on a projection method to compute lower dimensional features and hence any significant patterns occurring out of these forums would be captured without the requirement to manually filter and select particular forums. We note that unlike some related research using darkweb for cyber attack prediction which use large number of forums for obtaing signals for prediction \cite{palash}, we refrain from using forums with not enough data in the 1-year period of our study. This is to avoid the issues of missing data on days where we would need to predict attacks - an imputation measure for this is an active area of research \cite{okutan_2018} and we consider this as a step towards our future work. \\

	
	\subsection{CPE Groups}\label{sec:CPE groups}
	We gather the CPE data for all the vulnerabilities relevant to the darkweb discussions in our study from the publicly available repository of CPE data. In order to cluster the set of CVEs into a set of CPE groups, we use the set of CPE tags for each CVE from the NVD database maintained by NIST. For the CPE tags, we only consider the operating system platform and the application environment tags for each unique CPE. Examples of CPE would include: \textit{Micorsoft Windows\_95}, \textit{Canonical ubuntu\_linux}, \textit{Hp elitebook\_725\_g3}. The first component in each of these CPEs denote the operating system platform and the second component denotes the application environment and their versions.  Some of the CPE groups might be a parent cluster of another CPE group. For example, \textit{Microsoft Windows} would be a parent cluster for CPEs like \textit{Microsoft Windows\_8} or \textit{Microsft Windows\_10}. In this research, we do not consider any hierarchies in the CPEs for filtering out clusters, but as future research use, this can be considered. From our data we found that over the time period from April 2016 to September 2017, the top CPE groups having CVEs which are mentioned most widely in darkweb forum posts are ntp, php, adobe flash\_player, microsoft windows\_server\_2008, linux kernel,  microsoft windows\_7,  micorosft windows\_server\_2012, and canonical ubuntu\_linux. 
	

	\section{Prediction Problem} \label{sec:pred_problem}
	Before we describe our framework for using darkweb discussions in the forums for predicting external enterprise attacks, we formally describe our prediction task. Formally, given a target organization $E$, a set of unconventional (external) signals from the darkweb forums as features and a set of $A$ attack types for $E$, we solve a binary classification problem that investigates whether there would be an attack (0/1) of any type in $A$ for $E$ on a daily basis.
	
	\begin{figure}[!t]
		\centering
		\includegraphics[width=8cm, height=5cm]{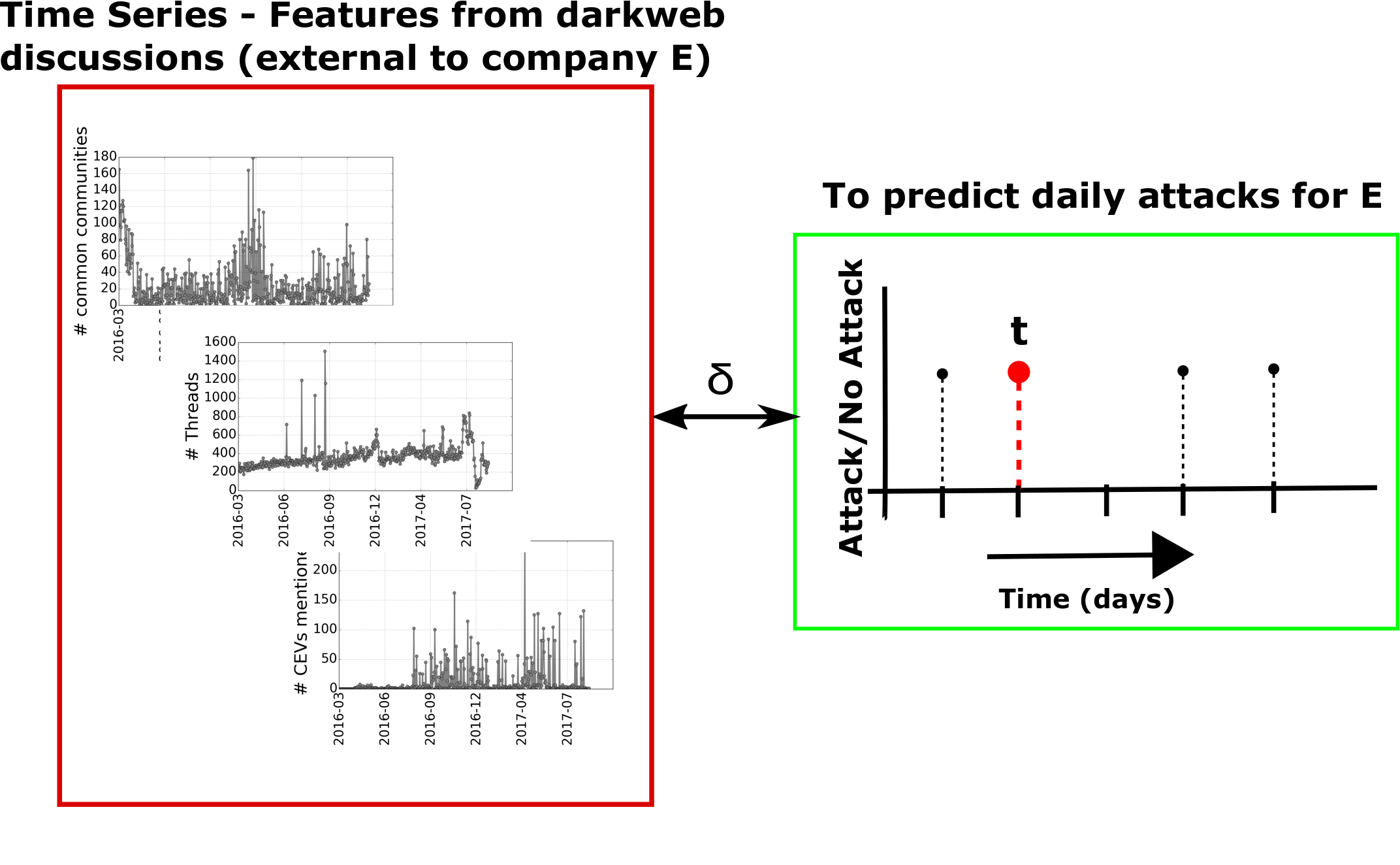}
		\caption{The prediction task. We use unconventional time series signals from the darkweb forum network to predict attack on a daily basis for company $E$.}
		\label{fig:pred_demo}
	\end{figure}
	
	The mechanism for attack predictions as shown in Figure~\ref{fig:pred_demo} can be described in 3 steps : (1) given a time point $t$ on which we need to predict an enterprise attack of a particular event type (2) we use features from the darkweb forums $\delta$ days prior to $t$ and, (3) we use these features as input to a machine learning model to predict attack on $t$. So one of the main tasks involves learning the attack prediction model, one for each event type. We describe the attack prediction framework in the following section.

	\section{Framework for Attack Prediction}\label{sec:framework}
	
	Since we attempt at building an integrated framework leveraging the network formed from the discussions in the forums as signals for predicting organization specific attacks, we segregate it into the three steps of any classic machine learning framework: 
	
	\begin{enumerate}
		\item \textit{Feature engineering}: As one of our contributions, we leverage the reply network formed from the thread replies in forums to build features for input to the model. To this end we build two kinds of features:
		\begin{itemize}
			\item \textit{Graph Based Features}: Here we identify features pertaining to the dynamics of replies from users with credible knowledge to regular posts - the intuition behind this is to see whether a post gaining attention from active and reputed users can be a predictive signal.
			\item \textit{Forum metadata}: We also gather some forum metadata as another set of features and we use them as baselines for our graph based features.
		\end{itemize}
		So as a first step towards achieving this, we devise an algorithm to create the reply network structure from the replies in the threads in this step prior to feature computation.
		
		\item \textit{Training (learning) models for prediction}: In this step, we first split the timeframe of our attack study into two segments: one corresponding to the training span and the other being the test span. However, unlike normal cross-validated machine learning models, we need to be careful about the time split, since we consider longitudinal networks for features and the training-test split should respect the forecasting aspect of our prediction - we use features $\delta$ days prior to the day we predict the attacks for. So instead of using cross-validation, we fix our training time span as the first few time points in our ground truth dataset (chronologically ordered) and the test span succeeding the training span. We build several time-series of individual features from step 1 using only forum discussions in the training span and use them as input along with the attack ground truth to a supervised model for learning the parameters (we build separate models for separate attack types and different attack organizations). This along with step 1 is shown in Figure~\ref{fig:workflow} on the left side under the training span stage.
		\begin{figure}[!t]
			\centering
			\includegraphics[width=13cm, height=7cm]{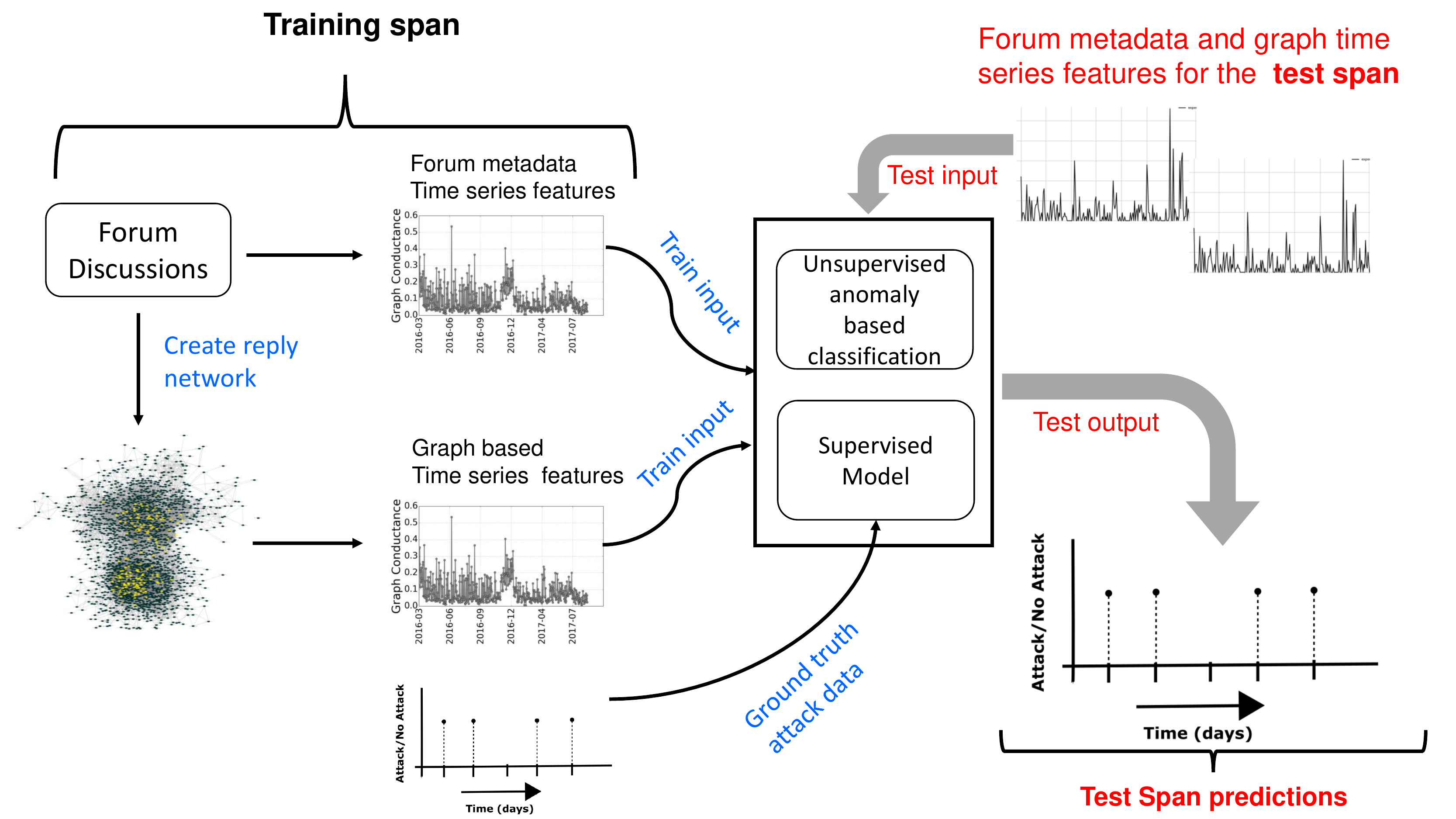}
			\caption{An overview of the framework used for attack prediction.}
			\label{fig:workflow}
		\end{figure}
		\item \textit{Attack prediction}: In this final step, we first compute the time series of the same set of features in the test span, instead that we now use the forum discussions in the test span ($\delta$ days prior to the prediction time point). We input these time series into the supervised model as well as an additional unsupervised model (that does not require any training using ground truth), to output attacks on a daily basis in the test span. This step is displayed in the right component of Figure~\ref{fig:workflow}.
		
	\end{enumerate}
	In the following sections, we explain the steps in details that also describes the intuition behind the approach used for attack prediction in our study.
	
	\subsection{Step 1: Feature Engineering }
	For the purposes of network analysis,  we assume the absence of global user IDs across forums\footnote{Note that even in the presence of global user IDs across forums, a lot of anonymous or malicious users would create multiple profiles across forums and create multiple posts with different profiles, identifying and merging which is an active area of research.} and therefore analyze the social interactions using networks induced on specific forums instead of considering the global network of all users across all forums. We denote the directed and unweighted reply graph of a forum $f \in F$ by $G^f$ = $(V^f, E^f)$ where $V^f$ denotes the set of users who posted or replied in some thread in forum $f$ at some time in our considered time frame of data and $E^f$ denotes the set of 3-tuple $(u_1, u_2, rt)$ directed edges where $u_1, u_2 \in V^f$ and $rt$ denotes the time at which $u_1$ replied to a post of $u_2$ in some thread in $f$, $u_1 \rightarrow u_2$ denoting the edge direction. We emphasize that this notation of the network discards links between users of 2 different forums as we did not connect or merge threads posted in two separate forums. We denote by $G_{\tau}^f$ = $(V_{\tau}^f, E_{\tau}^f)$, a temporal subgraph of $G^f$, $\tau$ being a time window such that $V_{\tau}^f$ denotes the set of individuals who posted in $f$ in that window and $E_{\tau}^f$ denotes the set of tuples $(v_1, v_2, rt)$ such that $rt \in \tau$, $v_1, v_2 \in V_{\tau}^f$. 
	
	\subsubsection{Constructing the reply network}
	
	We adopt an incremental analysis approach by splitting the entire set of time points in our frame of study (both for the training and test span) into a sequence of time windows $\Gamma$ = $\{\tau_1, \tau_2,  \ldots, \tau_{\mathcal{Q}} \}$, where each subsequence $\tau_i$, $i \in [1, \mathcal{Q}]$ is equal in time span and the subsequences are ordered by their starting time points for their respective span. This streaming aspect of the reply networks and the feature computation is based on our observation that the significance of users (in terms of \textit{important} posts in the forums) change very rapidly and for a one year span, computing features for a month based on historical information of users long time back is not convenient. From that perspective, we create evolving networks on a daily basis (but which incorporate historical knowledge), and compute features on a daily basis. However, in more realistic settings, the temporal resolution of these snapshots can be managed dynamically based on how often consecutive networks change significantly in terms of some distance metric as has been done in \cite{tang_res}. 
	
	Next we describe the operations: \textit{Create} - that takes a set of forum posts in $f$ within a time window $\tau$ as input and creates a temporal subgraph $G_\tau^f$  and \textit{Merge} - that takes two temporal graphs as input and merges them to form an auxiliary graph that incorporates historical information. To keep the notations simple, we would drop the symbol $f$ when we describe the operations for a specific forum in $F$ as context but which would apply for any forum $f$ $\in F$. We describe the two operations that describe how we map the features extracted from network structure $G^f$ to a time series, the analysis of which is the one of contributions of our research:  
	
	\begin{enumerate}
		\item \textit{CREATE:} In this step, we create the reply network based on individual threads within a forum $f$ on a daily basis. Let $h$ be a particular thread or topic within a forum $f$ containing posts by users $V_{h}^f$ = $\{u_1, \ldots, u_k \}$ posted at corresponding times $T_{h}^f$ = $\{t_1, \ldots , t_k\}$, where $k$ denotes the number of posts in that thread and $t_i \geq t_j $ for any $i > j$, that is the posts are chronologically ordered. Since we are considering a reply network on the forum posts, the lack of information as to who replied to whom necessitates the use of some heuristics to connect the users based on temporal and spatial information. We note that in situations where the data comes with the hierarchical reply structure of who-replies-to-whom, this step can be avoided and can be skipped to the next stage. A simple approach would be to consider either (i) a \textit{temporal constraint}: for each user $u_i$ of a post in a thread $h$ in forum $f$ at time $t_i$, we would create an edge $(u_i, u_k, t)$ such that $t_i - t_k$ $< thresh_{temp}$, $u_k$ denotes the user for the respective posts at time $t_k \in \tau $, $thresh_{spat}$ denoting a time threshold or (ii) a \textit{spatial constraint}: consider all edges $(u_i, u_k, t_i)$, where $u_k$ denotes the user of the $k^{th}$ post in the time ordered sequence of posts and $k - i \leq thresh$, $thresh$ denoting a count threshold. The idea behind reply edge construction based on the combination of these two constraints is the following: in a time interval where there are a lot of discussions, networks with the edges created from the condition bounded by $thresh_{temp}$ would be unduly over-dense. Thus the second condition bounds the number of posts (prior to its current post) that a user can reach to while replying using its current post. In a way, this ensures normalization since the hypothesis here is that a user can only reach/reply to a certain number of posts prior to the current time irrespective of how popular the discussions might be in a specific time intervals.
		
		We use both the constraints in the following way: for the $i^{th}$ post $p_{h, i}$ in the thread $h$ posted at time $t_i$, the objective is to create links from the user of this post to the posts prior to this as reply links. For this, we consider a maximum of $thresh_{spat}$ count of posts prior to $p_{h, i}$ (note the posts in the thread are considered chronologically ordered), that is all posts $p_{h, k}$ such that $k - i \leq thresh_{spat}$. The users for those respective posts would be the potential users to whom $u_{h, i}$ replied to (unidirectional links), which we denote by $\{u_{h, i\rightarrow k}\}$ and the corresponding set of posts $\{p_{h, i \rightarrow k}\}$. The next layer of constraints considering  temporal boundaries prune out candidates from $\{u_{h, k}\}$, using the following two operations:  
		\begin{figure}[!t]
			\centering
			\includegraphics[width=11cm, height=5cm]{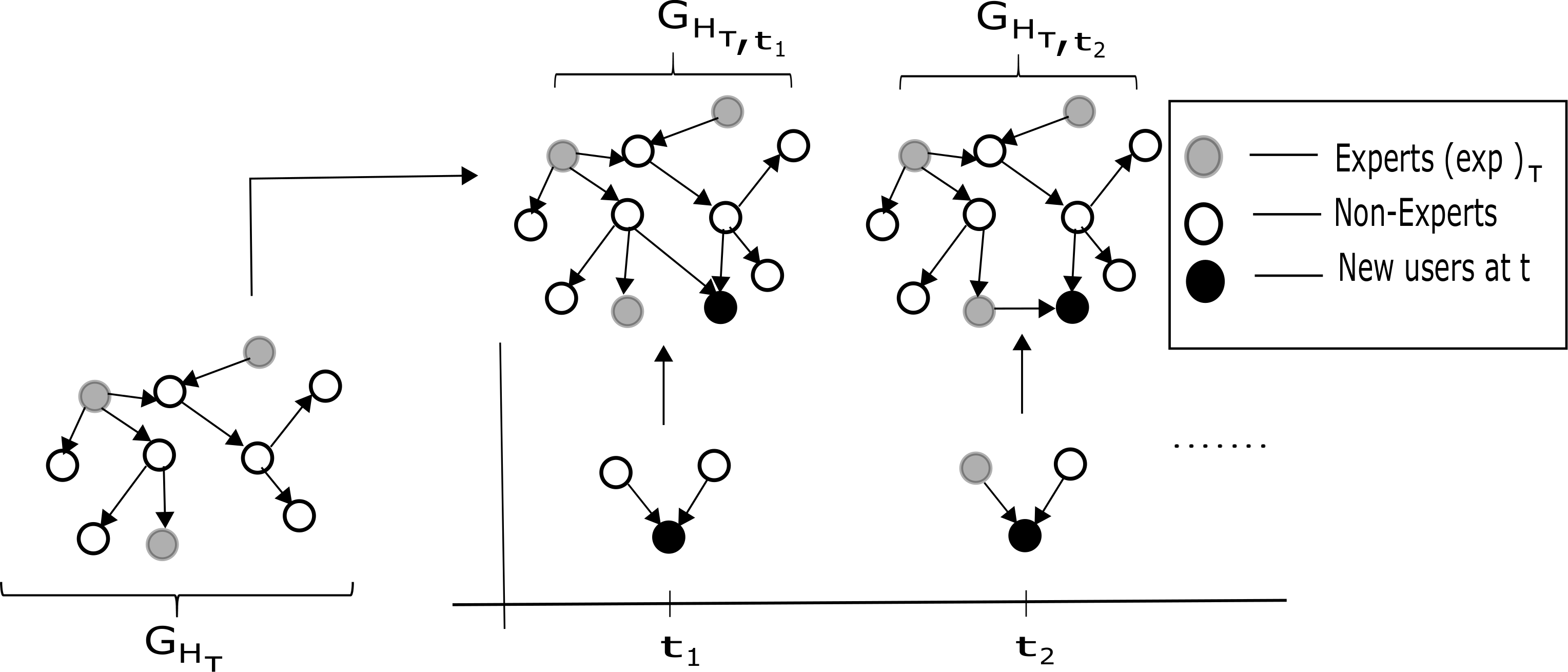}
			\caption{An illustration to show the \textit{Merge} operation: $G_{H_{\tau}}$ denotes the historical network using which the experts shown in gray are computed. $\{ G_{t_1}$, $G_{t_2}$, $\ldots \}$ denote the networks at time $t_1,  \ t_{2}, \ldots$ $\in \tau$, $\tau \in \Gamma$. Note that the experts are extracted only from $G_{H_\tau}$ and not on a regular basis. }
			\medskip
			\small
			
			\label{fig:create_demo}
		\end{figure}
		\begin{itemize}
			\item If $t_i - t_k < thresh_{temp}$, we form edges linking $u_{h, i}$ to all users in $\{u_{h, i \rightarrow k}\}$ (note the direction of reply). This takes care of the first few posts in $h$ where there might not be enough time to create a sensation, but anyhow the users might be replying as a general discussion in the thread. So we consider user of $i^{th}$ post replies potentially to all these users of $\{u_{h, i \rightarrow k}\}$ at one go whether it is at the beginning or whether it is in the middle of an ongoing thread discussion. 
			
			\item If $t_i - t_k >= thresh_{temp}$, we first compute the mean of the time differences between two successive posts in $\{p_{h, i \rightarrow k}\}$. We also denote the time difference between $t_i$ and the time of the last post in $\{p_{h, i \rightarrow k}\}$ considering the chronological ordering is maintained (this is the post prior to $i$), as $\Delta t_{i}$. If the computed mean is less than $\Delta t_{i}$, we form edges linking $u_{h, i}$ to all users in $\{u_{h, i \rightarrow k}\}$ (this is similar to the first constraint).  Else, as long as the mean is greater than $\Delta t_{i}$, we start removing the posts in $\{p_{h, i \rightarrow k}\}$ farthest in time to $t_i$ in order and recalculate the mean after removal of such posts. We repeat this procedure until at some iteration either the recomputed mean is less than $\Delta t_i$ or $t_i - t_k < thresh_{temp}$. This heuristic considers the case for posts that receive a lot of replies very frequently at certain time of the thread lifecycle, although it is not reasonable to consider posts which have been posted a while ago as being replied to by the current post in consideration. 
		\end{itemize}   
		
		Following this, $V^f$ = $ \cup_{h} V^f_h$ and $E^f$ = $\cup_{h} E_{h}^f$, that is we remove multiple interactions between the same set of users in multiple threads and without weighting these edges. As before, a temporal subgraph of $G^f$ would be denoted by $G_{\tau}^f$ where  $(u, v,  rt)$ $\in E_{\tau}$ denotes $u$ replied to $v$ at time $rt \in \tau$. Our objective after creating the reply network $G_{\tau}^f$ is to compute features from this network that could then be used as input to a machine learning model for predicting cyber attacks. These features would act as the unconventional signals that we have been addressing in this paper for predicting external enterprise specific attacks. In order to achieve that, we need to form time series of a feature $x$ (among a set of network features) denoted by $\mathcal{T}_{x, f}$ for every forum $f \in F$ separately: formally $\mathcal{T}_{x, f}$ is a stochastic process that maps each time point $t$ to a real number. \\
		
		\item \textit{MERGE:} In order to create a time series feature $\mathcal{T}_{x, f}$ for feature $x$ from threads in forum $f$, we use 2 reply networks: (1) a historical network $G_{H_{\tau}}$ which spans over time $H_{\tau}$ such that $\forall t' \in H_{\tau}$, and for any $t \in \tau$, we have $t' < t$, and (2) the network $G^f_t$ induced by user interactions between users in $E_t$, which varies temporally for each $t \in \tau$. We note that the historical network $G_{H_{\tau}}$ would be different for each subsequence $\tau$, so as the subsequences $\tau \in \Gamma$ progress with time, the historical network $G_{H_{\tau}}$ also changes, and we discuss the choice of spans $\tau \in \Gamma$ and $H_{\tau}$ in Section~\ref{sec:exp}. Finally, for computing feature values for each time point $t \in \tau$, we merge the 2 networks $G_{H_{\tau}}$ and $G_{t}$ to form the auxiliary network $G_{H_\tau, t}$ = $(V_{H_\tau, t}, E_{H_\tau, t})$, where $V_{H_\tau, t} = V_{H_\tau} \cup V_t$ and $E_{H_\tau, t}$ = $E_{H_\tau} \cup E_t$. A visual illustration of this method is shown in Figure~\ref{fig:create_demo}. At the end, we consider several network features over each $G_{H_\tau, t}^f$ and compute the feature values at time point $t$ to form $\mathcal{T}_{x, f}$ for feature $x$ and forum $f$.
	\end{enumerate}

	\subsubsection{Network based features}
	We leverage the network $G^f_{H_\tau}$ to compute features on a regular basis - the advantage is that this network contains  historical information but at the same time, this historical information does not update on a regular basis. For extracting network based features, we want to be able to focus on the interactions convened by users in forums with a knack towards posting credible information. The objective is to investigate whether any spike in attention towards posts on a day from such users with some \textit{credible reputation} translates to predictive signals for cyber attacks on an organization. This would also in a way help filter out  noisy discussions or replies from unwanted or naive users who post information irrelevant to vulnerabilities or without any malicious intent. We hypothesize that predictive signals would exhibit users in these daily reply networks whose posts have received attention (in the form of direct or indirect replies) from some ``expert" users - whether a faster reply would translate to an important signal for an attack is one of the novel questions we tackle here.  
	
	In order to be able to extract posts that receive attention on a daily basis, we first need to extract ``expert" users who attention we seek to gather.
	
	\paragraph{Expert Users.} For each forum $f$, we use the historical network $G_{H_{\tau}}^f$ to extract the set of $experts$ relevant to timeframe $\tau$, that is $exp_{\tau}^f$ $\in V_{H_{\tau}}^f$. First, we extract the top CPE groups $CP^{top}_{\tau}$ in the time frame $H_\tau$ based on the number of historical mentions of CVEs. These would be used as top CPEs for the span $\tau$. For this, we sort the CPE groups based on the sum of the CVE mentions that belong to the respective CPE groups and take the top 5 CPE groups by sum in each $H_{\tau}$.  Using these notations, the experts $exp_\tau^f$ from history $H_{\tau}$ considered for time span $\tau$ are defined as users in $f$ with the following three constraints:

	\begin{enumerate}
		\item Users who have mentioned a CVE in their post in $H_\tau$. This ensures that the user engages in the forums with content that is relevant to vulnerabilities. 
		\item Let $\theta(u)$ denote the set of CPE tags of the CVEs mentioned by user $u$ in his/her posts in $H_{\tau}$ and such that it follows the constraint: either $\theta(u)$ $\in CP_{\tau}^{top}$ where the user's CVEs are grouped in less than 5 CPEs or, $CP_{\tau}^{top}$ $\in \theta(u)$ in cases where a user has posts with CVEs in the span $H_{\tau}$, grouped in more than 5 CPEs.  This constraint filters out users who discuss vulnerabilities which are not among the top CPE groups in $H_\tau$.
		\item The in-degree of the user $u$ in $G_{H_\tau}$ should cross a threshold. This constraint ensures that there are a significant number of users who potentially responded to this user thus establishing $u$'s central position in the reply network. These techniques to filter out relevant candidates based on network topology has been widely used in the bot detection communities \cite{Nagaraja2010}.
	\end{enumerate}
	
	We avoid using other centrality metrics instead of using the in-degree in the third constraint since our focus here is not to judge the position of the user from the centrality perspective (for example, high betweenness would not denote the user receives multiple replies on its posts). Instead, we want to filter out users who receive multiple replies on their posts or in other words their posts receive attention. Essentially, these set of experts $exp_{\tau}$ from $H_{\tau}$ would be used for all the time points in $\tau$ as shown in Figure~\ref{fig:create_demo}. Our objective here is to not consider the degree as the proxy for user importance in any terms. Rather the degree indicates the number of replies it gets from other users.
	
	\paragraph{Why focus on experts?} To show the significance of these properties in comparison to other users, we perform the following hypothesis test: we collect the time periods of 3 widely known security events: the Wannacry ransomware attack that happened on May 12, 2017 and the vulnerability MS-17-010, the Petya cyber attack on 27 June, 2017 with the associated vulnerabilities CVE-2017-0144, CVE-2017-0145 and MS-17-010, the Equifax breach attack primarily on March 9, 2017 with vulnerability CVE-2017-5638. We consider two sets of users across all forums - $exp_{\tau}$, where $G_{H_\tau}$ denotes the corresponding historical network prior to $\tau$ in which these 3 events occurred and the second set of users being all $U_{alt}$ who are not experts and who fail either one of the two constraints: they have mentioned CVEs in their posts which do not belong to $CP^{top}$ or their in-degree in $G_{H_{\tau}}$ lies below the threshold.  We consider $G_{H_{\tau}}$ being induced by users in the last 3 weeks prior to the occurrence week of each event for both the cases, and we consider the total number of interactions ignoring the direction of reply of these users with other users. Let $\mathbf{deg_{exp}}$ denote the vector of count of interactions in which the \textit{experts} were involved and $\mathbf{deg_{alt}}$ denote the vector of counts of interactions in which the users in $U_{alt}$ were involved. We randomly pick number of users from $U_{alt}$ equal to the number of experts and sort the vectors by count. We conduct a 2 sample t-test on the vectors $\mathbf{deg_{exp}}$ and $\mathbf{deg_{alt}}$. The null hypothesis $H_0$ and the alternate hypothesis $H_1$ are defined as follows;
	\begin{equation*}
	H_0: \ \mathbf{deg_{exp}} \leq \ \mathbf{deg_{alt}}, \ \ \ \ H_1: \ \mathbf{deg_{exp}} > \ \mathbf{deg_{alt}}
	\end{equation*}
	
	The null hypothesis is rejected at significance level $ \alpha $ = 0.01 with $p$-value of 0.0007. This suggests that with high probability, experts tend to interact more prior to important real world cyber-security breaches than other users who randomly post CVEs. 
	
	Now, we conduct a second $t$-test where we randomly pick 4 weeks not in the weeks considered for the data breaches, to pick users $U_{alt}$ with the same constraints. We use the same hypotheses as above and when we perform statistical tests for significance, we find that the null hypothesis is not rejected at $\alpha$=0.01 with a $p$-value close to 0.05. This empirical evidence from the $t$-test also suggests that the interactions with $exp_{\tau}$ are more correlated with an important cyber-security incident than the other users who post CVEs not in top CPE groups and therefore it is better to focus on users exhibiting our desired properties as experts for cyber attack prediction. Note that the $t-test$ evidence also incorporates a special temporal association since we collected events from three interleaved timeframes corresponding to the event dates and we did not select any timeframe to show the evidence. 
	
	Next, we describe the following graph based features that we use to compute $\mathcal{T}_{x, f}[t]$ at time $t$, for which we also take as input the relevant experts $exp_\tau$. We describe 4 network features that capture this intuition behind the attention broadcast by these users - the idea is that a cyber-adversary looking to thwart the prediction models from working by curating similar reply networks using bots, would need to not only introduce such random networks  but would also have to get the desired attention from these experts which could be far challenging to achieve given that human attention is known to be different compared to bots. \cite{ferrara}.
	
	\paragraph{Graph Conductance.}
	As studied in \cite{Nagaraja2007,Danezis2009,cuts_randall}, social networks are fast mixing: this means that a random walk on the social graph converges quickly to a node following the stationary distribution of the graph.  Applied to social interactions in a reply network, the intuition behind computing the graph conductance is to understand the following: can we compute bounds of steps within which any attention on a post would be successfully broadcast from the non-experts to the \textit{experts} when a post closely associated with an attack is discussed \cite{rumor_cond}? One way of formalizing the notion of \textbf{graph conductance} $\mathbf{\phi}$ is: $\phi = \underset{X \subset V : \pi(X) < \frac{1}{2}}{\min} \phi_X$ where $\phi_X$, $X$ being the set of experts here is defined as: $\phi_{Experts} = \frac{\sum_{x \in exp_{\tau}} \sum_{y \in V_t \setminus exp_{\tau}} \pi(exp_\tau) P_{xy}}{\pi(exp_\tau)}$, and $\pi(.)$ is the stationary distribution of the network $G_{H_\tau, t}$. For subset of vertices $exp_\tau$, its conductance $\phi_{Experts}$ represents the probability of taking a random walk from any of the $experts$ to one of the users in $V_t \setminus exp_{\tau}$, normalized by the probability weight of being on an expert.
	
	Applied to the reply network comprising both experts and the regular users, the key intuition behind conductance as used here is: \textit{the mixing between expert nodes and the users of important posts is fast, while the mixing between expert nodes and regular nodes without important posts (in our view of importance as seeking attention) is slow.} So higher the value of conductance here, higher is the probability that the experts are paying attention to the posts and so there is a good chance that the conversations on those days could be reflective of a cyber attack in future.
	
	\paragraph{Shortest paths.}
	To understand the dynamics of distance between the non-experts and the set of experts prior to an attack, we compute the shortest distance metric between them as follows: \linebreak
	$SP(exp_\tau, V_t \setminus exp_{\tau}) = \frac{1}{|exp_{\tau}|} \ \sum_{e \in exp_{\tau}} \ \underset{u \in V_t \setminus exp_{\tau}}{\min} \ s_{e, u} 
	$, where $s_{e, u}$ denotes the shortest path in the graph $G_{H_\tau, d}$ from the expert $e$ to a user $u$ in the direction of the edges. Since the edges are formed in the direction of the replies based on time constraints, it also denotes how fast an expert replies in a thread that leads back in time to a post by $u$. Such distance metrics have been widely used in network analysis to understand the pattern of interactions \cite{Tang2009}. 
	
	\paragraph{Expert Replies.}
	To analyze whether \textit{experts} reply to users more actively when there is an important discussion going on surrounding any vulnerabilities or exploits, we compute the number of replies by an \textit{expert} to users in $V_t \setminus exp_{\tau}$. We calculate the number of out-neighbors  of $exp_\tau$ considering $G_{H_\tau, t}$. 
	
	\begin{algorithm}[!t]
		\KwIn{$exp_\tau$, $G_{H_{\tau}}$, $(V_t, E_{t})$}
		\KwOut{$CC(exp_\tau,  V_t \setminus exp_{\tau})$ - the number of communities shared by $V_t \setminus exp_{\tau}$ with $exp_\tau$ at $t$}
		
		$communities$ = $Louvain\_community(G_{H_\tau})$ \tcp*{dictionary storing node to community index mapping}
		
		$c_{expSet}$ $\leftarrow$ $()$ \;
		
		\ForEach{user u $\in exp_{\tau}$}{
			$c_{expSet}$.add($communities$[u]) \;
		}
		
		$V_{{H_\tau}, t}$ $\leftarrow$ $V_{H_{\tau}}$ $\cup$ $V_{t}$ \;
		$E_{{H_\tau}, t}$ $\leftarrow$ $E_{H_{\tau}}$ $\cup$ $E_{t}$ \;
		
		$CC(exp_\tau,  V_t \setminus exp_{\tau})$  $\leftarrow$ 0  \tcp*{stores count}
		\tcc{Iterate over the users in $V_t$ who have not been assigned communities from $H_\tau$}
		\ForEach{user u $\in V_{t}$}{
			\If{u $\in$ $V_{H_\tau}$  and communities(u) $\in$ $c_{expSet}$ }{
				$CC(exp_\tau,  V_t \setminus exp_{\tau})$ += 1\;
			}
			\Else{
				\ForEach{user v $\in exp_{\tau}$}{
					\tcc{Condition 1}
					\If{(v, u) $\in$ $E_{{H_\tau}, t}$ }{
						$CC(exp_\tau,  V_t \setminus exp_{\tau})$ += 1\;
						break \;
					}
					\tcc{Condition 2}
					\ForEach{user n $\in$ $inNeighbors(E_{{H_\tau}, t}, u)$}{
						\If{communities(n) $\in$ $c_{expSet}$ }{
							$CC(exp_\tau,  V_t \setminus exp_{\tau})$ += 1\;
							break \;
						}
					}
				}
			}
		}
		return $CC(exp_\tau,  V_t \setminus exp_{\tau})$
		
		\caption{Algorithm for computing Common Communities (CC) }
		\label{Algo:CCAlgo}
	\end{algorithm}
	
	\paragraph{Common Communities.}
	To evaluate the role of communities in the reply network and to assess whether experts engage with selected other users within a community when an information gains attention and could be related to vulnerability exploitation, we use community detection on the networks $G_{H_\tau}$. We use the Louvain method \cite{louvain} to extract the communities from a given network. Since it is not computationally feasible to compute communities in $G_{H_\tau, t}$ for all the time points $t \in \tau$, we first compute all the communities for the users in the historical network $G_{H_\tau}$. Following this, we use an approximation based on heuristics to compute the communities of new users $V_{new}$ = $V_{H_\tau, t} \setminus V_{H_\tau}$. Let $c_{experts}$ denote the set of communities that users in $exp_{\tau}$ belong to following the call to Louvain method in Line 1 of Algorithm~\ref{Algo:CCAlgo}. Let $c(u)$ denote the community index of a user $u$. We define the common communities measure as follows: $CC(exp_\tau,  V_t \setminus exp_{\tau})  = \{\mathcal{N}(c(u)) \ | \ c(u) \in c_{experts}  \wedge \ u \in  V_t \setminus exp_{\tau}\}$, that is it measures the number of non-experts at time $t \in \tau$ that share the same communities with $exp_\tau$.  We use 2 approximation constraints demonstrated in Lines 16-25 of Algorithm~\ref{Algo:CCAlgo} to assign a new user $u \in V_{new}$ to an expert community as follows:
	\begin{enumerate}
		\item \textit{Condition 1}: If an expert has an incoming edge to $u$, we increase the count of common communities by 1.
		\item \textit{Condition 2}: If $u$ has a incoming neighbor who shares a community in the set of communities of experts, we increase the count of common communities by 1. This is shown in Line 19 in the call to the $InNeighbors()$ method. 
	\end{enumerate}
	\begin{table*}[!t]
		\begin{tabular}{|p{2.3cm}|p{2.7cm}|p{8.5cm}|}
			\hline
			Group                                 & Features                 & Description                                                                                                                              \\ \hline
			\multirow{4}{*}{Expert centric}       & Graph Conductance        & \begin{tabular}[c]{@{}l@{}}$\tau_{x}[t]$ = $\frac{\sum_{x \in exp_{\tau}} \sum_{y \in V_t \setminus exp_{\tau} } \pi(exp_{\tau}) P_{xy} }{\pi(exp_{\tau})}$ \\ where $\pi(.)$ is the stationary distribution  of the network $G_{H_\tau, t}$, \\ $P_{xy}$ denotes the probability of  random walk  from vertices $x$ to $y$.\\ The conductance represents the probability of taking a random \\ walk  from any of the $experts$ to one of the users in $V_t \setminus exp_{\tau}$, \\ normalized by the probability weight  of being on an expert. \end{tabular} \\ \cline{2-3} 
			& Shortest Path            & \begin{tabular}[c]{@{}l@{}}$\tau_{x}[t]$ = $ \frac{1}{|exp_{\tau}|} \  \sum_{e \in exp_{\tau} } \  \underset{u \in V_t \setminus exp_{\tau}}{\min} \  s_{e,u} $  \\ where $s_{e, u}$ denotes the shortest path from an expert $e$ to user $u$ \\ following the direction of edges.\end{tabular}                                          \\ \cline{2-3} 
			& Expert replies           & \begin{tabular}[c]{@{}l@{}}$\tau_{x}[t]$ = $\frac{1}{|exp_{\tau}|} \  \sum_{e \in exp_{\tau}} \  |OutNeighbors(e)|$ \\ where $ OutNeighbors(.)$ denotes the out neighbors of user in the \\ network $G_{H_\tau, t}$.\end{tabular}                                                                                                             \\ \cline{2-3} 
			& Common Communities       & \begin{tabular}[c]{@{}l@{}}$\tau_{x}[t]$ = $ \{ \mathcal{N}(c(u) \ | \ c(u) \in c_{experts} \wedge \ u \in V_t \setminus exp_{\tau} \}$\\ where $c(u)$  denotes the community index of user $u$, $c_{experts}$ that \\ of  the experts and $\mathcal{N}(.)$  denotes a counting function. It counts \\ the number of users who share communities with experts. \end{tabular}                                      \\ \hline
			\multirow{4}{*}{\begin{tabular}[c]{@{}l@{}}Forum/User \\  Metadata \end{tabular}}      & Number of threads        & $\tau_{x}[t]$ = $|\{ h \  | \  \mbox{thread h was posted on t} \}|$                                                                                                                                                                                                                                                                          \\ \cline{2-3} 
			& Number of users          & $\tau_{x}[t]$ = $|\{ u \  | \  \mbox{user u  posted on t} \}|$                                                                                                                                                                                                                                                                             \\ \cline{2-3} 
			& Number of expert threads & $\tau_{x}[t]$ = $|\{ h \  | \  \mbox{thread h was posted on t by users u $\in$ experts} \}|$                                                                                                                                                                                                                                                         \\ \cline{2-3} 
			& Number of CVE mentions   & $\tau_{x}[t]$ = $|\{ CVE \  | \  \mbox{CVE was mentioned in some post on t} \}|$                                                                                                                                                                                                                                                      \\ \hline
		\end{tabular}
		\caption{List of features used for learning. Each feature $\tau_{x}$ is computed separately across forums. 	\label{tab:table_feat}}
	\end{table*}
	
	\subsubsection{User/Forum Metadata features}
	
	In addition to the network features, we compute the following forum based statistics for a forum $f$ at time point $t$: (1) The number of unique vulnerabilities mentioned in $f$ at time $t$, (2) The number of users who posted in $f$, (3) the number of unique threads in $f$ at time $t$, and (4) The number of threads in which there was at least one $expert$ post among all the posts in $f$ at $t$. \\
	
	\noindent A brief summary of all the features used in this study is shown in Table~\ref{tab:table_feat}.
	
	\subsection{Training models for prediction} \label{sec:learn_models}
	In this section, we explain how we use the time series features $\mathcal{T}_{x, f}$ across forums in $F$ described in the preceding section to predict an attack at any given time point $t$. We consider 2 models for our framework: (1) a supervised learning model in which the time series  $\mathcal{T}_x$ is formed by averaging $\mathcal{T}_{x, f}$ across all forums in $f \in F$ at each time point $t$ and then using machine learning models for the prediction task and, (2) an unsupervised learning model in which we take the time series $\mathcal{T}_{x, f}$ for each feature and each forum $f$ separately and then use dimensionality reduction techniques across the forums dimension. Following this, we use anomaly detection methods for the prediction task - this model does not use the training span ground truth attack data and directly works on features in the training and test span to predict attacks. However, in the supervised learning scenario we build separate prediction models for each attack type in $A$ and for each organization separately. We do not use the two learning models in conjunction nor do we combine data from different attack types together - we leave that as a future work to see how models built on one attack type could generalize to other types and whether we can use different attack types together as a multi-label classification problem although such models of synthesis have been used previously for attack prediction\cite{ai2}.  We treat the attack prediction problem in this paper as a binary classification problem in which the objective is to predict whether there would be an attack at a given time point $t$ (Refer Figure~\ref{fig:pred_demo}). Since the incident data in this paper contains the number of incidents that occurred at time point $t$, we assign a label of 1 for $t$ if there was at least one attack at $t$ and 0 otherwise. 
	
	\subsubsection{Supervised Learning} \label{sec:long_sparsity}
\begin{figure}[!t]
		\centering
		\includegraphics[width=5.5cm, height=2cm]{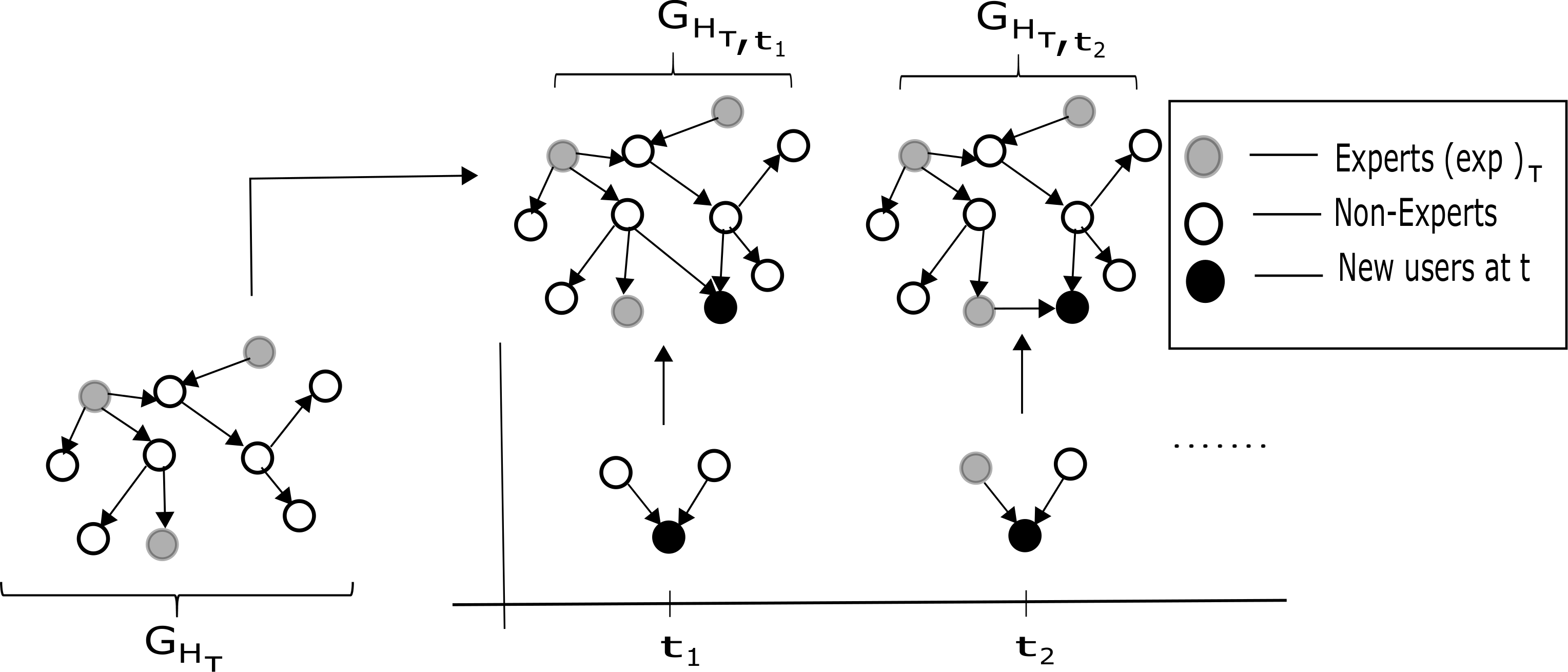}
		\caption{Temporal feature selection window for predicting an attack at time $t$}
		\label{fig:predict_des}
	\end{figure}

	We first discuss the technical details of the machine learning model that learns parameters based on the given training labels of different attack types in $A$ in the training span and uses them to predict whether an organization $E$ would be  vulnerable  to an attack of some type in $A$ at $t$ - we note again that we build different models for each attack type in $A$ for $E$, so predicting for each type means that we have to learn different models for the types, however the set of time series features gathered in the previous step as input is consistent across all models. In \cite{long_lasso,sparse_time_lag}, the authors studied the effect of longitudinal sparsity in high dimensional time series data, where they propose an approach to assign weights to the same features at different time spans to capture the temporal redundancy.  We use 2 parameters: $\delta$ that denotes the start time prior to $t$ from where we consider the features for prediction and $\eta$, the time span (window) for the features to be considered. An illustration is shown in Figure~\ref{fig:predict_des} where to predict an attack occurrence at time $t$, we use the features for each time $t$ $\in [t_{-\eta-\delta}, \  t_{-\delta}]$. We use logistic regression with longitudinal ridge sparsity that models the probability of an attack as follows:
	\begin{equation}
	\label{eq:log_prob}
	P(attack(t) = \ 1 | \ \mathbf{X}  ) = \frac{1}{1+e^{-(\beta_0 + \sum_{k=\eta +\delta}^{\delta} \beta_k \ x_{t-k}) } }
	\end{equation}
	The final objective function to minimize over $N$ instances where $N$ here is the number of time points spanning the attack time frame is : $l(\mathbf{\beta}) = -\sum_{i=1}^{N} (y_i(\beta_0 + \mathbf{x_i}^T\mathbf{\beta} ) - \mbox{log}(1 + \mbox{exp}^{\beta_0 + \mathbf{x_i}^T\mathbf{\beta}}) +  \lambda \mathbf{\beta}^T\mathbf{\beta}$.
	T
	
	To obtain the aggregate series $\mathcal{T}_x$ from individual forum features $\mathcal{T}_{x, f}$, we just average the values across all forums for each time point. Here we use each feature separately although later we discuss the combinations of features together with sparsity constraints in Section~\ref{sec:glasso}.

	\subsubsection{Unsupervised Learning}
	\label{sec:anom_sec}
	Now, we discuss the unsupervised learning model that directly takes as input the time series features in the training span as input and predicts the attacks for types in $A$ on an organization $E$ in the test span. However, unlike the supervised model, this model's prediction output does not depend on the type of attacks or the organization - $E$. It produces the same output for any attack - we try to see how do anomalies from such unconventional signals in the darkweb correlate with the attacks in the real world. Informally, anomalies are patterns in data that do not conform to a well defined notion of normal behavior. The problem of finding these patterns is referred to as anomaly detection \cite{Chandola_2009,Hodge_2004}. The importance of anomaly detection comes from the idea that anomalies in data translate to information that can explain actionable deviations from normal behavior thus leading to a cyber attack. We use subspace based anomaly detection methods that take as input, $\mathcal{T}_{x, f}$, aggregates them across all forums and finds anomalies in the cumulative time series for feature $x$. We derive motivation for this technique from the widely used projection based anomaly detection methods \cite{Lakhina_2004,Huang:2006} that detects volume anomalies from the time series of network link traffic. Additionally, there have been techniques in graph based anomaly detection that finds graph objects that are rare and considered outliers \cite{akoglu}. However, our motivation behind using anomaly detection does not lie from a feature analysis perspective or finding anomalous users but from a time series perspective - we observed that there could be spikes in time series of the same feature in different forums on different days. The question is how do we aggregate information from these spikes together instead of averaging them to an extent that the spikes die out in the aggregate. From that perspective, we find that the method used in \cite{Lakhina_2004} suits our framework - we want to be able to filter out the spikes from the same feature computed in different forums while projecting the dimension space of several forums to a 1-dimensional subspace. The overall procedure for detecting anomalies from the time series data on each feature has been described through the following steps. We will again drop the subscript $x$ to generalize the operations for all features.
	
	\begin{figure*}[!t]
		\centering
		\minipage{0.5\textwidth}
		\includegraphics[width=5cm, height=3cm]{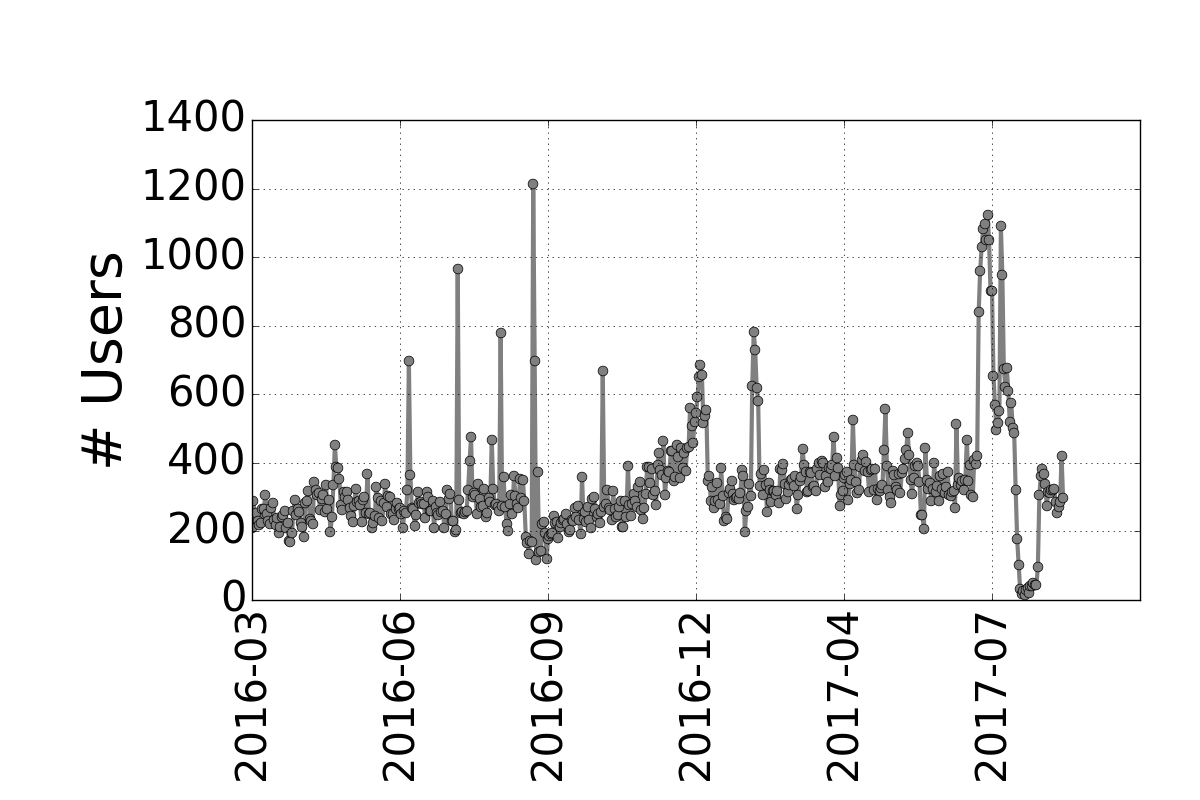}
		\subcaption{}
		\endminipage
		\hfill
		\minipage{0.5\textwidth}
		\includegraphics[width=5cm, height=3cm]{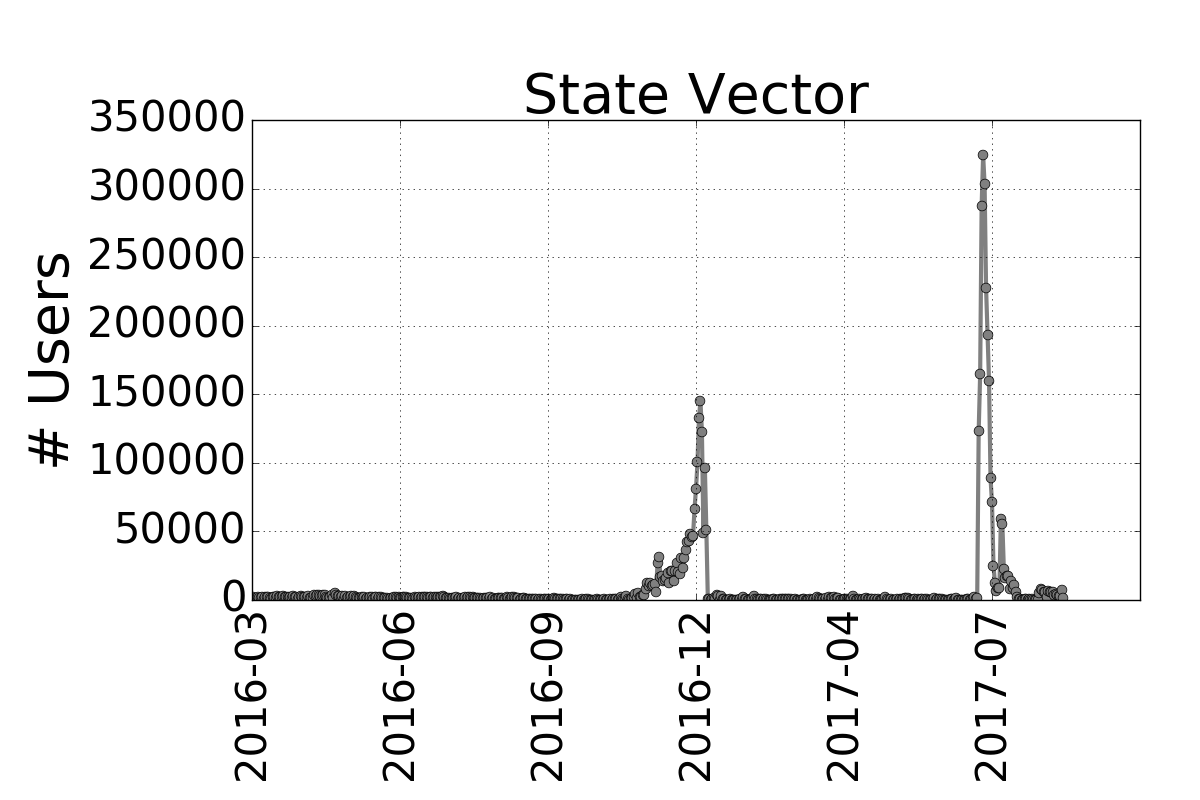}
		\subcaption{}
		\endminipage
		\hfill
		\\
		\minipage{0.50\textwidth}
		\includegraphics[width=5cm, height=3cm]{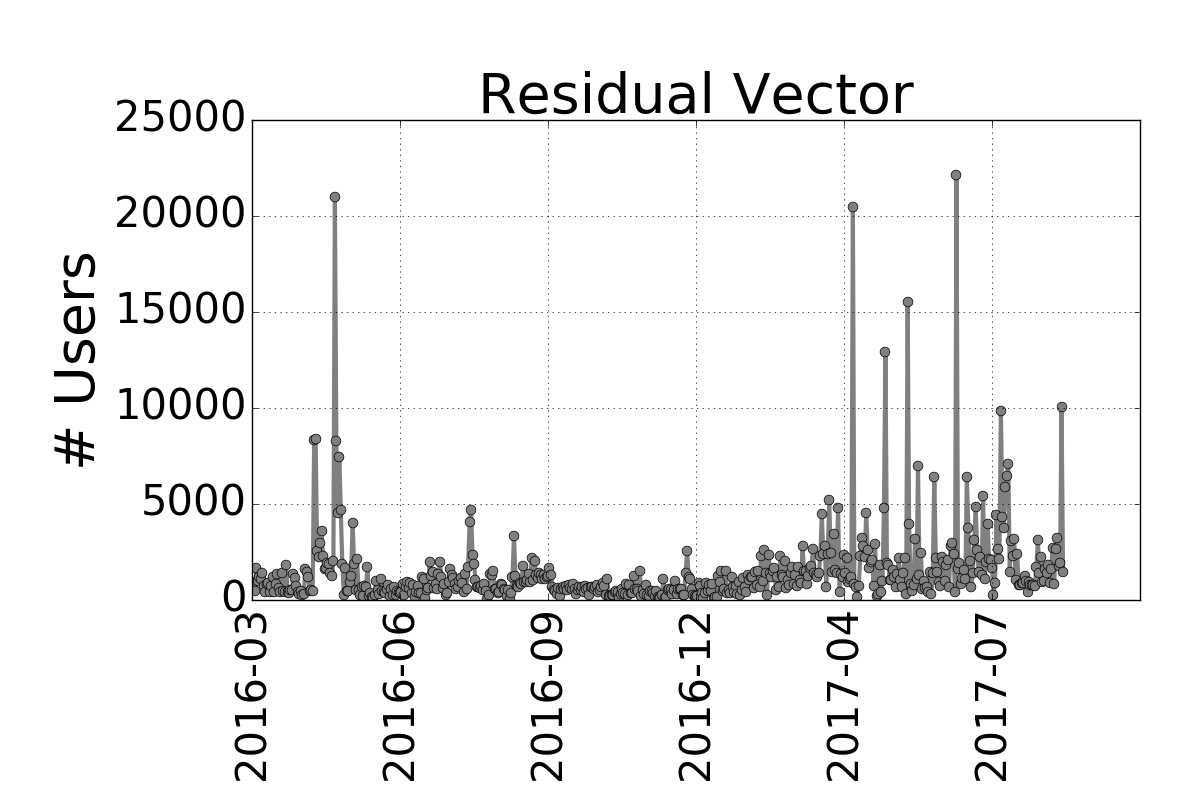}
		\subcaption{}
		\endminipage
		\caption{(a) The time series $\mathcal{T}$ for the number of users feature computed on a daily basis and averaged across all forums $F$ (b) The SPE $state$ time series vector after subspace separation (not averaged) (c) The SPE residual time series vector $\mathcal{R}$ after subspace separation (not averaged). }
		\label{fig:proj_sep}
	\end{figure*}
	
	\paragraph{Aggregating time series.}
	We create a matrix $\mathbf{Y}$ with dimensions ($\#$ time points) $\times$ (F), the rows denoting values at a single time step $t$ for forums $f \in F$. While $\mathbf{Y}$  denotes the set of measurements for all forums $F$, we would also frequently work with $\mathbf{y}$, a vector of measurements from a single timestep $t$.
	
	\paragraph{Subspace Separation.}
	Principal Component Analysis (PCA) \cite{pca} is a method to transform the coordinates of the data points by projecting them to a set of new axes which are termed as the principal components. We apply PCA on matrix $\mathbf{Y}$, treating each row of $\mathbf{Y}$ as a point in $\mathbb{R}^F$. Applying PCA to $\mathbf{Y}$ yields a set of $F$ principal components, $\{\mathbf{v}\}^F_{i=1}$. In general,the $kth$ principal component $\mathbf{v}_k$ is: $\mathbf{v}_k = \underset{\lVert \mathbf{v} \rVert = 1}{\mbox{arg max}} \ \ \lVert (\mathbf{Y}  - \sum_{i=1}^{k-1} \mathbf{Y v_i v_i^T})v \rVert$. We determine the \textit{principal axes (components)} by choosing the first few components that capture the maximum variance along their direction. Once these \textit{ principal axes }  have been determined, the matrix $\mathbf{Y}$ can be mapped onto the new axes leading to  as \textit{residual} or \textit{anomalous subspace}.
	
	For detecting anomalies, we need to separate the vectors $\mathbf{y}$ $\in \mathbb{R}^F$ at any timestep into normal and anomalous components. We will refer to these as the \textit{state} and \textit{residual} vectors of $\mathbf{y}$. The key idea in the subspace-based detection step is that, once $\mathcal{S}$ and $\tilde{\mathcal{S}}$ have been constructed, the separation can be done by projecting $\mathbf{y}$ onto these subspaces. We tend to decompose this $\mathbf{y}$ as: $\mathbf{y} = \hat{\mathbf{y}} + \tilde{\mathbf{y}}$. For this, we arrange the set of principal components corresponding to the normal subspace ($\mathbf{v}_1, \mathbf{v}_2, \ldots, \mathbf{v}_r$) as columns of a matrix $\mathbf{P}$ of size $f$ $\times$ $r$ where $r$ denotes the number of \textit{normal principal axes} determined from the previous step. We can then form 
	$\hat{\mathbf{y}}$ and $\tilde{\mathbf{y}}$ as:
	\begin{equation}\label{eq:proj_subs}
	\hat{\mathbf{y}} = \mathbf{P}\mathbf{P}^T\mathbf{y} = \mathbf{Cy} \ \ \ \ \mbox{and} \ \ \ \ \tilde{\mathbf{y}} = (\mathbf{I} - \mathbf{P}\mathbf{P}^T)\mathbf{y} = \tilde{\mathbf{C}}\mathbf{y}
	\end{equation}
	where the matrix $\mathbf{C} = \mathbf{P}\mathbf{P}^T $ represents the linear operator that performs projection onto the normal subspace, and $\tilde{\mathbf{C}}$ likewise projects onto the residual subspace.  Here $\hat{\mathbf{y}}$ is referred to as the state vector and $\tilde{\mathbf{y}}$ as the residual vector.
	
	\paragraph{Detection of anomalies.}
	\label{sec:anom_subs}
	The idea of anomaly detection is to monitor the residual vector that captures abnormal changes in $\mathbf{y}$.  As mentioned in \cite{Lakhina_2004,anom_stats}, there have been substantial research into designing statistical metrics for detecting abnormal changes in $\tilde{\mathbf{y}}$ using thresholding and we use one of the widely used metrics, the squared prediction error (SPE) on the residual vector: $SPE \  \equiv  \ \lVert \tilde{\mathbf{y}} \rVert  \ \equiv  \  \lVert \tilde{\mathbf{C}}\mathbf{y} \rVert^2$. This gives the SPE residual vector  and when combined over all time points gives us the residual vector time series denoted by $\mathcal{R}$. The SPE residual vector at any time point is considered normal if $SPE \leq \delta_{\alpha}^2$, where $\delta_{\alpha}^2$ denotes the threshold for the SPE at the 1 - $\alpha$ confidence level. We keep this threshold dynamic and would use it as a parameter for evaluating the anomaly based prediction models later on described in Section~\ref{sec:exp}. Figure~\ref{fig:proj_sep} demonstrates the decomposition of the time series into the SPE state and residual vectors. While Figure~\ref{fig:proj_sep}(b) captures most of the normal behavior,  the SPE residual time series in Figure~\ref{fig:proj_sep}(c) captures all the anomalies across all the forums. The key point of this anomaly detection procedure is that instead of monitoring the time series feature $\mathcal{T}_{x, f}$ separately across all forums in $F$ for predicting cyber attacks, we have reduced it to monitoring the SPE residual time series $\mathcal{R}_x$ for cyber attacks. 
	
	\subsection{Attack prediction}
	\paragraph{Anomaly detection to Attack prediction.}
	\label{sec:anom_subs}
	Following the subspace projection method to obtain $\mathcal{R}_{x}$ denoting the SPE residual vector, from the input time series feature $\mathcal{T}_{x,f}$  for all forums $f \in F$, we use threshold mechanisms on $\mathcal{R}_{x}$ to flag the time point $t$ as an anomaly if $\mathcal{R}_x[t]$ is greater than a threshold value. Given any test time point $t$ as the test instance, we first project the times series vector $\mathcal{T}_x[t_{-(\eta + \delta)} : t_{(- \delta)}]$ that contains the information of feature $x$ across all forums in $F$, on the anomalous subspace $\tilde{\mathbf{C}}$ = $\mathbf{I} - \mathbf{PP}^T$ given in Equation~\ref{eq:proj_subs}, if that time window is not already part of the training data. Following this, we calculate the the squared prediction error (SPE) that produces a 1-dimensional vector $\mathbf{y}_{test}$ of dimension $\mathbb{R}^{\eta \times 1}$. We count the number of anomalous time points $t_a$, denoted by $\mathcal{N}(t_a)$, with $t_a \in [t_{-(\eta + \delta)}, t_{(- \delta)}]$, time points that cross a chosen threshold. Finally, we flag an attack at $t$ if  $\mathcal{N}(t_a)$ $>=$ $max(1, \frac{\zeta}{7}) $. This metric gives a normalized count threshold over a week for any $\zeta$ and for this window parameter $\zeta$ being less than a week, we just count whether there is at least one anomaly in that time gap. The fact that we avoid the attack ground truth data to learn event based parameters has some pros and cons : while in the absence of sufficient data for training supervised models, such anomaly detectors can serve a purpose by investigating various markers or features for abnormal behavior leading to attack, the disadvantage is such methods cannot be tailored to specific events or specific attack types in organizations.
	\paragraph{Supervised model prediction.} For the logistic regression model, we first create the features time series $\mathcal{T}_{x}$ for the test span and use it to calculate the probability of attack in Equation~\ref{eq:log_prob}. When the probability is greater than 1, we output a positive attack case else we predict a no-attack case.
	
	\section{Experiments and Results}
	\label{sec:exp}
	
	\subsection{Parameter settings}
	In our work, the granularity for each time index in the $\mathcal{T}$ function is 1 day, that is we compute feature values over all days in the time frame of our study. For incrementally computing the values of the time series, we consider the time span of each subsequence $\tau \in \Gamma$ as 1 month, and for each $\tau$, we consider $H_\tau$ = 3 months immediately preceding $\tau$. That is, for every additional month of training or test data that is provided to the model, we use the preceding 3 months to create the historical network and compute the corresponding features on all days in $\tau$. As mentioned earlier, this streaming nature of feature generation ensures we engineer the features relevant to the timeframe of attack prediction. For choosing the experts with an in-degree threshold, we select a threshold of 10 (we tried the values in the list [5, 10, 15, 20]) to filter out users having in-degree less than 10 in $G_{H_\tau}$ from $exp_{\tau}$. We obtain this threshold by manually investigating a few experts in terms of their content of posts  and we find that beyond a threshold of 10, a lot of users get included whose posts are not relevant to any malicious information or signals.
	
	For the reply network construction, we have 2 parameters: $thresh_{spat}$ and $thresh_{temp}$ corresponding to the spatial and temporal constraints. For setting both these constraints, we used a 2D grid search over these parameters by constructing the reply network using pairwise combinations of these 2 parameters. Following this, for each combination we fit the in-degree distribution to power law with an exponent of 1.35. We fix the power law exponent based on a study \cite{power_law} done where the authors found that a reply network which was created when the thread reply hierarchy was known in 2 forums, 
	was best fit to a power law (in-degree distribution) when the exponents were in the range $[1.35, 1.75]$. We take the pair combination which gives us the minimum difference when we calculate the error arising from our degree distribution and $p(k) \sim k^{-1.35}$. Using this procedure we found $thresh_{spat}$=10 (posts) and $thresh_{temp}$= 15 (minutes) to have the best fit in terms of the reply network we created. 
	
	The hyper-parameters for the logistic regression model $\eta$ and $\delta$ have been selected using a cross validation approach which we discuss briefly in the Results section. Similarly for detection of anomalies, the threshold parameter for the residual vector $\delta_\alpha^2$ mentioned in Section~\ref{sec:anom_subs}, we test it on different values and plot the ROC curve to test the performance. For the choice anomaly count threshold parameter $\zeta$, such that we tag a cyber attack on $t$ when the count of anomalies in the selected window $t_{-\eta-\delta}, t_{-\delta}$ crosses $\zeta$, we set it to 1. The reason behind this is from manual observation where we find very days on which there are spikes and therefore, as a simple method, we just attribute an attack to a day if there was at least one anomaly in the time window prior to it. We do realize that this parameter needs to be cross-validated but our observations suggest that there would be very low precision in the performance when $\zeta$ is set to a high value. 
	
	\subsection{Results}
	To demonstrate the effectiveness of the features on real world cyber attacks, we perform separate experiments with the learning models described in Section~\ref{sec:learn_models}: while for the anomaly detection based prediction, we use the same set of features as the only input for attack prediction across different attack types, for the supervised model, we build different learning models using the ground truth available from separate attack types in $A$. Additionally we only perform supervised classification for the $malicious-email$ and the $endpoint-malware$ attack types leaving out $malicious-destination$ due to lack of sufficient training data.
	\begin{table}[!t]
		\begin{tabular}{|p{3cm}|p{1.7cm}|p{1.7cm}|p{1.7cm}|p{1.7cm}|}
			\hline
			& \textbf{Train positive sample} & \textbf{Train negative samples} & \textbf{Test positive samples} & \textbf{Test negative samples} \\ \hline
			\textbf{Malicious email}     & 65                             & 178                             & 32                             & 60                             \\ \hline
			\textbf{Endpoint Malware}    & 49                             & 134                             & 31                             & 92                             \\ \hline
			\textbf{Malware Destination} & 7                              & 115                             & 8                              & 84                             \\ \hline
		\end{tabular}
		\caption{Statistics of the training and test samples from Armstrong.}
		\label{tab:stats_arm}
	\end{table}
	As mentioned in Section~\ref{sec:learn_models}, we consider a binary prediction problem in this paper - we assign an attack flag of 1 for at least 1 attack on each day and 0 otherwise have the following statistics: for \textit{malicious-email}, out of 335 days considered in the dataset, there have been reported attacks on 97 days which constitutes a positive class ratio of around 29\%, for \textit{endpoint-malware} the total number of attack days are 31 out of 306 days of considered span in the training dataset which constitutes a positive class ratio of around 10\%, and for \textit{endpoint-malware} we have a total of 26 days of attack out of a total of 276 days considered in the training set that spanned those attack days constituting a positive class ratio of 9.4\%. Table~\ref{tab:stats_arm} shows the statistics of the training and test data for the 3 cyber attacks types from Armstrong. Although we did not use remedial diagnostics in our learning models to account for this class imbalance, the absence of a large training dataset and the missing attack data information accounting for irregularities make a strong case for using sampling techniques to address these issues which we leave as a future research direction for cyber attack focused studies. One of the challenges in remedial diagnostics for imbalances in classes is that here we need to take into account the temporal dependencies while incorporating any sampling techniques as remedies. However, we run a complementary experiment using SMOTE sampling as a simple measure for introducing synthetic samples into the training dataset which we discuss in Section~\ref{sec:super_pred}.
	
	For evaluating the performance of the models on the dataset, we split the time frame of each event into 70\%-30\% averaged to the nearest month separately for each $event-type$. That is we take the first 70\% of time in months as the training dataset and the rest 30\% in sequence for the test dataset. We avoid shuffle split as generally being done in cross-validation techniques in order to consider the consistency in using  sequential information when computing the features. As shown in Figures~\ref{fig:types_events}, since the period of attack information provided varies in time for each of the events, we use different time frames for the training model and the test sets. For the event \textit{malicious email} which remains our primary testbed evaluation event, we consider the time period from October 2016 to May 2017 (8 months) in the darkweb forums for our training data and the period from June  2017 to August 2017 (3 months) as our test dataset, for the $endpoint-malware$, we use the time period from April 2016 to September 2016 (6 months) as our training time period and June 2017 to August 2017 (3 months) as our test data for evaluation.
	
	\subsubsection{Unsupervised model prediction performance}
	Here we use the subspace projection method described in Sections~\ref{sec:anom_sec}, to filter out anomalies from the SPE residual time series vector $\mathcal{R}_x$. We then use these anomalies to predict the attacks as described there and try to see the tradeoffs between the number of true alerts and the number of false alerts obtained. We consider the first 8 principal components among the 53 forums that we considered. Among them we used the first 3 as the \textit{normal axes} and the rest 5 as our \textit{residual axes} based on empirical evidence that shows these 3 components capture the maximum variance. 
	
	\begin{figure*}[!t]
		\minipage{0.5\textwidth}
		\includegraphics[width=5.5cm, height=3.5cm]{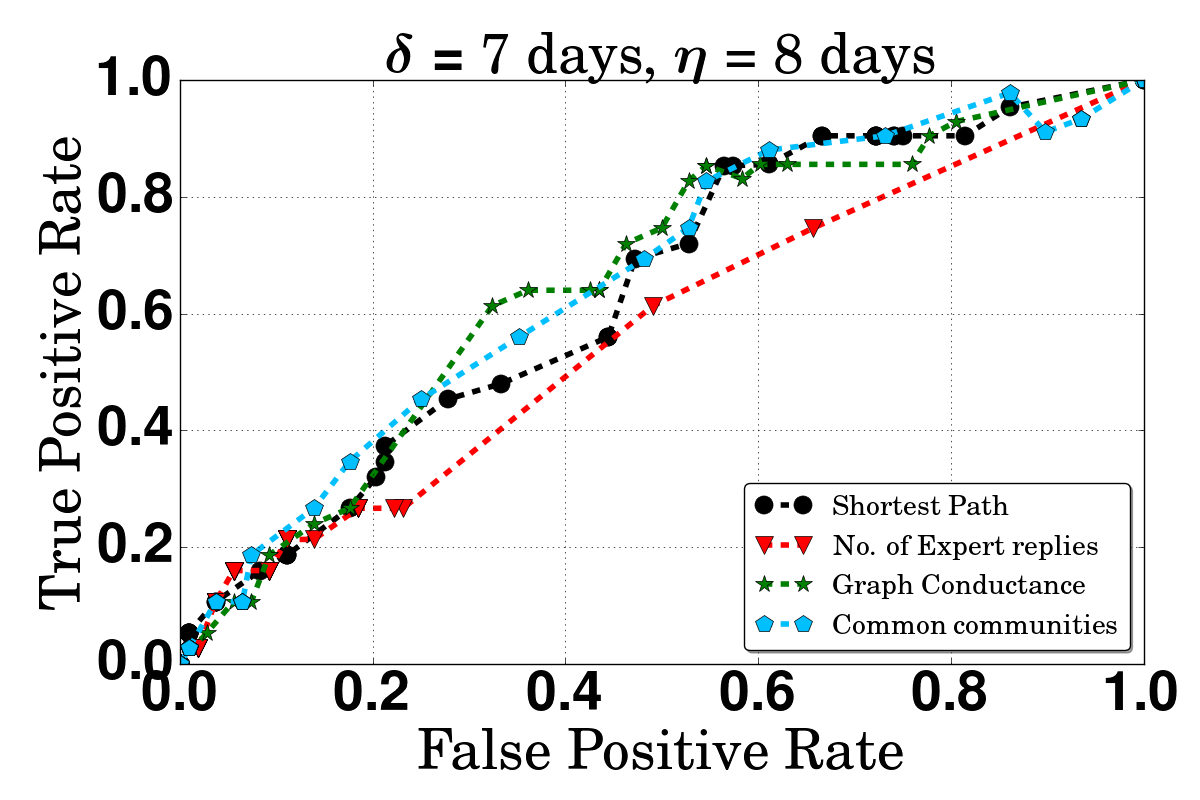}
		\subcaption{malicious-email}
		\endminipage
		\hfill
		\minipage{0.5\textwidth}
		\includegraphics[width=5.5cm, height=3.5cm]{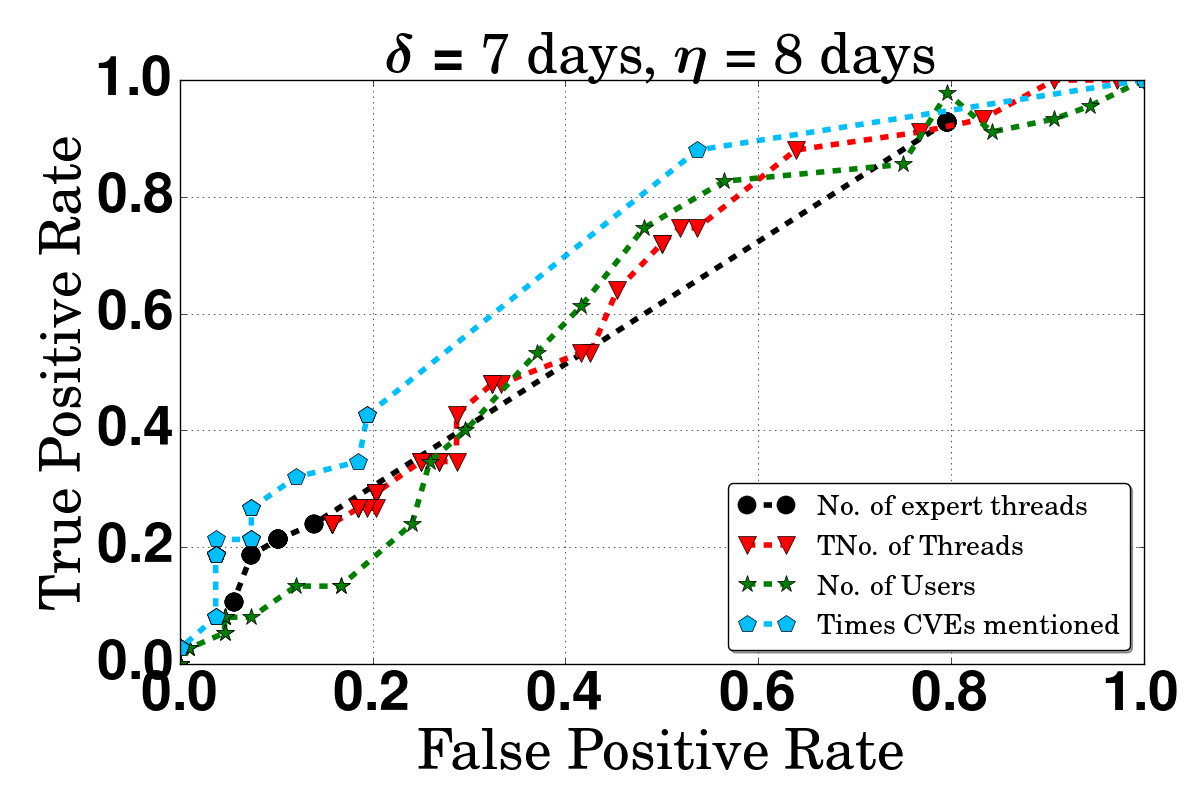}
		\subcaption{malicious-email}
		\endminipage
		\hfill
		\\
		\minipage{0.5\textwidth}
		\includegraphics[width=5.5cm, height=3.5cm]{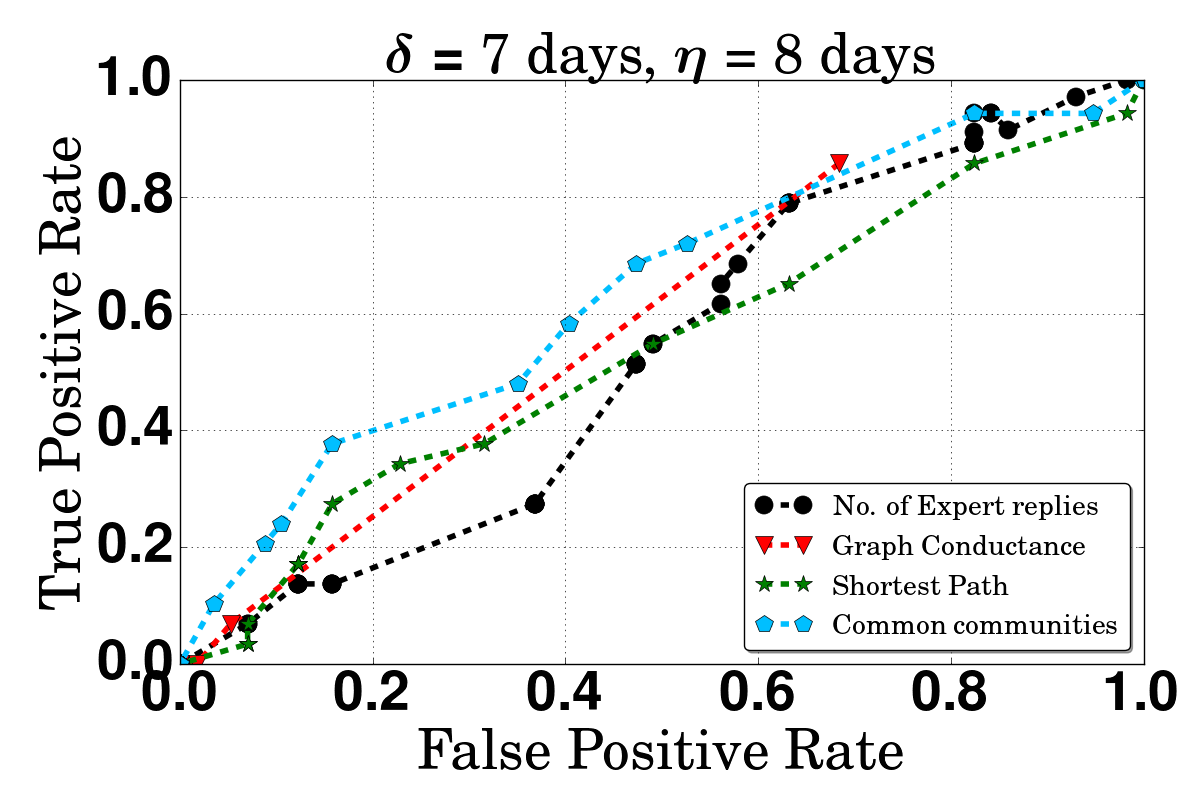}
		\subcaption{endpoint-malware}
		\endminipage
		\hfill
		\minipage{0.5\textwidth}
		\includegraphics[width=5.5cm, height=3.5cm]{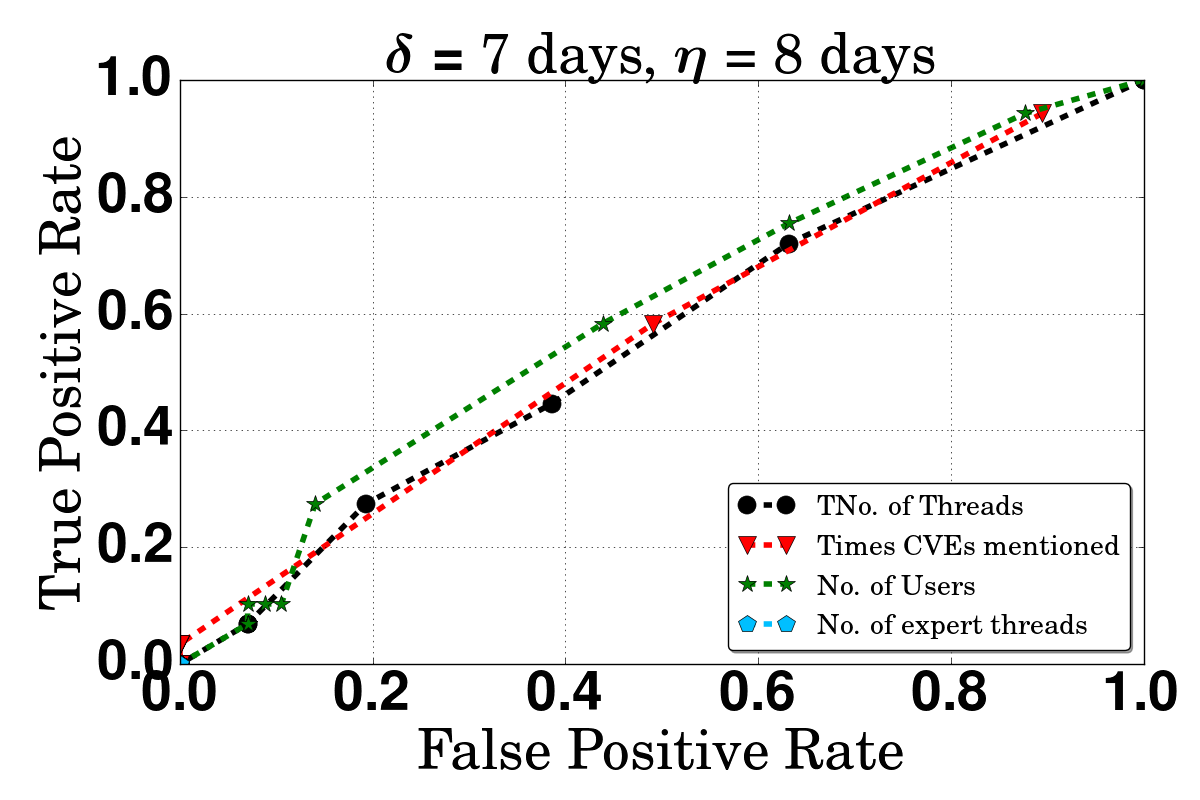}
		\subcaption{endpoint-malware}
		\endminipage
		\hfill
		\\
		\minipage{0.5\textwidth}
		\includegraphics[width=5.5cm, height=3.5cm]{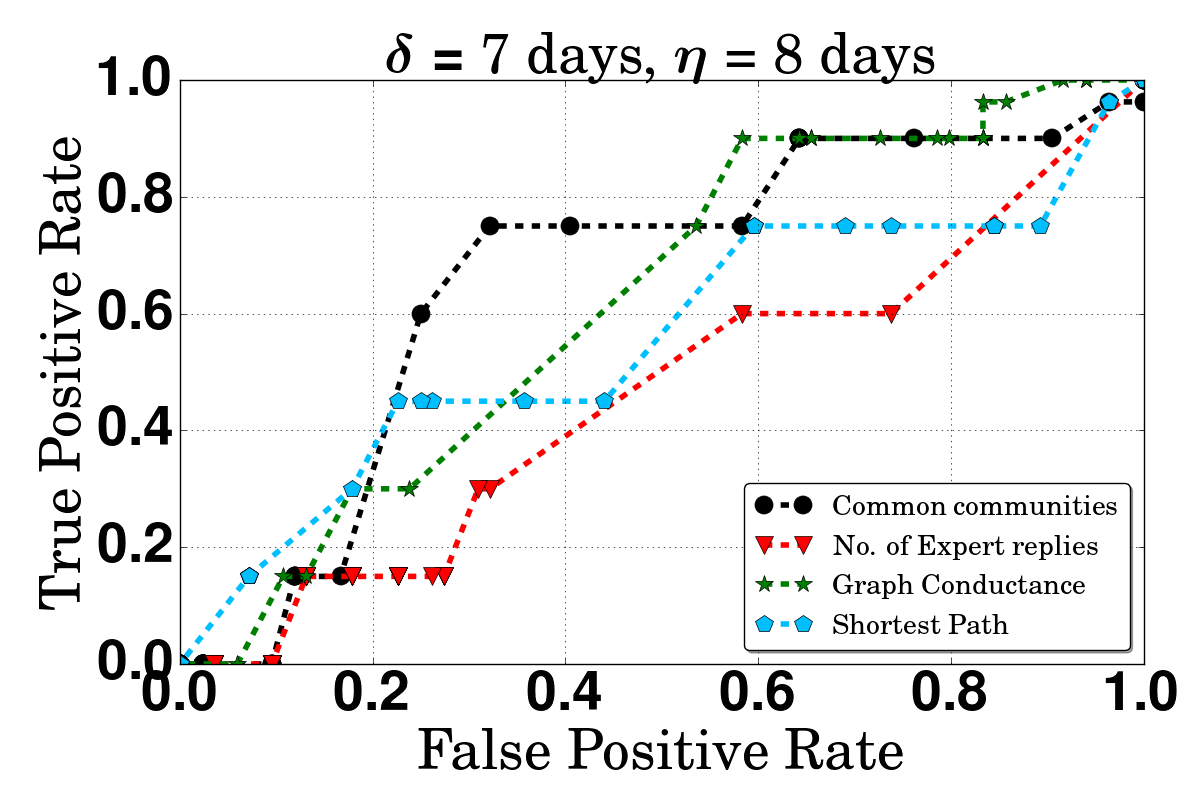}
		\subcaption{malicious-destination}
		\endminipage
		\hfill
		\minipage{0.5\textwidth}
		\includegraphics[width=5.5cm, height=3.5cm]{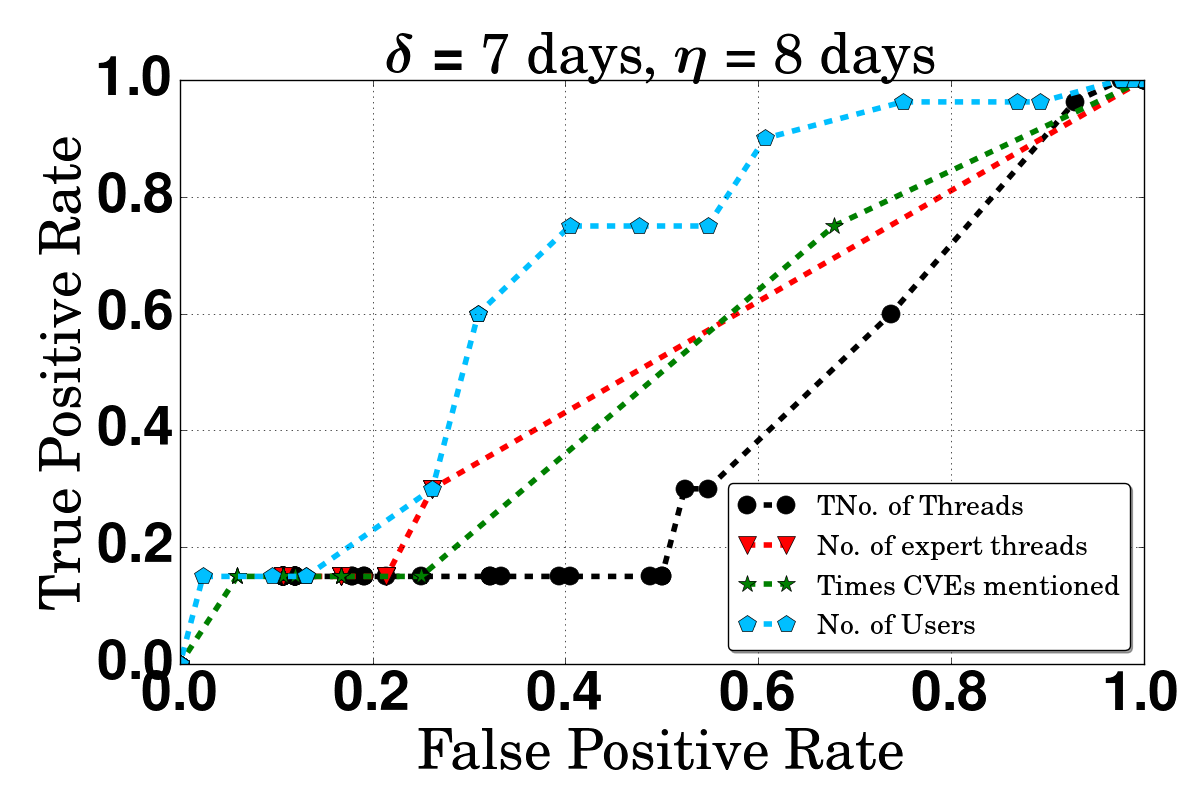}
		\subcaption{malicious-destination}
		\endminipage
		\hfill
		\caption{ROC curves for prediction using unsupervised anomaly detection methods: $\delta$ = 7 days, $\eta$=8 days }
		\label{fig:anom}
	\end{figure*}

	For evaluating the prediction performance, we examined the ROC (Receiver Operating Characteristic) curves for the features over different spans of $\delta$ and $\eta$ but we present our keys findings from the case where we set $\eta$=8 days and $\delta$=7 days shown in Figure~\ref{fig:anom} although we did not find general conclusions over the choices of the parameters $\eta$ and $\delta$ from the results. Each point in these ROC curves denotes a threshold among a set of values chosen for flagging a point in the vector obtained from the squared prediction error of the projected test input $\mathbf{y}$, that crosses the threshold as an anomaly.  We present the results in each plot grouped by the $event-type$ and the feature classes: forum statistics and graph based statistics. From the Figures~\ref{fig:anom}(a) and (b), for the event type $malicious-email$, we obtain the best \textit{AUC (Area Under Curve)} results of 0.67 for the vulnerability mentions by users feature among the forum statistics groups and an AUC of 0.69 for graph conductance among the set of graph based features. For the event type $malicious-destination$, we obtained a best AUC of 0.69 for the common community count feature among the set of graph based features and a best AUC of 0.66 on the number of users at $t_d$ among the forum statistics. For the event-type $endpoint-malware$, we obtain  a best AUC of 0.69 on the number of users stats and 0.63 on the common communities \textit{CC} feature. Empirically, we find that among the network features examined that rely on the set of \textit{experts}, it is not sufficient to just look at how these experts reply to other users in terms of frequency, shown by the results where they exhibit the least AUC in the unsupervised setting that we considered. The fact that common communities and the graph conductance turn out to be better predictors than just the shortest path distance or the number of replies by experts, suggest that \textit{experts} tend to focus on posts of a few individuals when any significant post arises and hence, focusing on individuals who are close to these users in terms of random walks and communities would be favorable.

	One of the reasons behind the poor performance of the detector on the $malicious-destination$ type of attacks compared to $malicious-email$ although the total number of incidents reported for both of them are nearly the same is that the average number of incidents for any week of attack for the 3 attack-types are: for $malicious-email$, we have an average of 2.9 attacks per week, for $endpoint-malware$, we have an average of 3.6 attacks per week and for $malicious-destination$, there are an average of 1.52 attacks per weeks. So although the number of incidents are similar, the number of days of attacks on which the attack occurs is lesser for $malicious-destination$ attacks and  which is important for the binary classification problem considered here.

	\begin{figure*}[!t]
		\minipage{0.5\textwidth}
		\includegraphics[width=4.2cm, height=2.8cm]{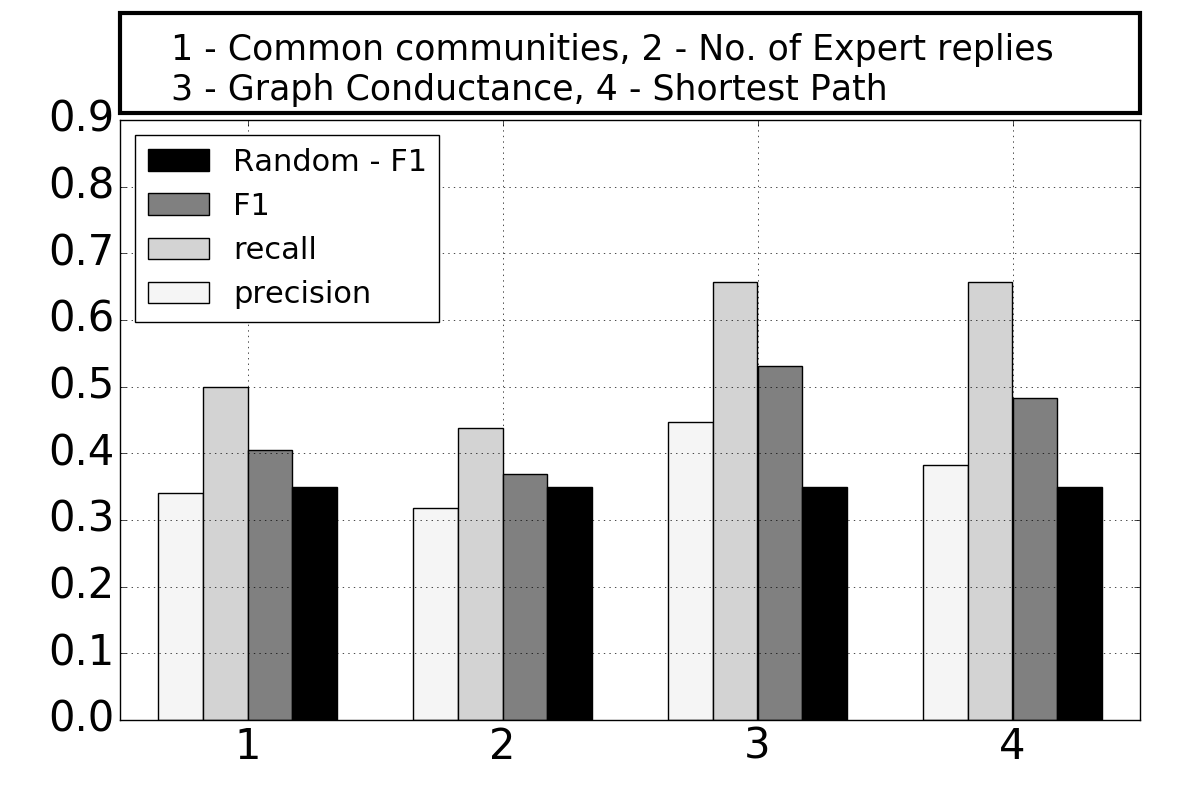}
		\subcaption{malicious-email}
		\endminipage
		\hfill
		\minipage{0.5\textwidth}
		\includegraphics[width=4.2cm, height=2.8cm]{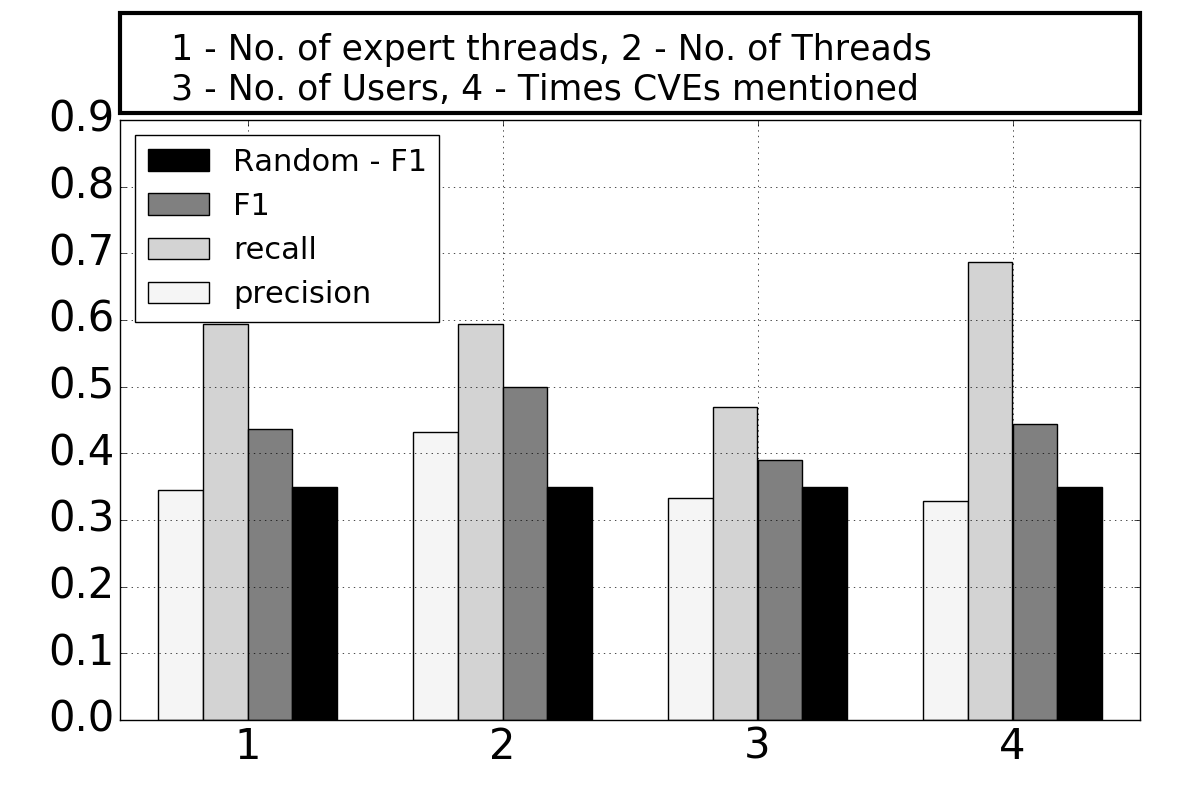}
		\subcaption{malicious-email}
		\endminipage
		\hfill
		\\
		\minipage{0.5\textwidth}
		\includegraphics[width=4.2cm, height=2.8cm]{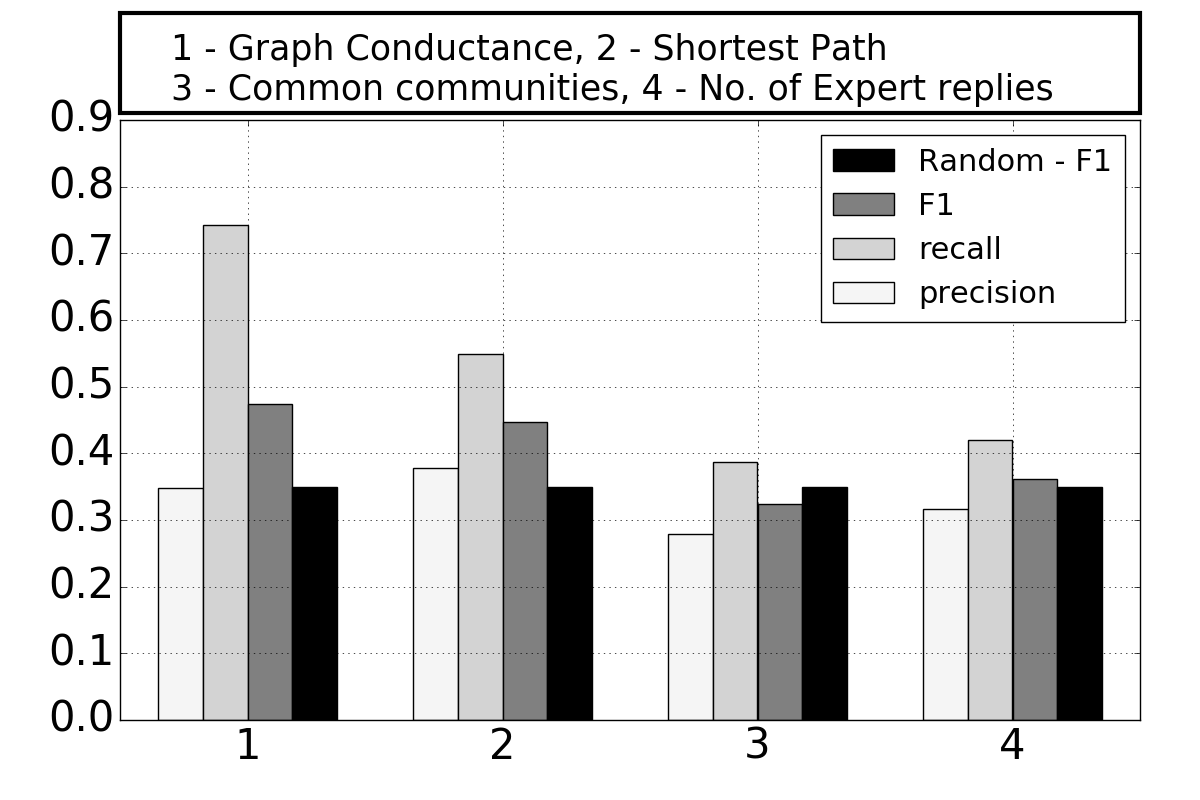}
		\subcaption{endpoint-malware}
		\endminipage
		\hfill
		\minipage{0.5\textwidth}
		\includegraphics[width=4.2cm, height=2.8cm]{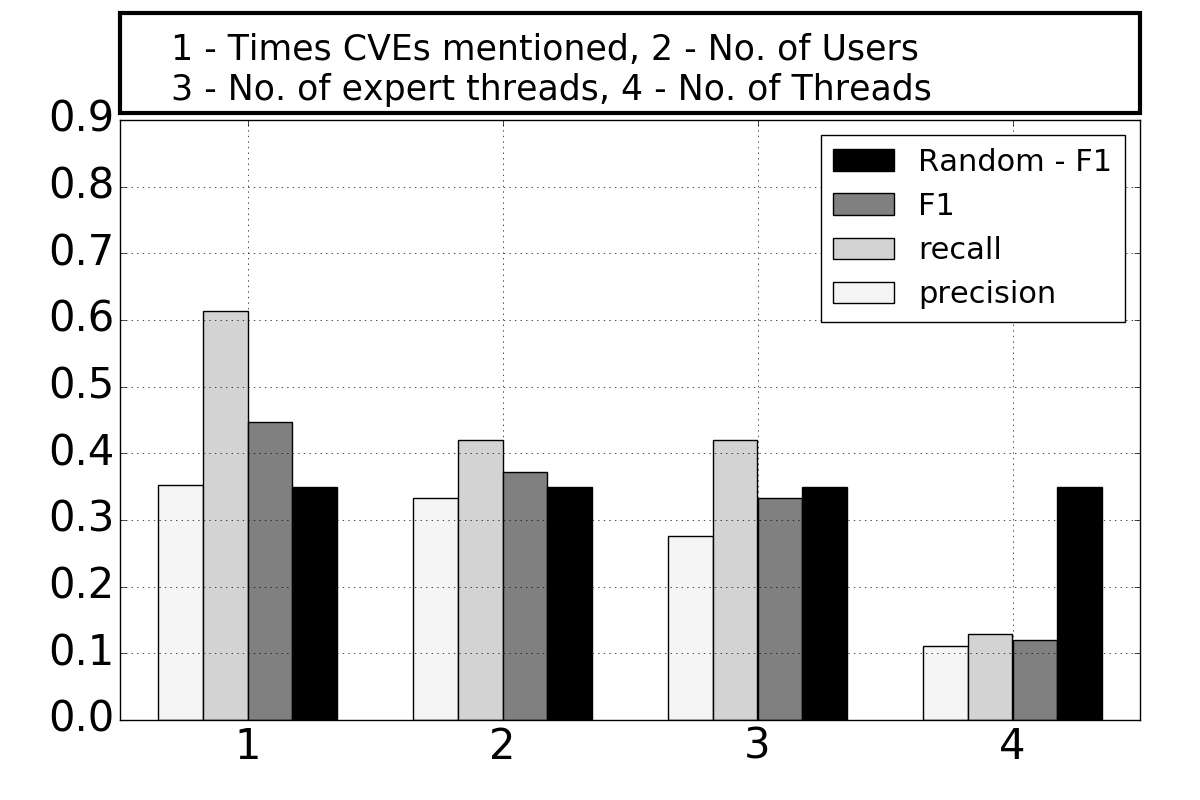}
		\subcaption{endpoint-malware}
		\endminipage
		\hfill
		\caption{Classification results for the features considering the supervised model: $\delta$ = 7 days, $\eta=8$ days.}
		\label{fig:class_7}
	\end{figure*}

	\subsubsection{Supervised model prediction performance} \label{sec:super_pred}
	For the logistic regression model, we consider a span of 1 week time window $\eta$ while keeping $\delta$ = 8 days similar to the unsupervised setting. Due to absence of sufficient positive examples, we avoid using this model for predicting attacks of type $malicious-destination$. From among the set of statistics features that were used for predicting $malicious-email$ attacks shown in Figure~\ref{fig:class_7}(b), we observe the best results using the number of threads as the signal for which we observe a precision of 0.43, recall of 0.59 and an F1 score of 0.5 against the random F1 of 0.34 for this type of attacks. From among the set of graph based features, we obtain the best results from graph conductance with a precision of 0.44, recall of 0.65 and an F1 score of 0.53 which shows an increase in recall over the number of threads measure. Additionally, we observe that in case of supervised prediction, the best features in terms of F1 score are graph conductance and shortest paths whereas number of threads and vulnerability mentions turn out to be the best among the statistics. For the attacks belonging to the type $endpoint-malware$, we observe similar characteristics for the graph features where we obtain a best precision of 0.34, recall of 0.74 and an F1 score of 0.47 against a random F1 of 0.35, followed by the shortest paths measure. Howeve,r for the statistics measures we obtain a precision of 0.35, recall 0.61 and an F1 score of 0.45 for the vulnerability mentions followed by the number of threads which gives us an F1 score of 0.43. Although the common communities features doesn't help much in the overall prediction results, in the following section we describe a special case that demonstrates the predictive power of the community structure in networks. The challenging nature of the supervised prediction problem is not just due to the issue of class imbalance, but also the lack of large samples in the dataset which if present, could have been used for sampling purposes. As an experiment, we also used Random Forests as the classification model, but we did not observe any significant improvements in the results over the random case, suggesting the LR model with temporal regularization  helps in these cases of time series predictions. 
	
	\begin{figure*}[!t]
		\minipage{0.1\textwidth}
		\includegraphics[width=6cm, height=3.5cm]{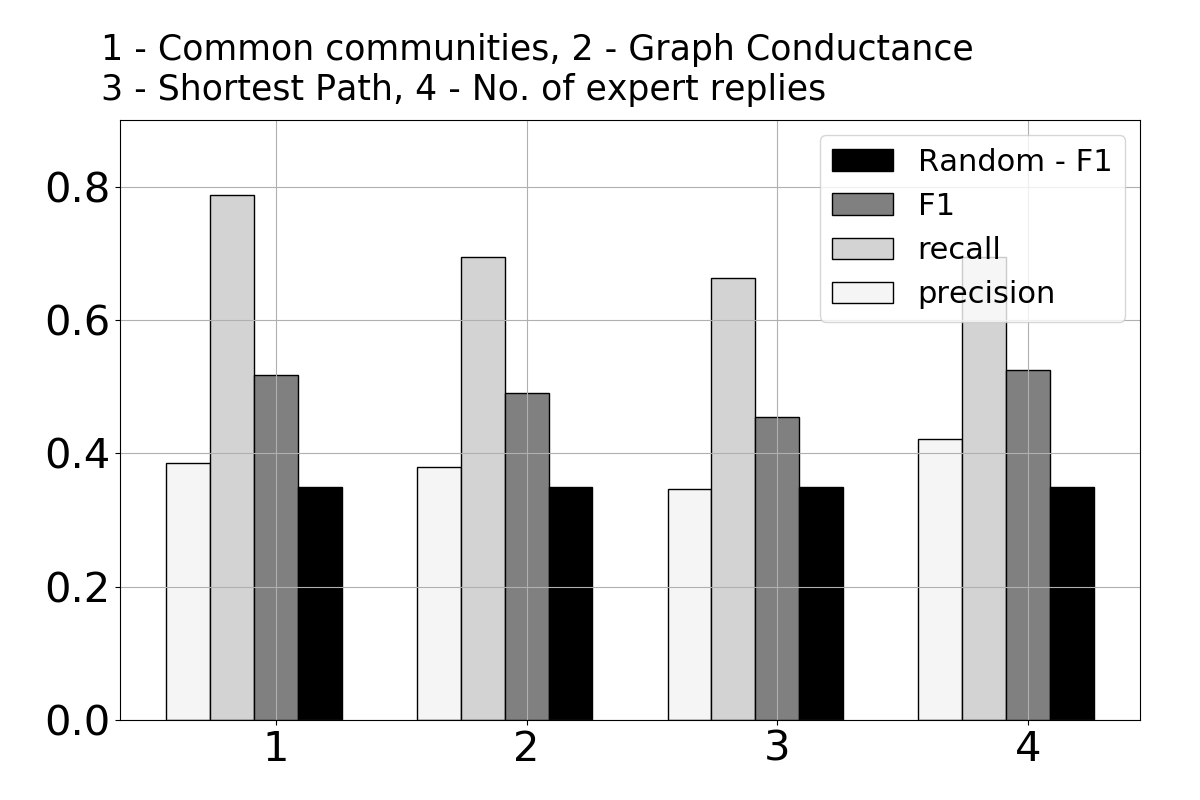}
		\endminipage
		\hfill
		\minipage{0.5\textwidth}
		\includegraphics[width=6cm, height=3.5cm]{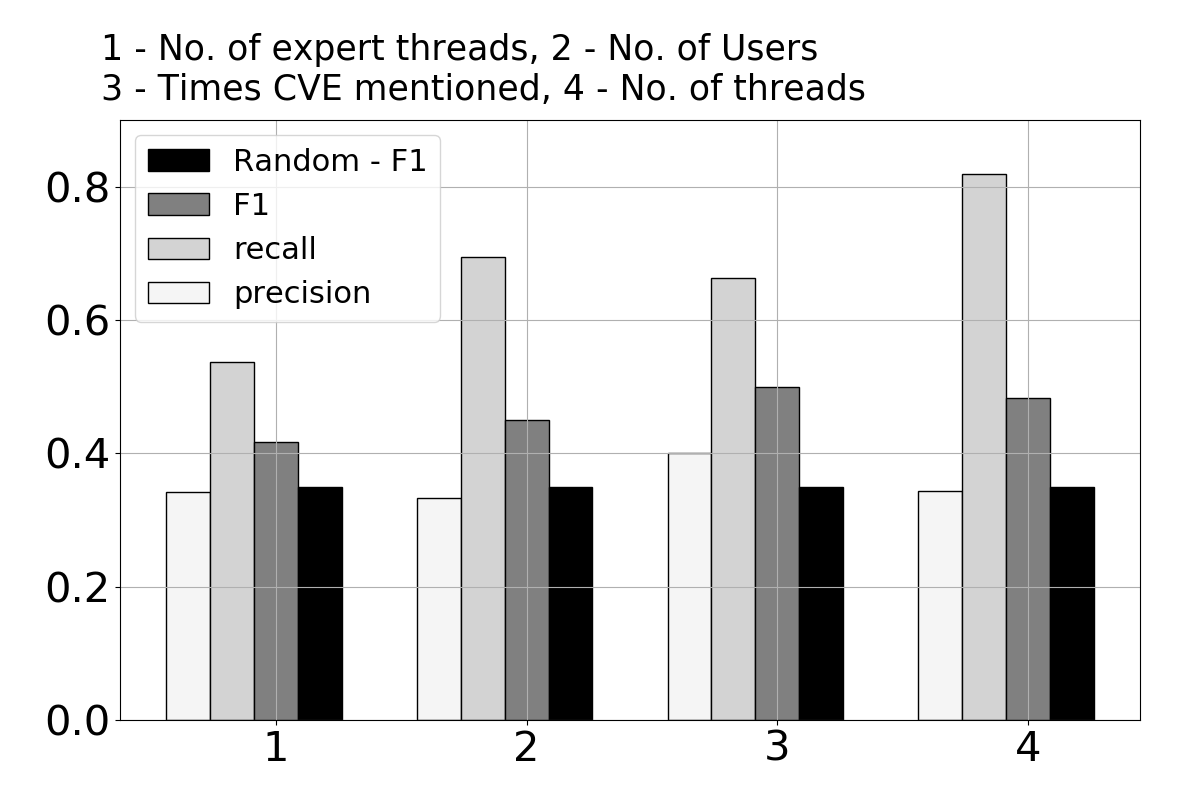}
		\endminipage
		\hfill
		\caption{Classification results for the malicious-email attack dataset using SMOTE sampling on top of the supervised learning model}
		\label{fig:class_SMOTE}
	\end{figure*}
	
	Additionally, we use SMOTE to deal with the class imbalance and we plot the results for the malicious email attacks in Figure\ref{fig:class_SMOTE} - from the results and comparing them with those in Figure~\ref{fig:class_7}, we find that while for all features the recall increases, the precision drops substantially. We find that among the graph features, both graph conductance and the number of expert replies perform equally well with an F1 score of 0.52 while the number of threads with CVE mentions achieves the best results with an F1 score of 0.49.
	
	\subsubsection{Model with feature combinations} \label{sec:glasso}
	One of the major problems of the dataset is the imbalance in the training and test dataset as will be described in Section~\ref{sec:exp}. The added complexities arise from the fact that if we consider all features over the time window of feature selection, then the total number of features $z$ (variables) for the learning models is: $z$ = $\#$ features $\times$ $(\eta)$. In our scenario, this would typically be almost equal to the number of data points we have for training depending on $\eta$ and also depending on whether we consider different variations of the features in Table~\ref{tab:table_feat}, which might result in overfitting.  So in order to use all features in each group together for prediction, we use 3 additional regularization terms in the longitudinal regression model : the L1 penalty, the L2 penalty and the \textit{Group Lasso} regularization \cite{group_lasso}. We adapt this framework of regularization to our set of features following previous studies on lasso for longitudinal data \cite{long_lasso} and the final objective function can be written as:
	
	\begin{equation}\label{eq:glasso}
	l(\mathbf{\beta}) = -\sum_{i=1}^{N} \log(1 + e^{-y_i (\beta^T \mathbf{x_i}) }) + \frac{m}{2} \lVert \beta \rVert^2_2  + l\lVert \beta \rVert_1 + g.GL(\beta)
	\end{equation} 
	where $m$, $l$ and $g$ are the hyper-parameters for the regularization terms and the $GL(\beta)$ term is 
	$\sum_{g=1}^G \lVert \beta_{\mathcal{I}_g} \rVert_2$, where $\mathcal{I}_g$ is the index set belonging to the $g^{th}$ group of variables, $g = 1 \ldots G$. Here each $g$ is the time index $t_h$ $\in [t_{-\eta-\delta}, \  t_{-\delta}]$, so this group variable selection selects all features of one time in history while reducing some other time points to 0.  It has the attractive property that it does variable selection at the temporal group level and is invariant under (group-wise) orthogonal transformations like ridge regression. We note that while there are several other models that could be used for prediction that incorporates the temporal and sequential nature of the data like hidden markov models (HMM) and recurrent neural networks (RNN), the logit model allows us to transparently adjust to the sparsity of data, specially in the absence of a large dataset. For the model with the Group lasso regularization in Equation~\ref{eq:glasso}, we set the parameters $m, l, g$ and 0.3, 0.3 and 0.1 based on a grid search on $m$ an $l$ and keeping $g$ low so that most time points within a single feature is set to 0 for avoiding overfitting.

	\begin{figure}[!t]
		\centering
		\minipage{0.1\textwidth}
		\includegraphics[width=6cm, height=3.5cm]{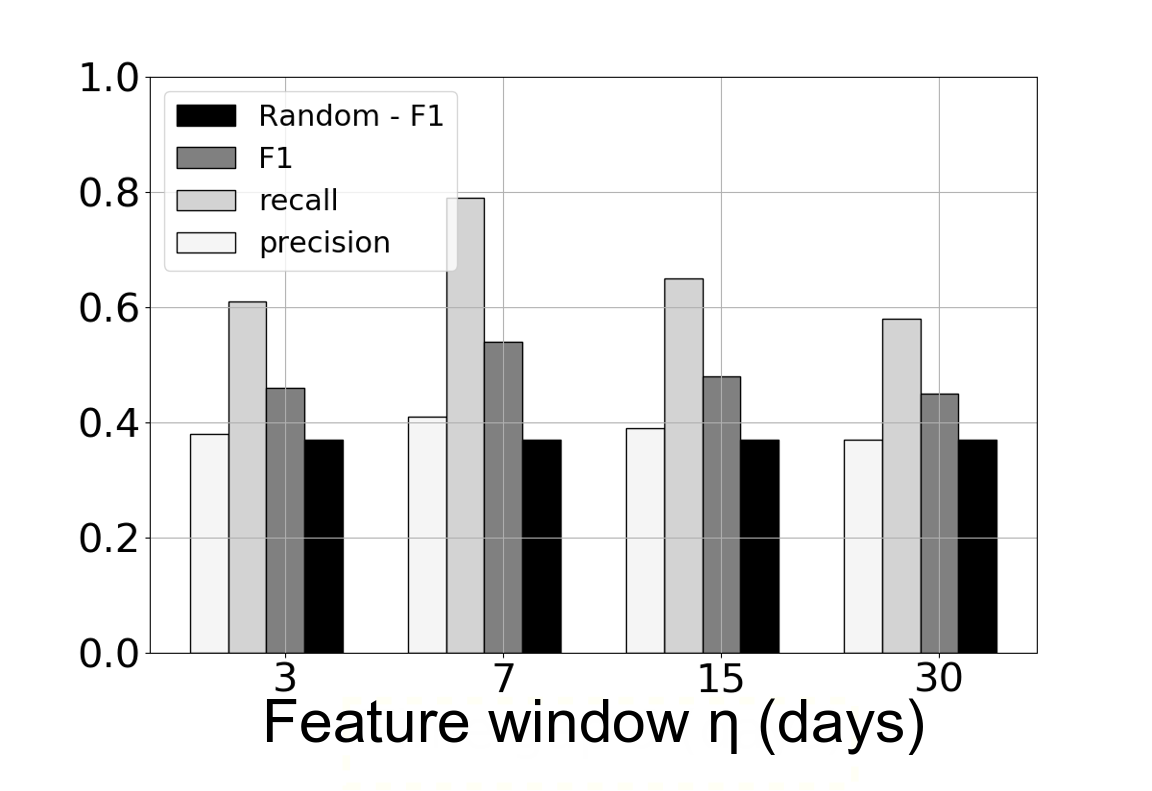}
		
		\endminipage
		\hfill
		\minipage{0.5\textwidth}
		\includegraphics[width=6cm, height=3.5cm]{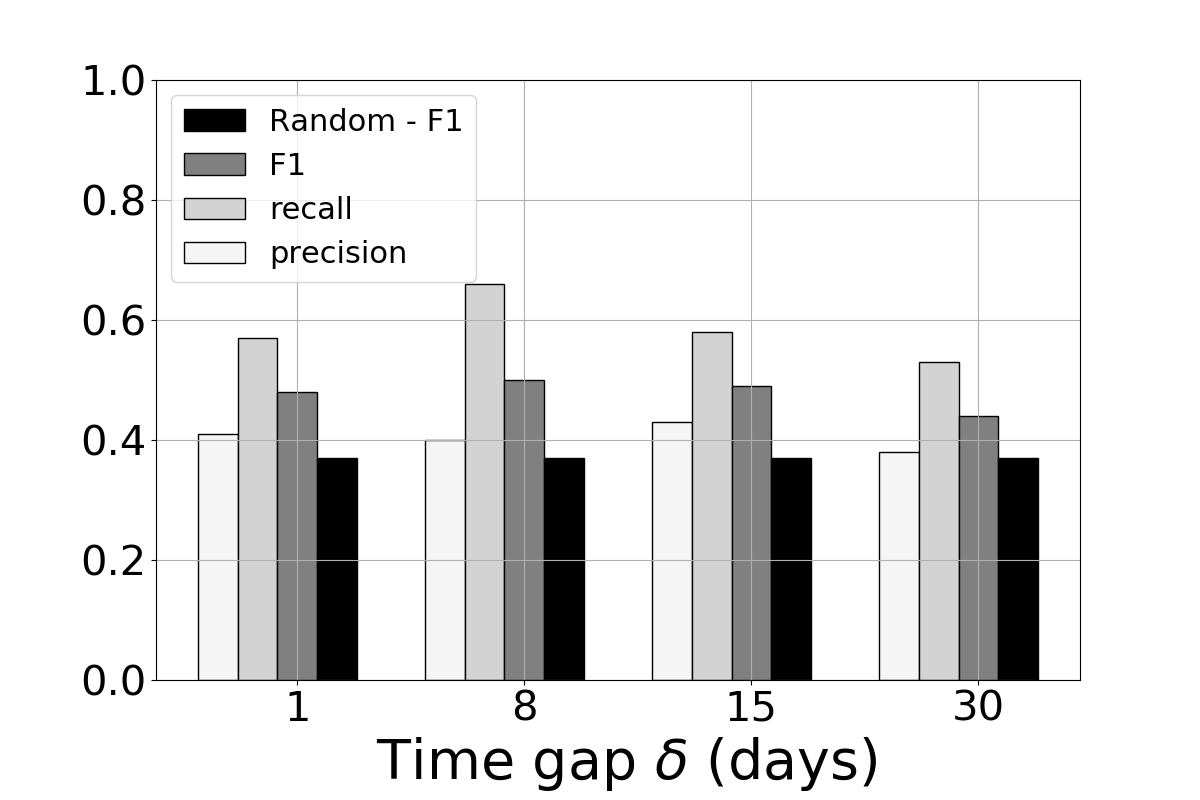}
		
		\endminipage
		\hfill
		\caption{Classification results for malicious email with feature combination and considering group lasso: m=0.3, l=0.3, g=0.1. Refer to Equation~\ref{eq:glasso} for the model used for this prediction.}
		\label{fig:glassost}
	\end{figure}
	
	We cross validated this model on the 2 hyper-parameters: $\eta$ and $\delta$ and we found that while the recall increases for all combinations of hyper-parameters for all features compared to results shown in Figure~\ref{fig:class_7}, the precision remains the same across different values of the hyper-parameters. We test on different $\eta$ keeping $\delta$ fixed at 8 days and we test on different $\delta$ keeping $\eta$ fixed at 7 days. We obtain the best results predicting attacks for  the malicious-email type using $n=7$ and $\delta=8$ days - we get a best F1 value of 0.56 (using $eta$ = 7 days and keeping $delta$ fixed at 8) using this feature combination model against the best F1 score of 0.53 obtained from using single features without regularization.  
	
	\section{Discussions}
	\label{sec:discuss}
	
	As with most machine learning models and setups that attempt binary and multiclass classification including neural networks, the features attributed to the predictions can in most situations explain correlation - the causation needs more controlled studies like visualization by projecting features onto a lower dimensional space, ablation studies or understanding feature importance and using regularization techniques for ensuring sparsity for some features or eliminating redundancy \cite{lime}. To this end, we try to investigate whether our framework with the signals from the darwkeb discussions correlate to real world events or to other types of attacks. We present 3 controlled studies that show the extent to which the results of our framework are interpretable.  
	
	\subsection{Prediction in High Activity Weeks} 
	One of the main challenges in predicting external threats without any method to correlate them with external data sources like darkweb or any other database is that it is difficult to validate which kinds of attacks are most correlated with these data sources. To this end, we examine a controlled experiment setup for the $malicious-email$ attacks in which we only consider the weeks which exhibited high frequency of attacks compared to the overall timeframe: in our case we consider weeks having more than 5 attacks in test time frame. These high numbers may be due to multiple attacks in one or few specific days or few attacks on all days. The main idea is to see how well does the supervised model perform in these weeks of interest compared to the random predictions with and without prior distribution of attack information. We run the same supervised prediction method but evaluate them only on these specific weeks.
	
	\begin{figure}[!h]
		\centering
		\minipage{0.1\textwidth}
		\includegraphics[width=6cm, height=3.5cm]{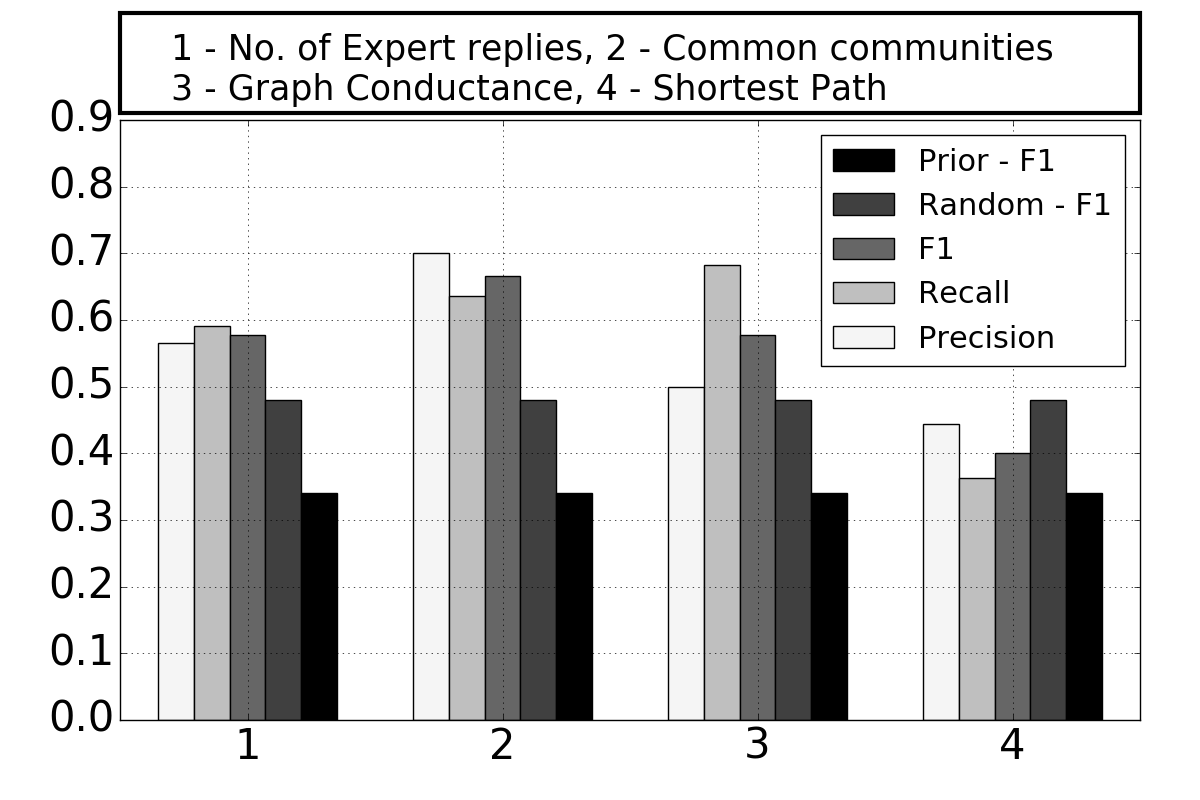}
		
		\endminipage
		\hfill
		\minipage{0.5\textwidth}
		\includegraphics[width=6cm, height=3.5cm]{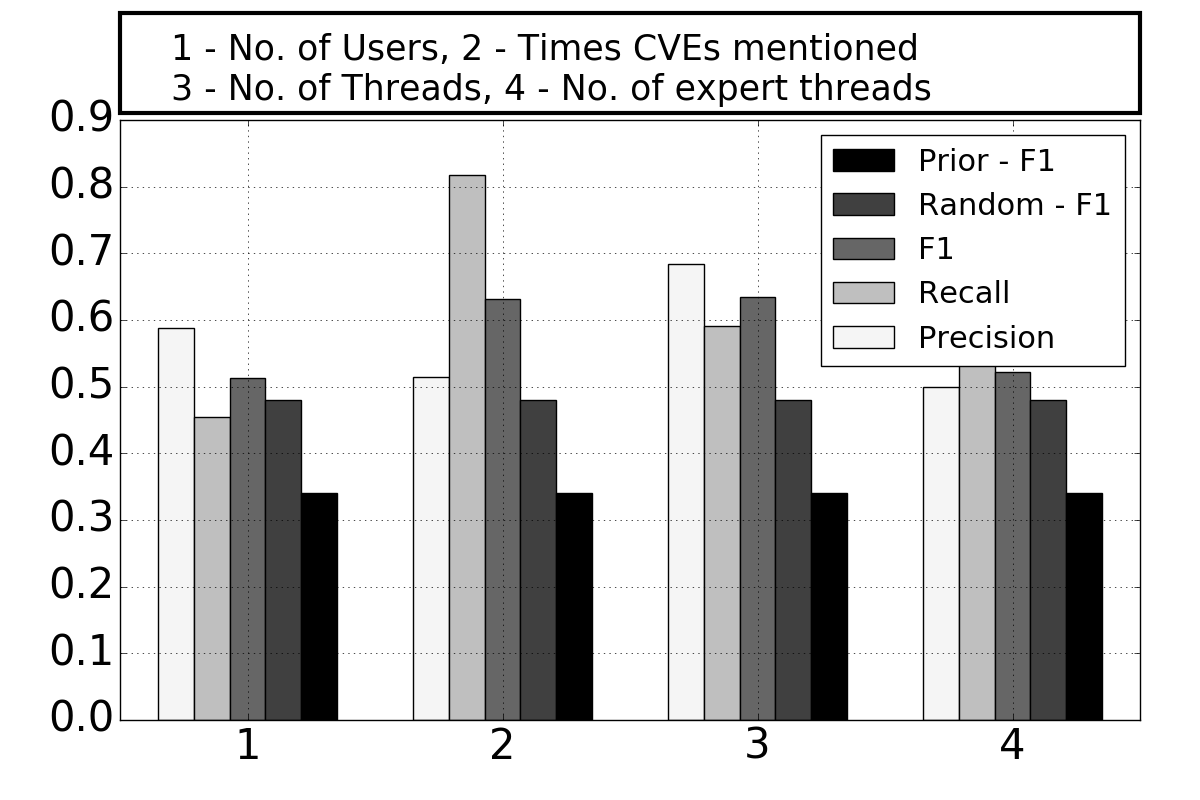}
		
		\endminipage
		\hfill
		\caption{Classification results for $malicious-email$ attacks in high frequency weeks, $\delta$ = 7 days and $\eta$ = 8 days.}
		\label{fig:class_best}
	\end{figure}
	
	From the results shown in Figure~\ref{fig:class_best}, we find that the best results were shown by the common communities feature having a precision of 0.7 and a recall of 0.63 and an F1 score of 0.67 compared to the random (no priors) F1 score of 0.48 and a random (with priors) F1 score of 0.34 for the same time parameters. Among the statistics measures,  we obtained a highest F1 score of 0.63 for the vulnerability mentions feature. Additionally, we find unlike the results over all the days, for these specific weeks, the model achieves high precision while maintaining comparable recall emphasizing the fact that the number of false positives are also reduced during these periods. This empirically suggests that for weeks that exhibit huge attacks, looking at Darwkeb sources for vulnerability mentions and the network structure analytics can definitely help predict cyber attacks.
	
	\begin{figure}[!t]
		\centering
		\includegraphics[width=10cm, height=5cm]{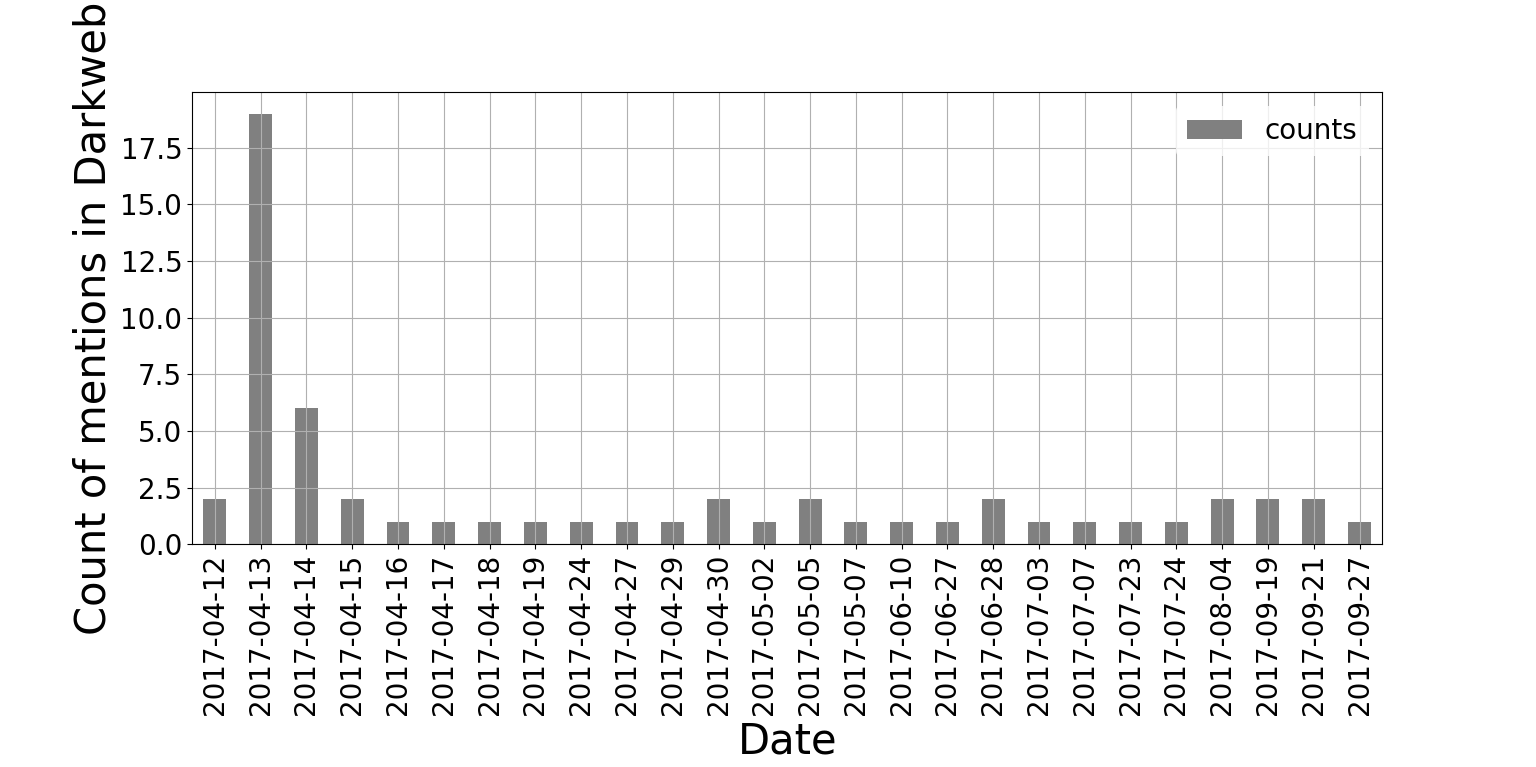}
		\caption{Lifecycle of darkweb forum mentions of the vulnerability CVE-2017-0199.}
		\label{fig:cve_0199}
	\end{figure}
	
	\subsection{Real World Attacks}
	In order to assess whether the features and the learning model are predictive of \textit{vulnerability exploitation based cyber attack} incidents in the real world, we manually collected one case of vulnerability exploitation that led to real world attacks and which had discussions on the darkweb associated with those vulnerabilities. Since our main evaluations were reported on the malicious email incidents and as mentioned before, the \textit{malicious-email} events are caused by malicious email attachments which when downloaded could cause a malicious script to run and execute its code thus intruding the host systems. 
	
	\paragraph{CVE-2017-0199.}
	This vulnerability is exploited through malicious Microsoft Office RTF documents that allows a malicious actor to download and execute a Visual Basic Script when the user opens the document containing the exploit. As reported in several documents \footnote{https://www.fireeye.com/blog/threat-research/2017/04/cve-2017-0199-hta-handler.html, https://portal.msrc.microsoft.com/en-US/security-guidance/advisory/CVE-2017-0199}, the document can be sent through an email or a link attachment and therefore is an example of malicious-email breach. This vulnerability has a CVS severity score of 7.8 which is considered high by NIST \footnote{https://nvd.nist.gov/vuln/detail/CVE-2017-0199}. There were reports of systems being exploited several months even following the patched date of this vulnerability. In this respect, this vulnerability captured a lot of attention due to the widespread damage that it created. The lifecycle of that vulnerability in the darkweb is shown in Figure~\ref{fig:cve_0199}.
	
	\begin{figure}[!t]
		\centering
		\minipage{0.7\textwidth}
		\includegraphics[width=9cm, height=4.5cm]{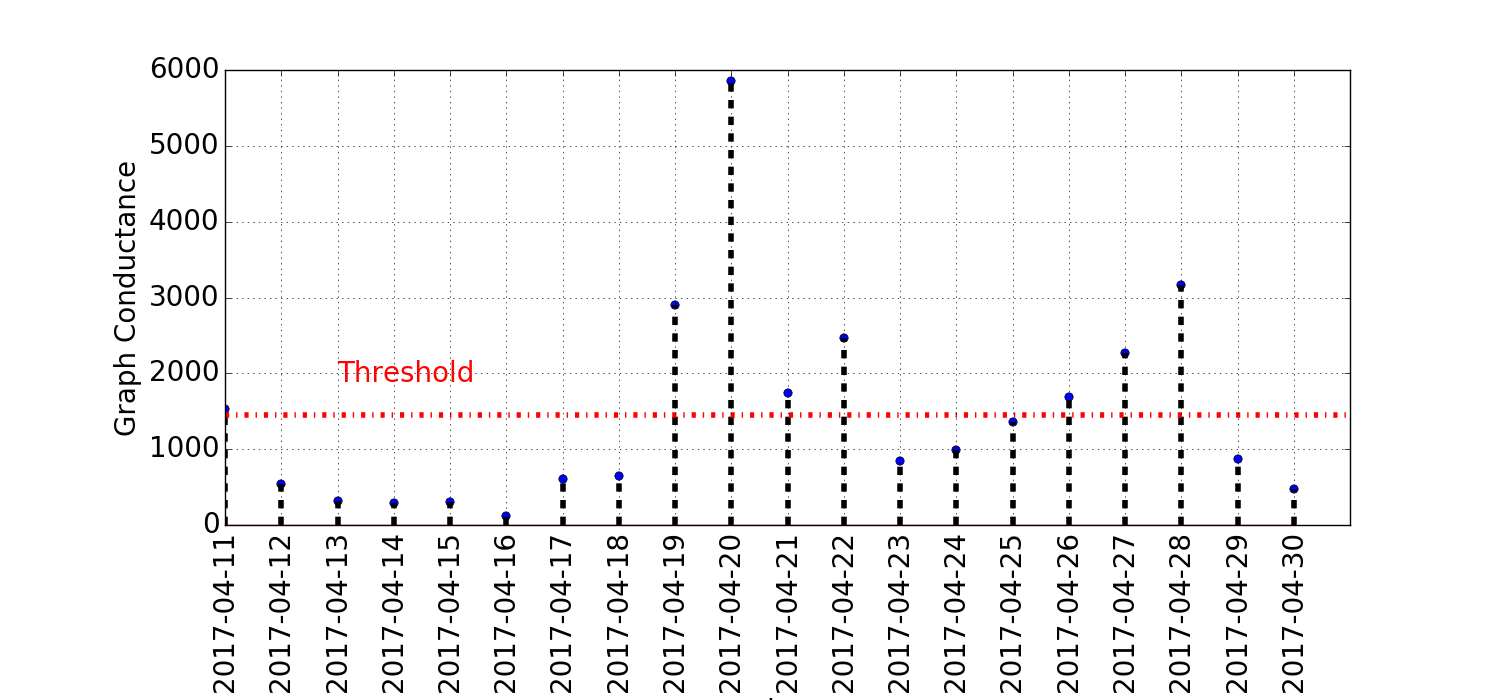}
		\subcaption{}
		\endminipage
		\hfill
		\\
		\minipage{0.7\textwidth}
		\includegraphics[width=9cm, height=4.5cm]{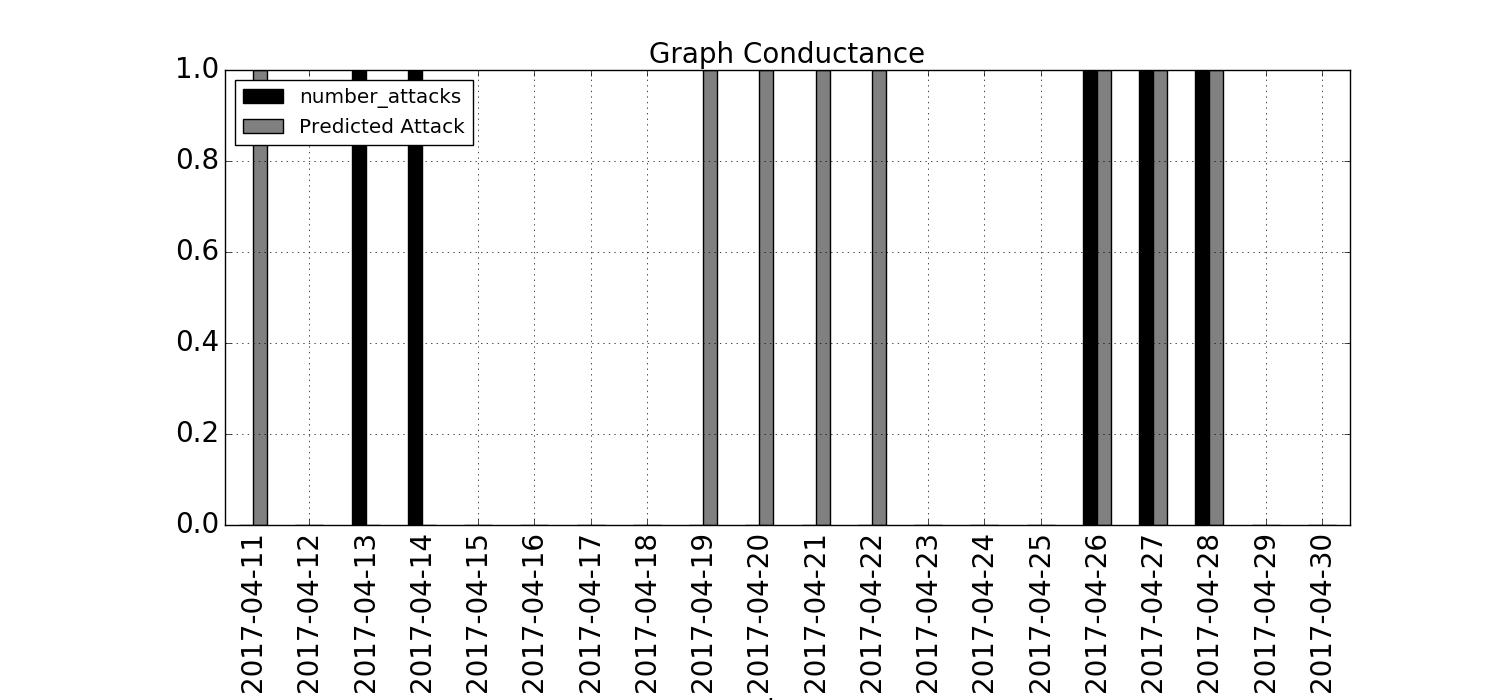}
		\subcaption{}
		\endminipage
		\hfill
		\caption{(a) The graph conductance measure plotted for the weeks from April 11 2017 to April 2017. The red line denotes the threshold $\delta_{\alpha}$ above which an anomaly is flagged for that day. (b) The actual attack vector and the predicted attacks in the same weeks.}
		\label{fig:attacks_pred}
	\end{figure}
	
	Although Microsoft released the patch on April 11, 2017 \footnote{https://blog.talosintelligence.com/2017/04/cve-2017-0199.html}, discussions started as early as April 12 on the darkweb and there were 18 discussions mentioning the vulnerability on April 13, 2017. When we looked at  the content of the discussions on April 13, 2017, we found that most of the discussions surrounding users trying to execute the exploit - whether with malicious intentions or not is a research of sentiment analysis which is also conducted in this domain \cite{al2015bisal,chen2008sentiment}. When we looked at the attacks in the same and following weeks from Armstrong's malicious email incidents dataset, we found that the first attack occurred on April 13, 2018 and in the following week there were attacks on 3 consecutive days April, 26, 27 and 28 as shown in Figure~\ref{fig:attacks_pred}(b).  The period contained a total of 5 days of reported malicious-email incidents in the span of 20 days considered. 
	
	\begin{figure*}[!t]
		\centering
		\minipage{0.5\textwidth}
		\includegraphics[width=6cm, height=4cm]{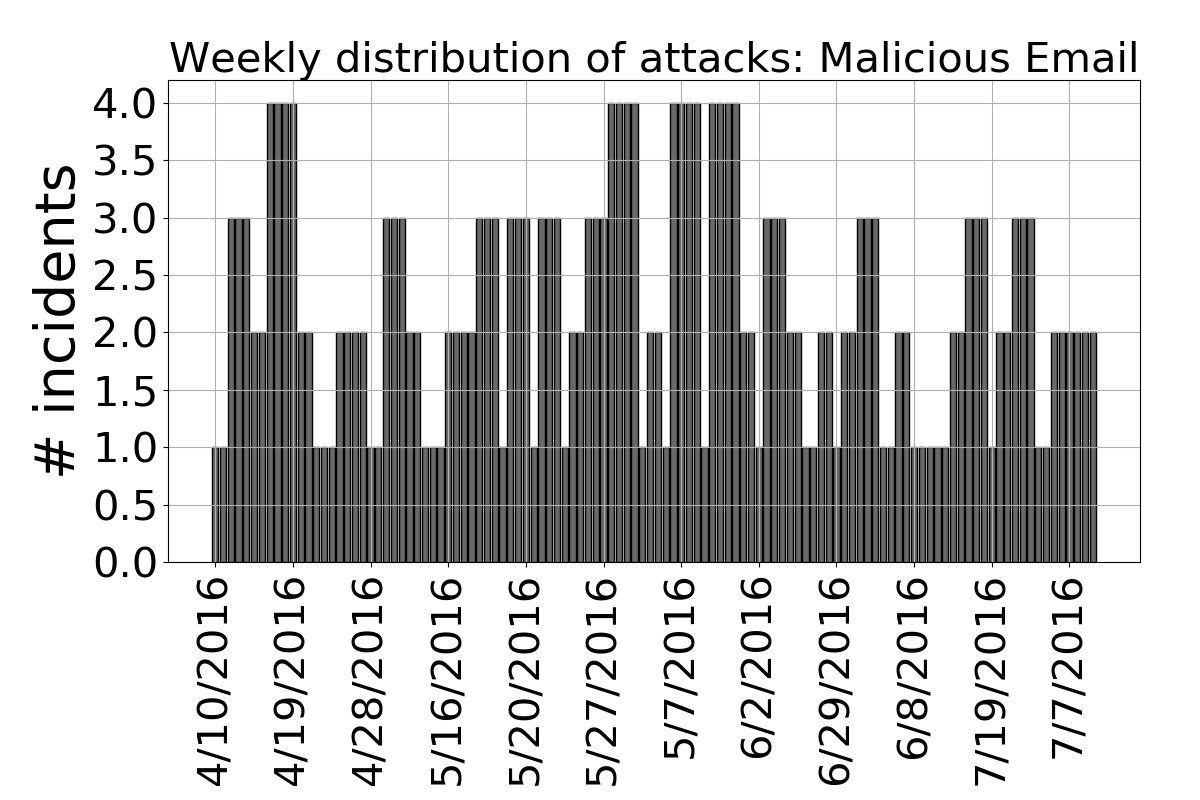}
		\subcaption{}
		\endminipage
		\hfill
		\minipage{0.5\textwidth}
		\includegraphics[width=6cm, height=4cm]{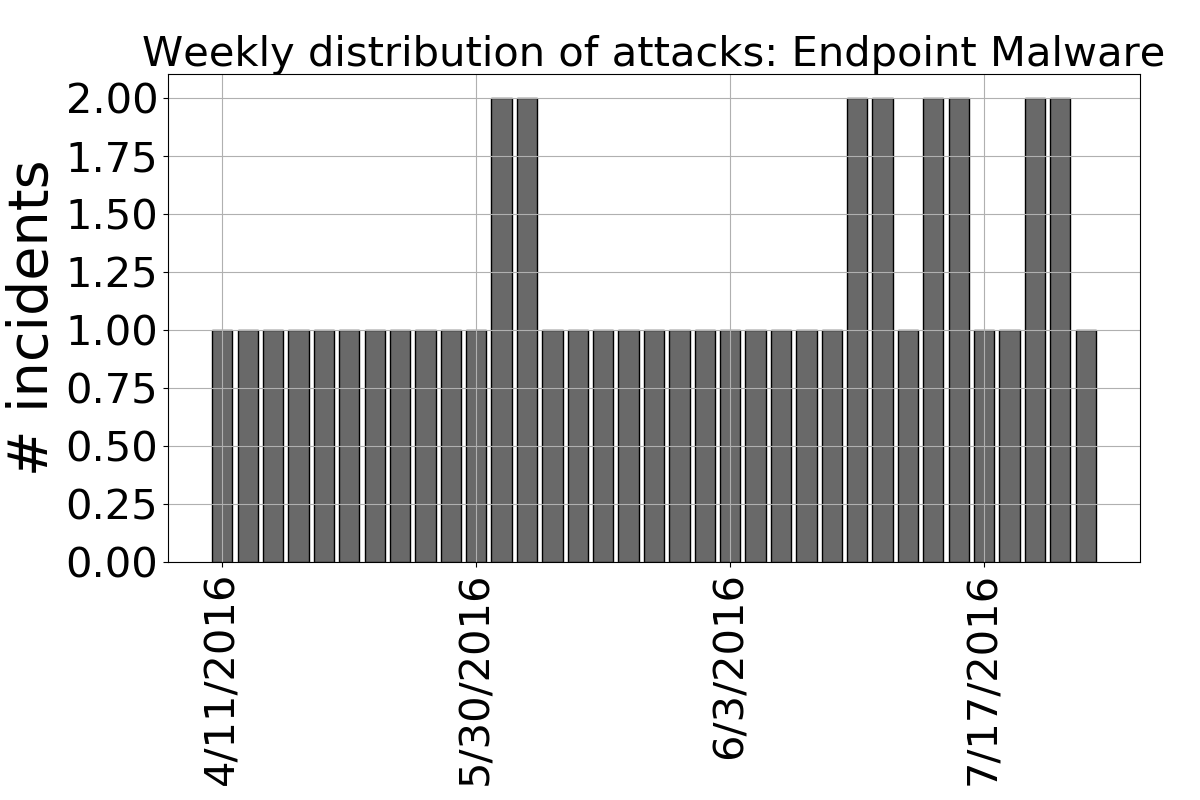}
		\subcaption{}
		\endminipage
		\hfill
		\\
		\minipage{0.5\textwidth}
		\includegraphics[width=6cm, height=4cm]{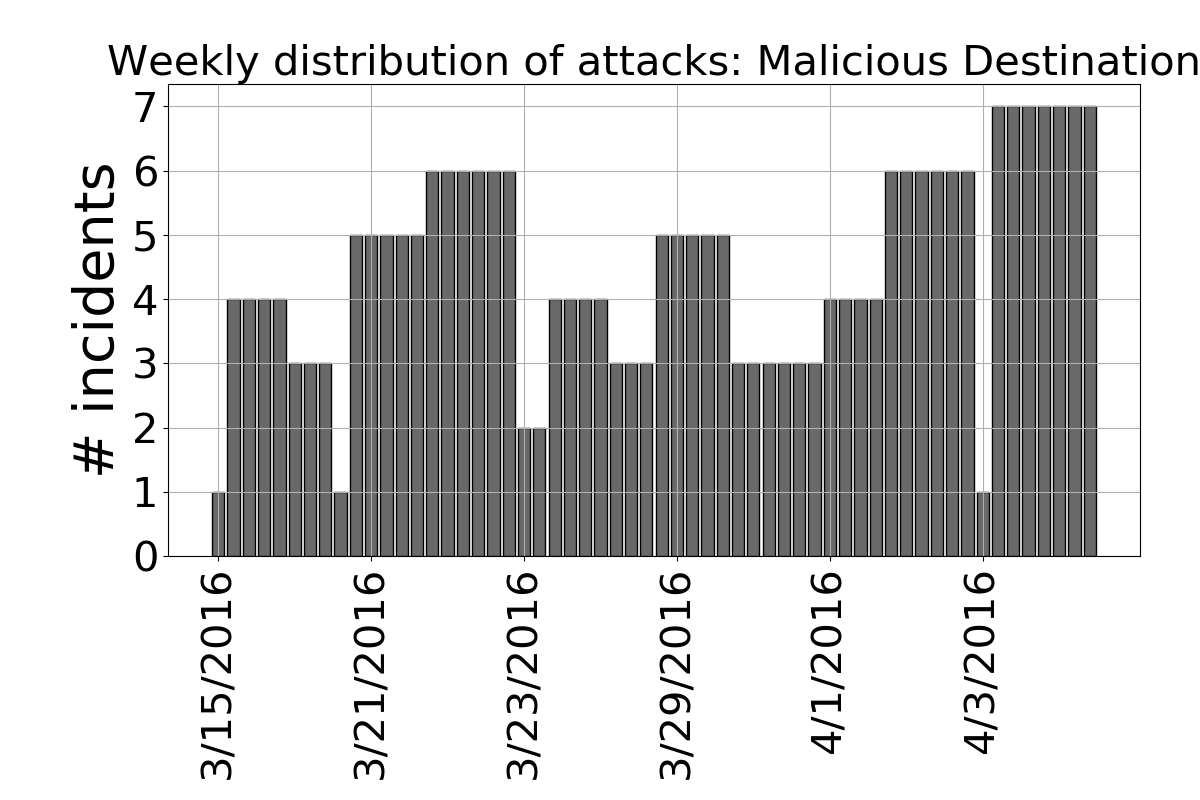}
		\subcaption{}
		\endminipage
		\hfill
		\caption{Weekly occurrence of security breach incidents for Dexter of different types (a) Malicious email (b) Endpoint Malware (c) Malicious destination}
		\label{fig:types_events_dexter}
	\end{figure*}
	We use  $\eta$=7 days, and $\tau$=8 days for the features (the same parameters used in the previous experiments) and we set $\zeta$=7, that is we flag a day $t$ as an anomaly if $\mathcal{N}(t) \geq 1$, or in other words if there is at least one anomaly flagged in the time period $[t_{-\eta-\delta}, \  t_{-\delta}]$. For setting the thresholds that captures whether a particular day has an anomaly in terms of the feature values, we kept the threshold to the mean of the feature values obtained from the training dataset for the respective features. Here we show the feature \textit{Graph Conductance} for the weeks in Figure~\ref{fig:attacks_pred}(a), the  red line denoting the mean of the training data. We flag any day $t$ as having an anomaly if the graph conductance on that day crosses the red line. This setup was able to predict the attacks on days April 26, 27 and 28 successfully while missing the attacks on April 13 and April 14. This led to a precision of 0.26 and recall of 0.6 and an F1 score of 0.46 in those 20 days. We have two important observations: first it is clear that the predicted attacks on the 3 days were due to the anomalies raised in the previous 2 weeks as shown in Figure~\ref{fig:attacks_pred}(b) and secondly, although the CVE mentions shown in Figure~\ref{fig:cve_0199} does not show any spikes on April 19, 20, 21 , 22 and our feature anticipated some anomaly on those days which caused the alerts in the following weeks.

	\begin{figure*}[!t]
		\minipage{0.5\textwidth}
		\includegraphics[width=4.2cm, height=2.8cm]{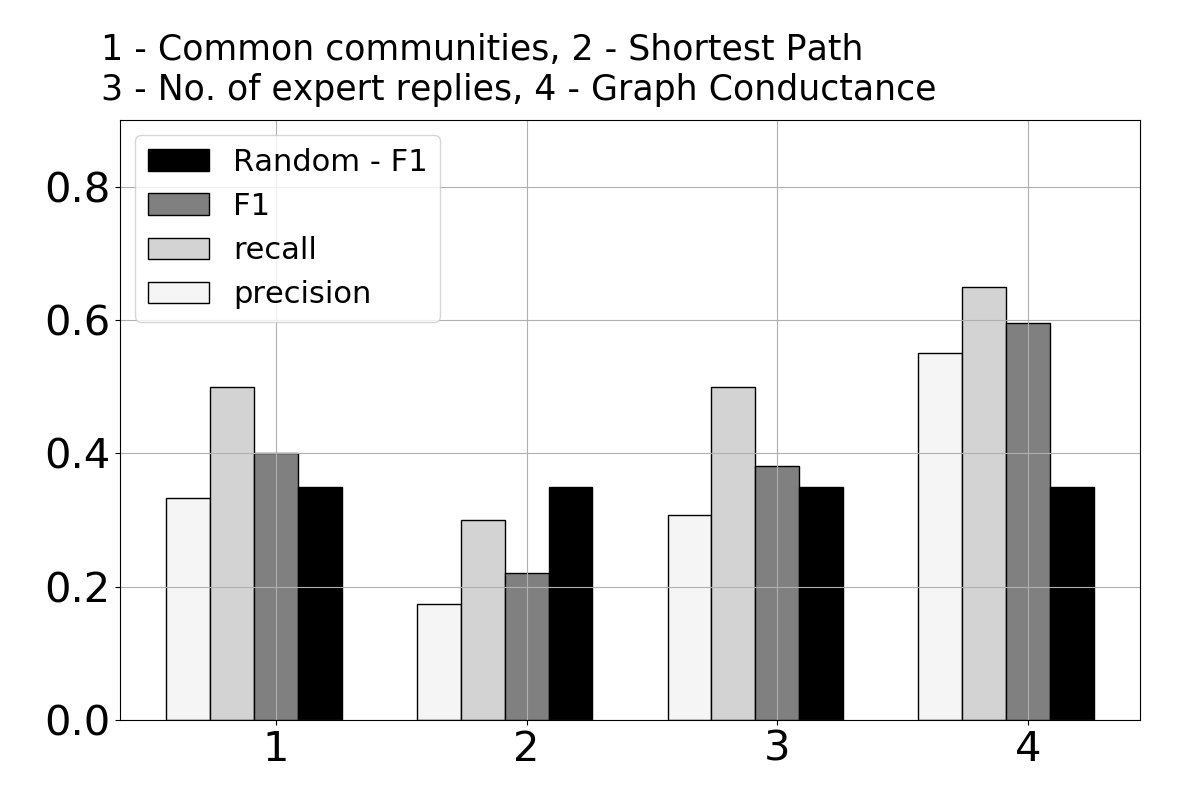}
		\subcaption{malicious-email}
		\endminipage
		\hfill
		\minipage{0.5\textwidth}
		\includegraphics[width=4.2cm, height=2.8cm]{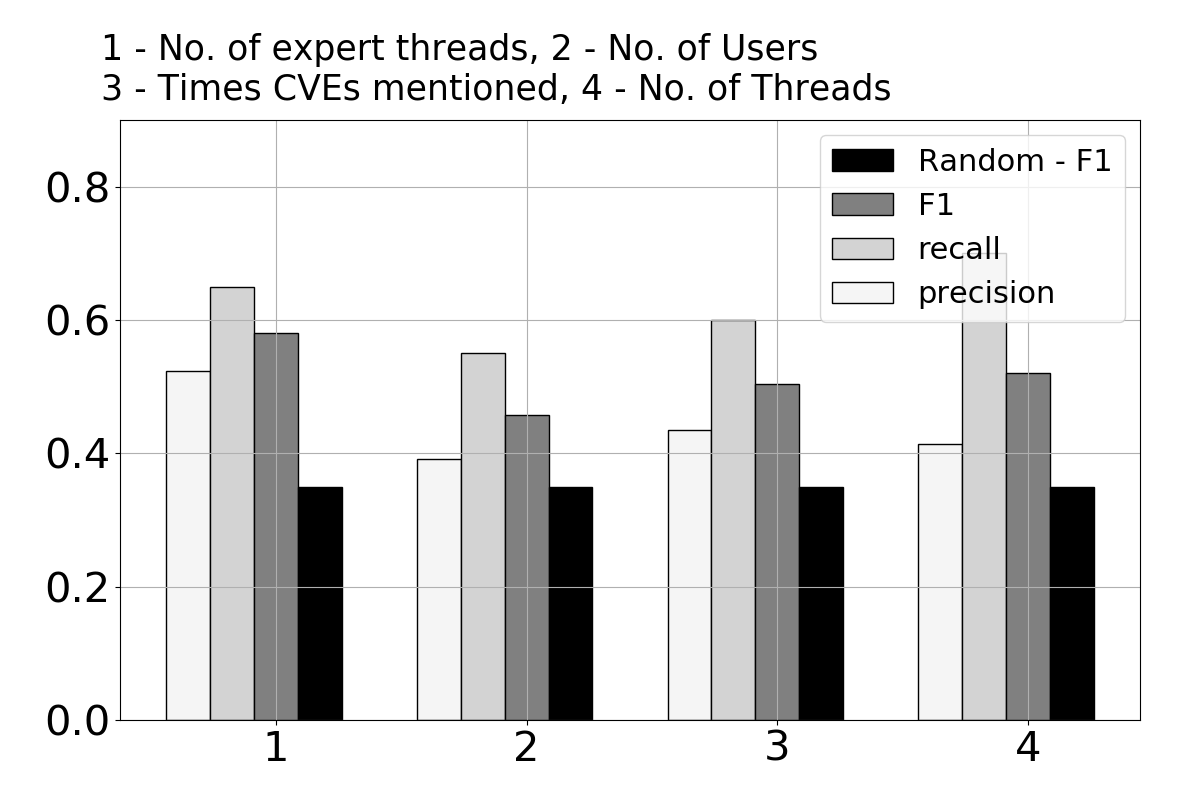}
		\subcaption{malicious-email}
		\endminipage
		\hfill
		\\
		\minipage{0.5\textwidth}
		\includegraphics[width=4.2cm, height=2.8cm]{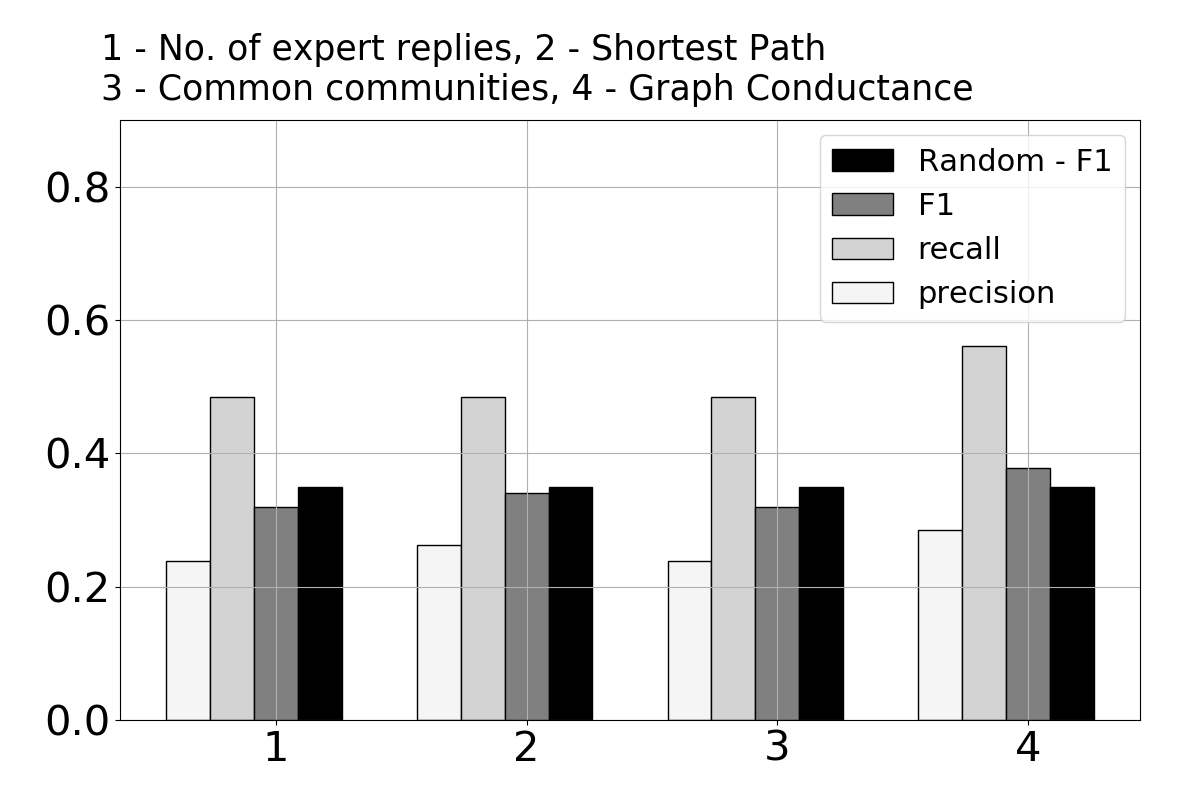}
		\subcaption{endpoint-malware}
		\endminipage
		\hfill
		\minipage{0.5\textwidth}
		\includegraphics[width=4.2cm, height=2.8cm]{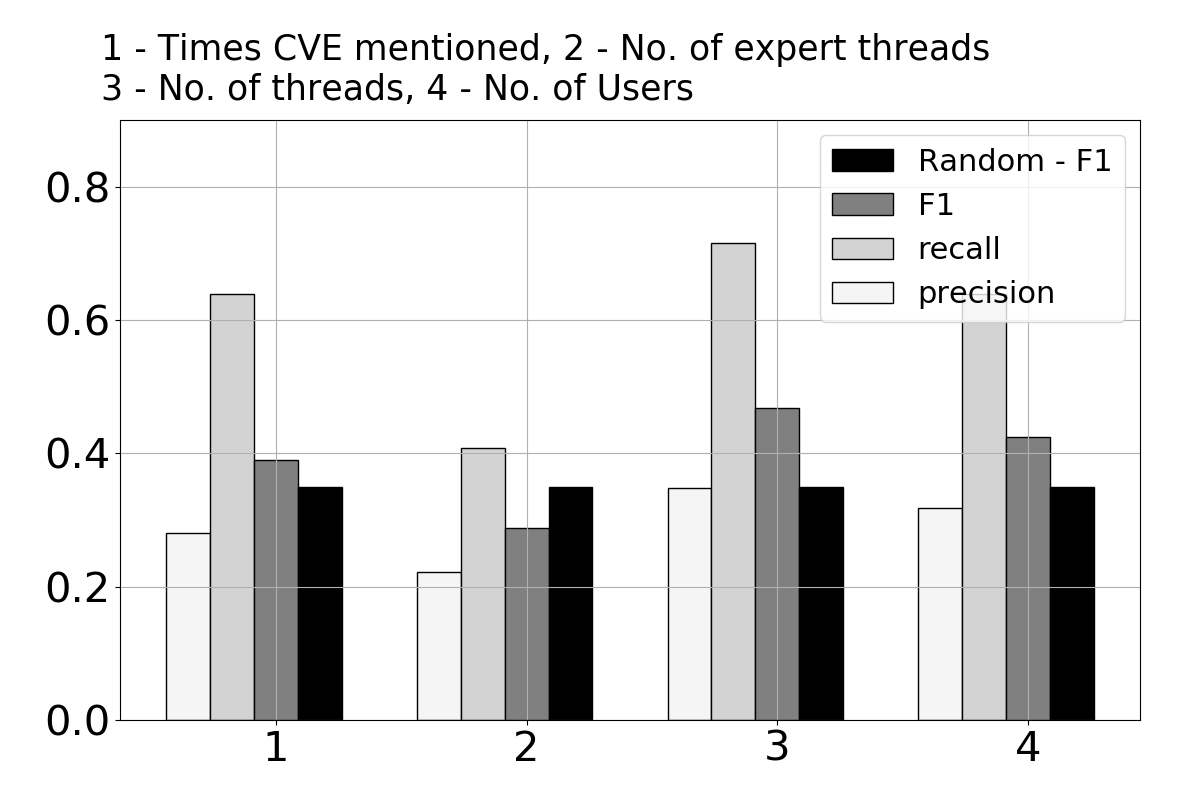}
		\subcaption{endpoint-malware}
		\endminipage
		\hfill
		\caption{Classification results on Dexter events for the features considering the supervised model: $\delta$ = 7 days, $\eta=8$ days.}
		\label{fig:class_dexter}
	\end{figure*}
	
	\subsection{Experiments with another security breach dataset}
	One of the reasons behind using Armstrong dataset as our ground truth data is the length of the time frame over which the attack data was available - not just the number of attack cases reported (one could have a lot of attack cases reported for only a few days). Since we are attempting a binary classification problem, the more spread the attacks are, the more training point we have for our models and test points for evaluation. However, as a complementary experiment on the learnability of the model parameters specific to companies, we test the prediction problem on a dataset of security incidents from another company named Dexter.As shown in Figure~\ref{fig:types_events}, the distribution of attacks over time is different for the events. We observe that compared to the Armstring dataset, the time span for which the attack ground truth data is available is much shorter - we obtained around 5 months of attack data for the 3 events shown in Figure~\ref{fig:types_events_dexter}, starting from April 2016 to August 2016. We have 58 distinct days with at least one incident tagged as \textit{malicious-destination}, 35 distinct days tagged as \textit{endpoint-malware} and 114 distinct days for \textit{malicious-email} events. We had a total of 565 incidents (not distinct days) over a span of 5 months that were considered in our study which is twice the number reported for Armstrong. However, compared to the data spread over 17 months obtained from Armstrong, we have only 4 months to train and test using Dexter data.
	
	We use the same attack prediction framework for predicting the attacks on Dexter, the results of which are shown in Figure~\ref{fig:class_dexter} - we obtain the best F1 score of 0.6 on the malicious email attacks using the graph conductance measure and an F1 score of 0.59 using the expert threads statistics forum metadata feature (refer to Table~\ref{tab:table_feat}) against a random F1 score of 0.37. This suggests that the network features which on how experts reply to posts from regular users can be useful in obtaining improved results over other features which do not consider this reply path structure.
	
	\section{Conclusions and Future Work}
	In this study, we attempt to empirically argue whether the reply network structure from the darkweb discussions could be leveraged to predict external enterprise threats. We try to leverage the network and interaction patterns in the forums to understand the extent to which they can be used as useful indicators. Our method achieved a best F1 score of 0.53 for one type of attacks against class imbalanced attack data using Logistic Regression models while being able to maintain high recall. Using an unsupervised anomaly detector, we are able to achieve a maximum AUC of 0.69 by the leveraging the network structure. The main premise of this work is based on using two different datasets to correlate attacks and user interactions - the limitations clearly lie in being precisely able to infer the path to the attack through discussions. This would require some additional mechanisms on leveraging the content to check whether the discussions catered to a particular exploit that caused the attack. But we believe that our framework caters to the general understanding of how user interaction patterns can be mined using attributes related to vulnerabilities and how they can be leveraged to create a framework for attack prediction.

	\section*{Acknowledgment}
	
	Some of the authors are supported through the AFOSR Young Investigator Program (YIP) grant FA9550-15-1-0159, ARO grant W911NF-15-1-0282, and the DoD Minerva program grant N00014-16-1-2015.

\section*{Appendix} 

The outline for the algorithm for creating the social graph $G$ has been described in Algorithm~\ref{alg:create_graph}. \\

\begin{algorithm}[!t]
	\KwIn{Forum posts $P^f$ for forum $f$, time spans $\Gamma = \{\tau_1, \ldots \tau_k \}$, $\mathcal{H} = \{ H_{\tau_1}, \ldots H_{\tau_k}\}$}
	\KwOut{Time series function , $\mathcal{T}^f$ mapping the points in $\Gamma$ to a real value.}

	\For{each $\tau$ in $\Gamma$} {
		$G_{H_{\tau}}$ $\leftarrow$ \textit{Create}($P^f$, $H_{\tau}$) \tcp*{create the historical network using posts from time span $H_{\tau}$} 
		\For{each time index $t$ in $\tau$} {
			$G_{t}$ $\leftarrow$ \textit{Create}($P^f$, $t$) \tcp*{create the curren network using posts from time span $t$} 
			$G_{H_\tau, t}$ $\leftarrow$ \textit{Merge}($G_{H_\tau}$, $G_{t}$) \tcp*{Create the auxiliary network for $t$}
			
			$\mathcal{T}^f[t]$ $\leftarrow$ Feature value for time $t$ considering  $G_{H_\tau, t}$  \;
		}
	}
	return $\mathcal{T}^f$
	
	\caption{Computing the time series function $\mathcal{T}$} 
	\label{alg:create_graph}
\end{algorithm}

\end{document}